\documentclass[onecolumn]{pasjph}
\usepackage{graphicx}

\begin{document}
\Received{2017/07/03}
\Accepted{2017/11/13}


\title{Inflight Calibration of the Hitomi Soft X-ray Spectrometer (2) Point Spread Function}

\author{
Yoshitomo Maeda\altaffilmark{1,2},
\email{ymaeda@astro.isas.jaxa.jp}
Toshiki Sato\altaffilmark{3},
Takayuki Hayashi\altaffilmark{4,5},
Ryo Iizuka\altaffilmark{1},
Lorella Angelini\altaffilmark{4},
Ryota Asai\altaffilmark{3},
Akihiro Furuzawa\altaffilmark{6},
Richard Kelley\altaffilmark{4,7},
Shu Koyama\altaffilmark{1},
Sho Kurashima\altaffilmark{3},
Manabu Ishida\altaffilmark{1,2},
Hideyuki Mori \altaffilmark{4},
Nozomi Nakaniwa\altaffilmark{3},
Takashi Okajima\altaffilmark{4}, 
Peter J. Serlemitsos\altaffilmark{4},
Masahiro Tsujimoto\altaffilmark{1},
Tahir Yaqoob\altaffilmark{4,7}
}
\altaffiltext{1}{Institute of Space and Astronautical Science, Japan Aerospace Exploration Agency, 3-1-1 Yoshinodai, Chuo-ku, Sagamihara, Kanagawa 252-5210}
\altaffiltext{2}{The Graduate University for Advanced Studies, 3-1-1 Yoshinodai, Chuo-ku, Sagamihara, Kanagawa 252-5210}
\altaffiltext{3}{Tokyo Metropolitan University, 1-1 Minami-Osawa, Hachioji, Tokyo 192-0397}
\altaffiltext{4}{NASA/Goddard Space Flight Center, Greenbelt, MD 20771 USA}
\altaffiltext{5}{Nagoya University, Furo-cho, Chikusa-ku, Nagoya 464-8602}
\altaffiltext{6}{Fujita Health University, 1-98 Dengakugakubo, Kutsukake-cho, Toyoake, Aichi 470-1192}
\altaffiltext{7}{Department of Physics, University of Maryland Baltimore County, 1000 Hilltop Circle, MD 21250, USA}

\maketitle

\begin{abstract}

We present results of inflight calibration of the point spread function (PSF)  of the Soft X-ray Telescope (SXT-S) that focuses X-ray onto the pixel array 
of the Soft X-ray Spectrometer system (SXS). We make a full array image of a point-like source by extracting a pulsed component of the Crab nebula emission.
Within the limited statistics afforded by an exposure time of only 6.9~ksec and the limited knowledge of the systematic uncetainties, we find that the raytracing
model of 1\farcm2 half-power-diameter (HPD) is consistent with an image of the observed event distributions across pixels.
The ratio between the Crab pulsar image and the raytracing  
shows scatter from pixel to pixel that is 40\% or less in all except one pixel. 
The pixel-to-pixel ratio has a spread of 20\%, on average, for the 15 edge
 pixels, with an averaged statistical error of 17\% ($1\sigma$). In the central 
16 pixels, the corresponding ratio is 15\% with an error of 6\%. 
\end{abstract}

\section{Introduction}

\label{sec:intro}  

The international X-ray observatory  ASTRO-H (\cite{2016SPIE.9905E..0UT}), later renamed Hitomi, was launched on February 17, 2016(JST). 
Two Soft X-ray Telescopes (SXTs) were mounted on the Hitomi observatory (\cite{2014SPIE.9144E..28S}, \cite{2016SPIE.9905E..0ZO}). 
 The SXT is composed of two independent units called SXT-I and SXT-S, which focus X-rays onto
detectors of the CCD camera of  the Soft X-ray Imaging system (SXI) and of the SXS calorimeter's pixel array 
respectively.  The SXS boasts an unprecedented high energy resolution up to energies of a few tens keV (\cite{2016SPIE.9905E..0VK}). 

A knowledge of the angular response of the SXT-S is crucially important to perform high-resolution spectroscopy of extended sources such as clusters of galaxies.
The angular response of the telescope (1\farcm2 HPD) is broad compared to the size of an SXS pixel, which is 
$0\farcm5\times0\farcm5$. 
Therefore, X-ray photons are spilled out of the pixel and into other pixels. 
The spill also occurs out of the field of view (FOV).
For example, in one of the Hitomi targets, the Perseus cluster, the spill-over photons from adjacent pixels, or regions out of the FOV, significantly contaminate the spectra of the sky region of interest (references in this special issue).  
The calibration of the PSF is needed to estimate the relative flux contributions from surrounding regions and to address the dynamics and plasma temperature of the hot gas in the Perseus cluster (e.g., \cite{2016Natur.535..117H}). 

It is necessary to validate the ground calibration with in-orbit data in case of unexpected systematics in the ground measurements.  
During the ground measurements, we found that the PSF of the SXT-S is mainly dominated by the figure errors of the reflectors and the errors in their positioning (\cite{2016SPIE.9905E..0ZO}). Data for estimating the errors on ground were taken with the X-ray calibration facility at ISAS/JAXA (\cite{2015JATIS...1d4004H}). The errors were examined with ``spot scan" measurements (\cite{2014SPIE.9144E..59S}, \cite{2016JATIS...2d4001S}), in which we illuminate
regions all over the aperture of each quadrant with a square beam 8 mm on a side.  Based on the scan results,  ``maps" of image blurring and the focus position were made. The former and the latter reflect the figure errors and positioning errors, respectively, of the foils that are within the incident 8 mm $\times$ 8 mm beam. As a result, these
errors in a quadrant were estimated to be $\sim$0\farcm9 to 1\farcm0 and $\sim$0\farcm6 to 0\farcm9, respectively.  It is found that the larger the positioning
error in a quadrant is, the larger its HPD is. The HPD map, which manifests itself as local image blurring, is very
similar from quadrant to quadrant, but the map of the focus position is different from location to location in each
telescope. 
These errors are incorporated into the input calibration files for the raytracing software  {\texttt xrtraytrace} (\cite{2016SPIE.9905E..14A}, \cite{Yaqoob2017a}).

Due to the unexpected short life of the Hitomi satellite,
the gate valve is placed at the entrance of the detector Dewar and was still kept closed during the observations. It consists of a beryllium window of a thickness of about 260~$\mu$m, which is covered by a stainless mesh  with a 0.2
mm thick with 71 \% open fraction. The window is supported by an opaque support structure that is approximated by a criss cross with a surrounding ring. 
Since the ground-based measurements were made without the gate valve, we cannot compare the measurements with the inflight data directly.
We then modify the results of the raytracing program by incorporating the gate valve obstruction
structures, and calculate the image at the focus.
We make an inflight calibration by comparing the inflight image with images calculated with the raytracing code.

Here, we present inflight PSF calibration results for the SXT-S. See \citet{Nakajima2017a} for the SXT-I.  The effective area calibration of the SXT-S is presented separately in \citet{EA2017a}. 

\medskip
\section{Observations} 
\label{sec:obs}

Inflight calibration of  the PSF was carried out with the data from the Crab nebula  (catalogue designations M1, NGC~1952, Taurus A, SN~1054) . 
The Crab nebula was observed on 2016 March 25 from 12:35 to 18:01 UT. It was imaged at around the best focus (i.e., the optical axis) of the $6\times6$ pixel array of the SXS.  
The total observation time span was 21.5 ks,
whereas the total on-source time was 9.7 ks. 

The observation was made in a normal mode but with the gate valve closed.  
The gate valve reduces the numbers of
incoming photons that can reach the focal plane with an energy dependence, the
attenuation being strongest below $\sim 4$~keV due to the beryllium absorption, and above $\sim 10$~keV due to the structure blocking. The actual incoming rate onto the detector array in the 0.3--30 keV band was 13\% of what we would expect if the Crab nebula was observed with the gate valve open.


\section{Method \& Analysis}\label{dataana}

The PSF of a focusing telescope system is usually constructed using a bright and point-like source. 
The brightness reduces the uncertainty of the background subtraction in the PSF tail so
that the dominant errors are due to Poisson statistics, 
whereas the point-like nature of the source eliminates broadening of the sharp core due to
source spatial structure.   
For example, the calibration of the Suzaku XRT-I was made using SS Cyg, which has a brightness
on the order of 10 milli Crab (\cite{1995PASJ...47..105S}).  
However, due to the limited number of targets observed by the Hitomi mission, 
no bright and point-like standard source was observed. 
It forced us to calibrate the PSF using the pulsed component of the Crab nebula, i.e., the Crab pulsar. 
Located close to the center of the nebula, the Crab pulsar has a spin rate of $\sim34$ milliseconds and,
is viewed as a point-like source. 
By just picking up the pulsed component, we can extract an image of the point-like source, ``the Crab pulsar". 
We calibrate the SXS PSF by using the Crab-pulsar image. 

During the ground calibration, the image of the SXT-S was over-sampled by the CCD detector with a 22.5$\mu$m pixel size,
which corresponds to $\sim$0\farcs7 (\cite{2016JATIS...2d4001S}). 
Using the image, the Encircled Energy Fraction, or EEF, was calculated 
as a function of radius, by deriving the energy enclosed by a circle of a given radius centered on the centroid, as a fraction of the energy enclosed within some normalising radius. The centroid was defined as the image peak. The HPD of 1\farcm2 was derived using this EEF. 
However, the SXS pixel has a size of 800~$\mu$m, which corresponds to 0\farcm5. The image is under-sampled in the flight data. 
The HPD is not easy to calculate using the inflight data. We thus do not use the HPD or EFF to characterize 
the inflight PSF, but instead using the image directly (in other words, 
we examine the pixel-by-pixel probability distribution, or normalized intensity distribution). 


The usage of the pulsed component of the Crab has two particular merits for the PSF calibration. One is negligible systematic uncertainty for background subtraction, and the other is elimination of a dust-scattering tail. 
The background contains both the detector and sky contributions. The sky background can vary as a function
of the direction of the pointing due to the flux fluctuation across the sky. 
The detector background can vary with time since it can be changed with the orbit etc. 
Therefore, the outer region of the detector of the same observation 
(i.e., at the same time and at the same sky position) is sometimes chosen as a background. 

The SXS is a narrow-field instrument, in which the square 
array only covers an area 3.\farcm05 wide, whereas the HPD of the SXT-S measured on ground is 1\farcm2, or
about one-third of the array size. A spill-over flux beyond the edges of the array is still not negligible. 
The over-subtraction of the PSF tail can be a concern for its calibration if we use a corner pixel as a background. The array of the SXS is only six by six pixels.  Picking up the corner pixels as a background region,  therefore, sacrifices three of the 35 pixels in the array (about 10\%).  

Our approach is to use the Crab off-pulse component as a background. The off-pulse component contains the instrument background and sky background that is exactly equivalent to that in the on-pulse component. 
We can just remove the instrument background, and sky background as well as the off-pulse component by simply subtracting the off-pulse component of its own pixel. The phase-resolved analysis allows us to make a point-source image for the whole array of the SXS without sacrificing any pixels.  

As for the elimination of the dust scattering tail, 
the dust scattering is known to make a halo structure around the source with 
a size that is an arcmin wide or wider (\cite{1965ApJ...141..864O}). Once the emission from the pulsar is scattered by the interstellar dust/grain in front, it experiences a time lag due to a detour of the arrival path, and
this makes the emission unpulsed. Therefore, the pulsed component is almost free from the dust scattering halo and should be a purely point-like source. 
Although standard PSF measurements using bright point-like sources would also work for the SXS calibration if they would have been made, the measurements using the Crab pulsar give us a good opportunity to calibrate the SXS PSF. 

The data of the Crab nebula are first screened with the standard criteria.  In addition to the standard screening, we have two effects to be addressed for the PSF calibration. 
 One is a pixel live-time (or exposure map) and the other is the satellite pointing stability. Both effects make the image distorted.  
 Note that the pixel live-time correction applied here is not directly supported by the standard software when an SXS image is made (\cite{2016SPIE.9905E..14A}). However, the exposure map file made by the tool ``{\it ahexpmap}'' does contain an exposure time map and list of pixel fractional exposures.
 The detailed procedure of the pixel live-time correction is presented in \citet{EA2017a}. 
 We describe the corrections and screening with which we extracted a point-source image. 
 
The total throughput of the detector array for bright sources such as the Crab nebula is determined by the CPU processing speed of the onboard digital electronics called the Pulse Shape Processor (PSP; \cite{2016SPIE.9905E..3TI}). The PSP consists of two identical units (PSP-A and PSP-B), and each unit has one FPGA and two CPU boards. A total of 36 pixels is processed independently with no priority for a particular pixel.
Inner pixels close to the PSF center receive more counts than the
 surroundings, and the pixels with more counts suffer a larger fraction of the dead
 time. This causes an image distortion. All dead times are recorded, thus this can be
 corrected.

Dead times are only significant in the 10 pixels close to the center. In these pixels, events can be lost by dead times or inappropriate flagging due to overlapping of multiple pulses. The fraction of lost events, after correcting for the dead time, is estimated to be $\sim$10\% for these pixels. Other pixels have live time fractions greater than 99\%, thus the effects in those pixels are negligible.

\begin{figure}
 \begin{center}
\ \ \ \ \ \ \ \  (a) Raw	 \hspace{5.5cm} (b) Live-time corrected \\
\includegraphics[width=160mm,bb=42 257 568 533]{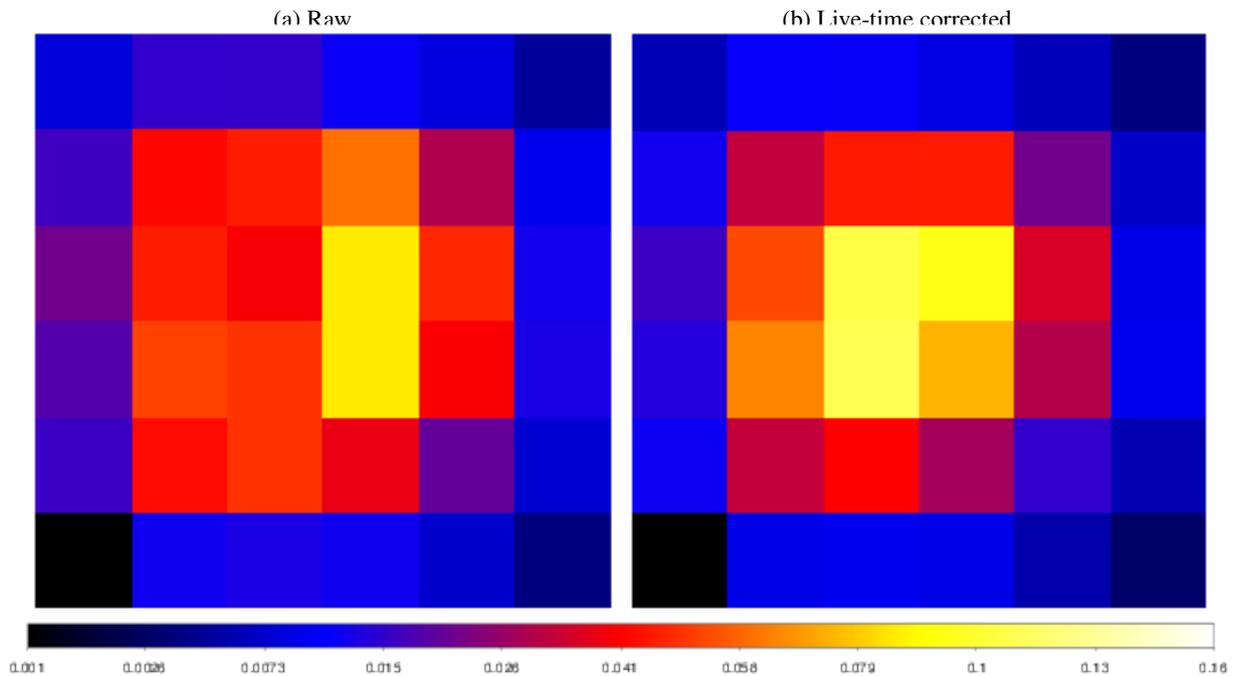}
 \end{center}
  \caption{SXS images of the Crab nebula in DET coordinates. (a) the raw image. (b) the live-time corrected image. Each pixel value is normalized by the total count rate. The bottom-left corner is identified with the calibration pixel and is blackened. }
\label{f05}
\end{figure}

 By using the events of all grades, we made a raw image in detector coordinates (``DET system'':Figure~\ref{f05}). By applying the pixel-by-pixel live-time correction to the image, we made the flux image that is also shown in Figure~\ref{f05}. 
It was found that in the raw image the events in the inner $4\times4$ pixels have
a roughly flat spatial distribution.  
After the live-time correction is applied, the peak then appears at around the center of the array. 



Even after the  live-time correction was made, the event distribution actually fluctuated 
because of the satellite pointing
stability, which primarily depends on whether or not the Star Tracker (STT) was used. 
To suppress the effect of the telescope pointing fluctuation, we only used events occurring during intervals
in which the telescope was controlled by the STT (6.9 out of 9.7~ks exposure). Consequently, a higher
stability of 4\farcs7 at a 1 $\sigma$ level was achieved\footnote{For details, see Hitomi-MEMO-2016-001 at https://heasarc.gsfc.nasa.gov/docs/hitomi/calib/hitomi\_caldb\_docs.html.}.

\begin{figure}
 \begin{center}
  \includegraphics[width=120mm,bb=0 0 842 595]{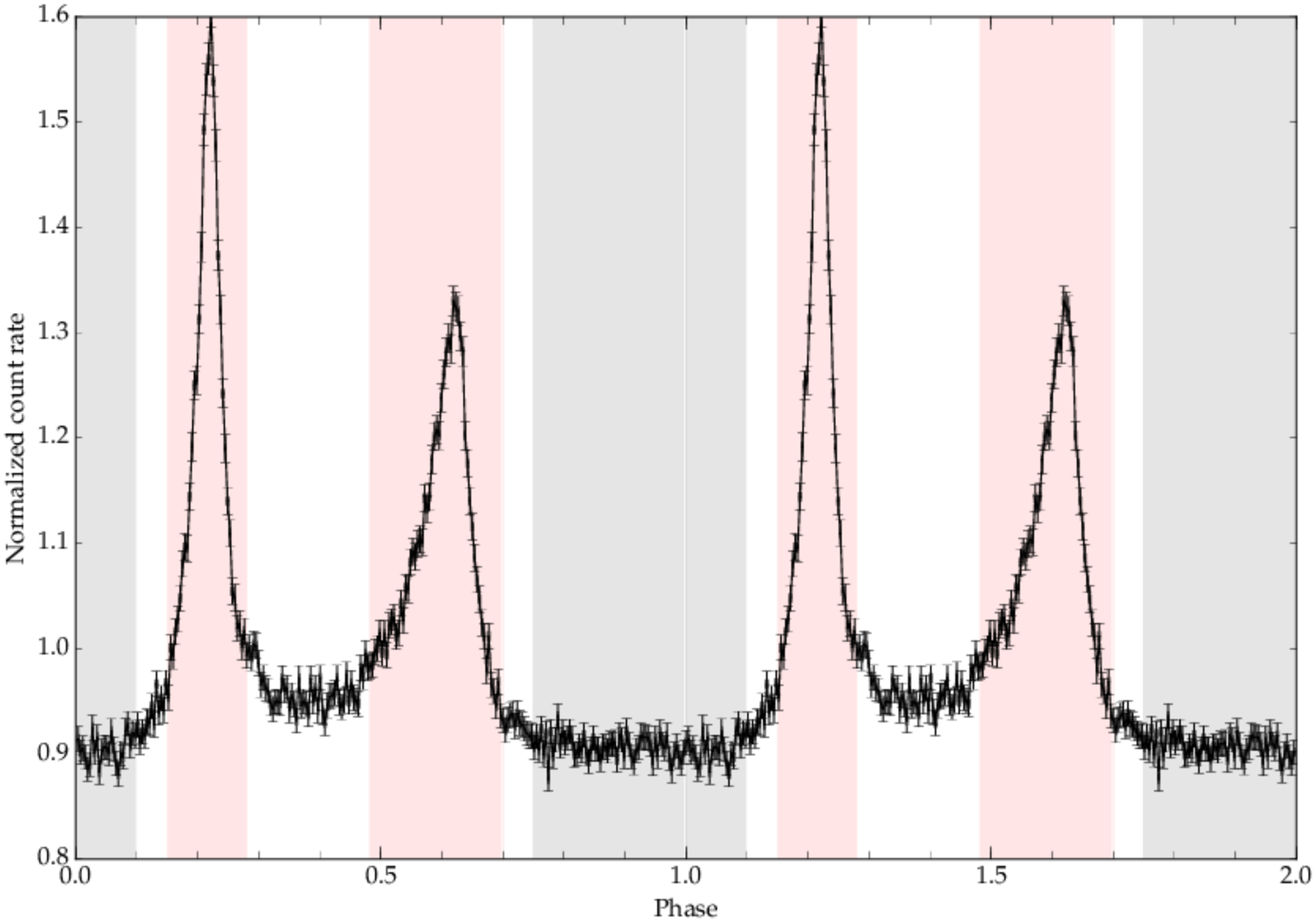}
 \end{center}
 \caption{Folded light curve with a 33.7204626~ms period, using events in the
 2--20~keV band. Red and grey shades define the on- and off-pulse phases respectively, each of
 which occupies 35\% of a phase.}
 \label{f04}
\end{figure}

We fold the events with a period of 33.7204626 ms at the epoch of 17472 days (TJD). 
Figure~\ref{f04} shows the light curve folded with the pulse period after a barycentric
correction. Events with all grades in the 2--8~keV were used. 
The pulses are clearly detected in the folded lightcurve.
The relative timing
accuracy is as high as $<$80~$\mu$s for the SXS  (\cite{Koyama2017a} in prep.).
All pixels suffering dead times have a dead
time duration longer than 5 s (see details in \cite{EA2017a} in prep.). Therefore, the 34 ms pulse profile
is not distorted by the dead time.



Figure~\ref{f07} shows the on-minus-off pulse image of the Crab nebula, i.e., the Crab pulsar,
and corresponds to the in-orbit PSF.  The pulse-on phase was defined as 35\% of a cycle
containing the first and second peaks, whereas the pulse-off phase was defined outside
of the two peaks and the bridge between them. The number of events in the 2--8~keV band, of
all grades in each pixel, was computed during the on- or off-pulse phases, and the
difference of the two was divided by the exposure time of the pixel. The exposure time
has no dependence on the pulse phase and the average can be used for both phases.

\section{Discussion}\label{s5}

\begin{figure*}
 \begin{center}
(a) The Crab pulsar \hspace{5cm} (b) Ground measurements  \ \ \ \ \ \ \ \ \ \ \ \\
\includegraphics[width =80mm, bb=0 0 418 426]{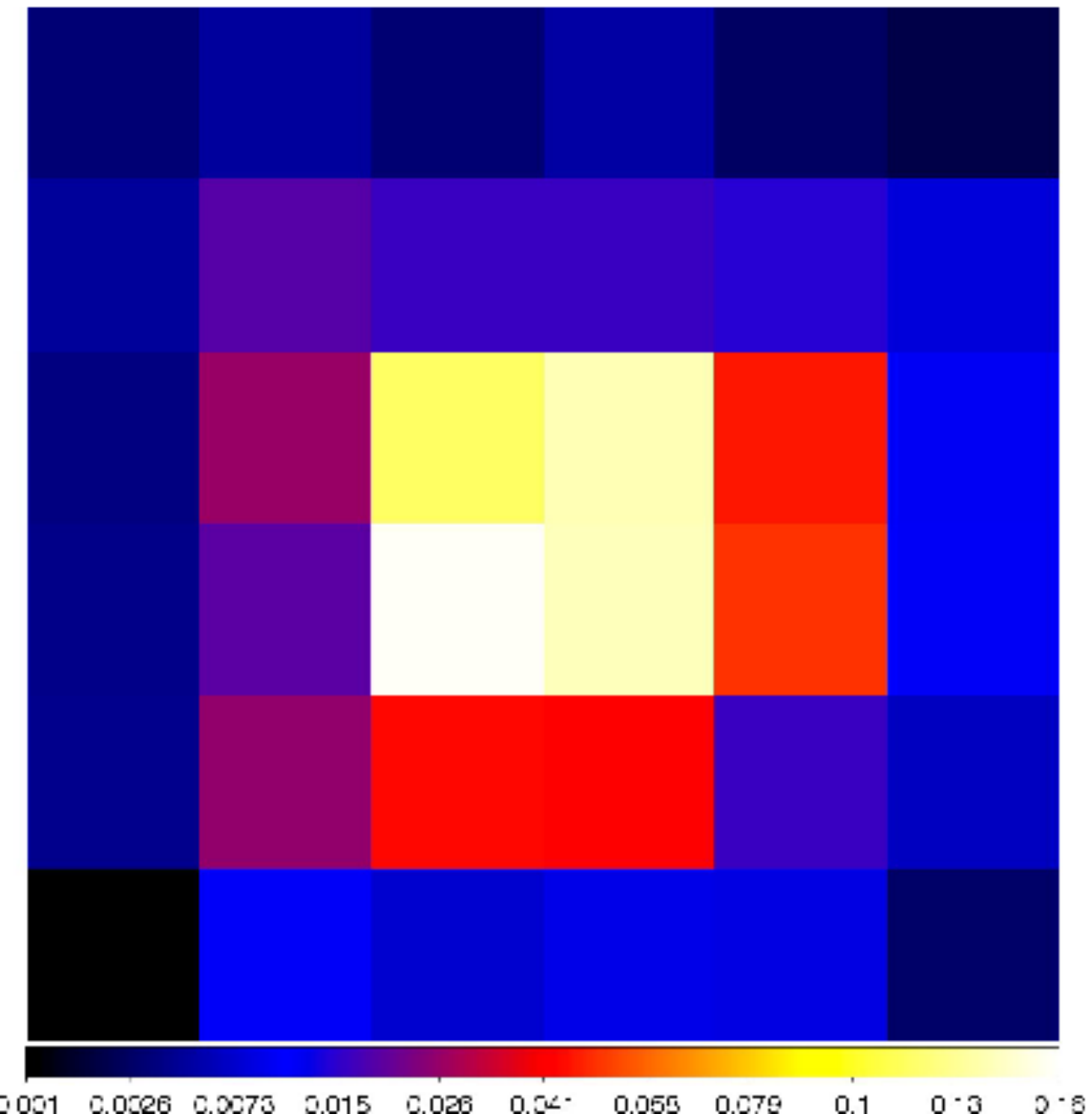} \ 
\includegraphics[width =80mm, bb=0 0 418 426]{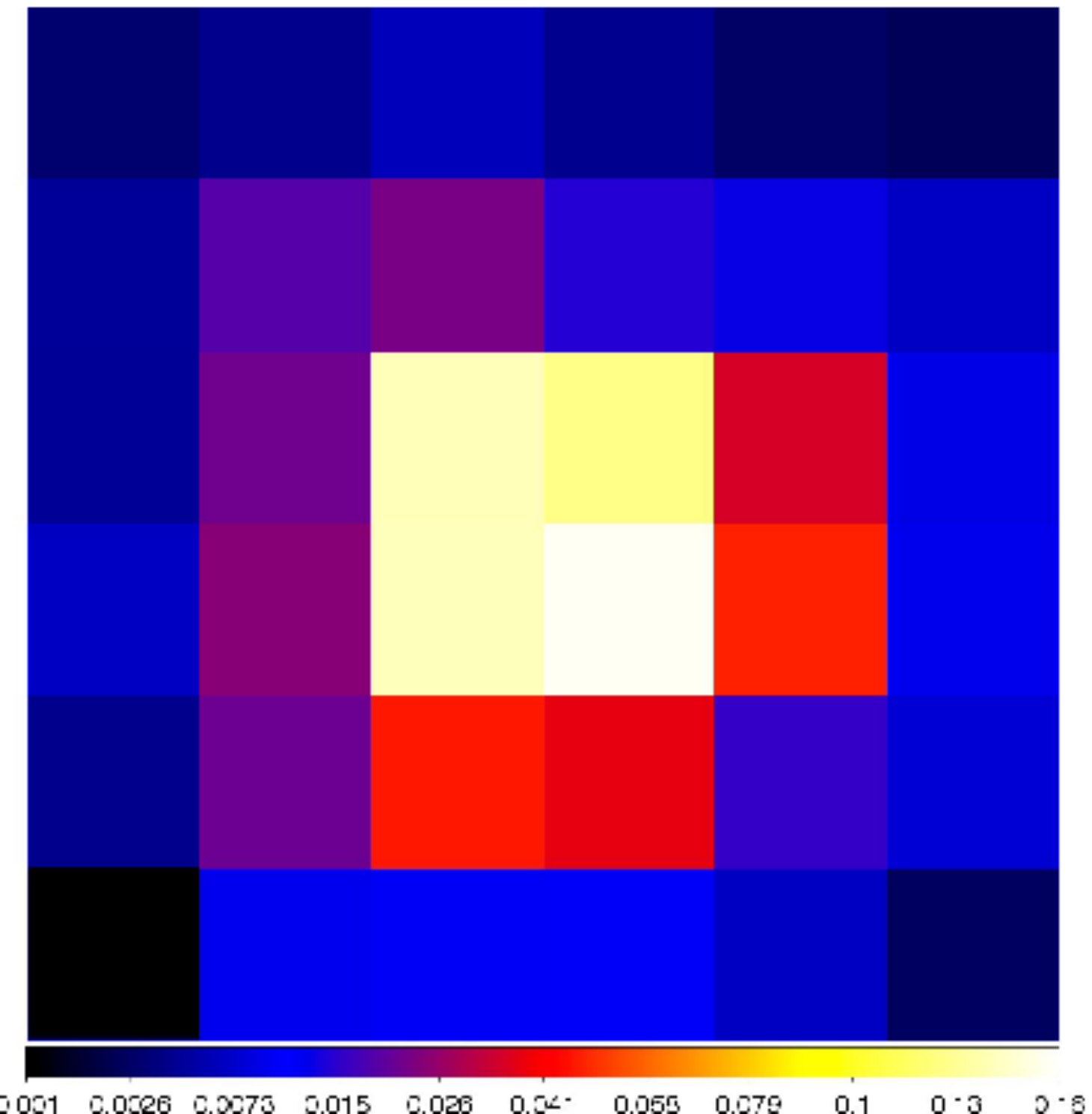}\\
\vspace{1cm}
\ \ \ \ \ \ (c) The raytracing model with the GV  \hspace{1.5cm} (d) The raytracing model without the GV   \ \ \ \ \ \ \ \ \ \\
\includegraphics[width =80mm, bb=0 0 418 426]{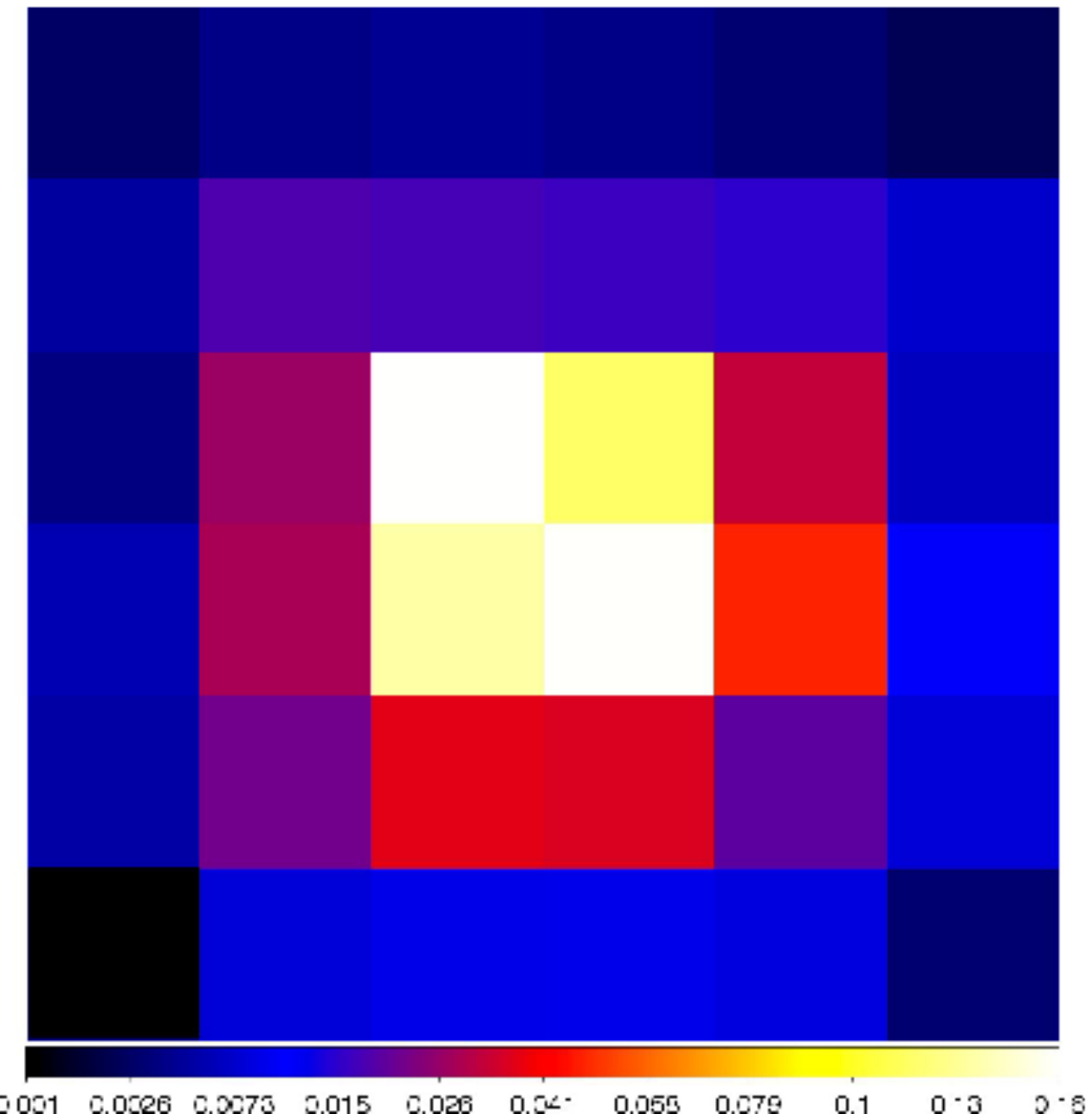}\
\includegraphics[width =80mm, bb=0 0 418 426]{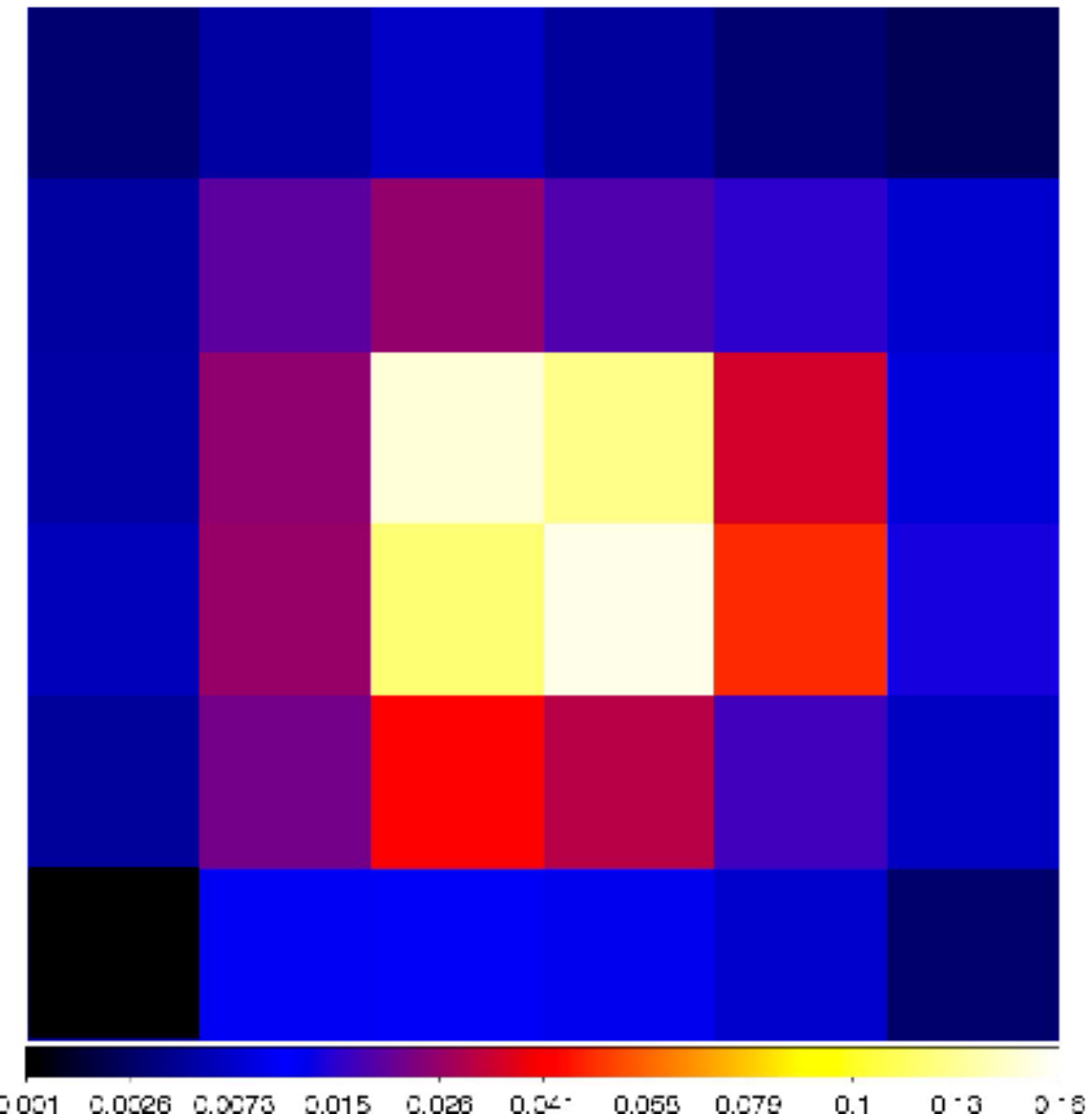} \\
 \end{center}
 \caption{(a) Crab pulsar image from the SXS pixel array. (b) Point source image measured on ground. (c) Raytracing image
 with the gate valve structure. (d) Same as (c) but without the gate valve. Each pixel value is normalized by the total count rate.}
\label{f07}
\end{figure*}

Owing to the high brightness of the Crab pulsar, the pulsed component provides us a full-array 
image for inflight calibration of the PSF (figure~\ref{f07}).  
We here compare the Crab-pulsar image with the computed PSF by the raytracing program {\texttt xrtraytrace} (\cite{Yaqoob2017a} submitted). The {\texttt xrtraytrace} tool is called by the ftool {\texttt aharfgen}, which is used for PSF calculation for the scientific analysis of data, and for calculating the telescope and detector efficiency portions of the response function (Ancillary Response File : ARF)  (\cite{2016SPIE.9905E..14A}).  

In order to calculate the PSF with the raytracing software {\texttt xrtraytrace}, we made three assumptions as follows. The first concerns the tilt of the telescope, i.e., the orientation of the on-axis direction. We assumed that the optical axis is aligned to measurements made in ground experiments. This assumption is confirmed to be valid since the effective area of the SXS at the nominal position is consistent with that expected from the on-axis assumption (\cite{2017SPIEmaeda}).   

\begin{figure}
 \begin{center}
\includegraphics[width=160mm,bb=0 0 1024 768]{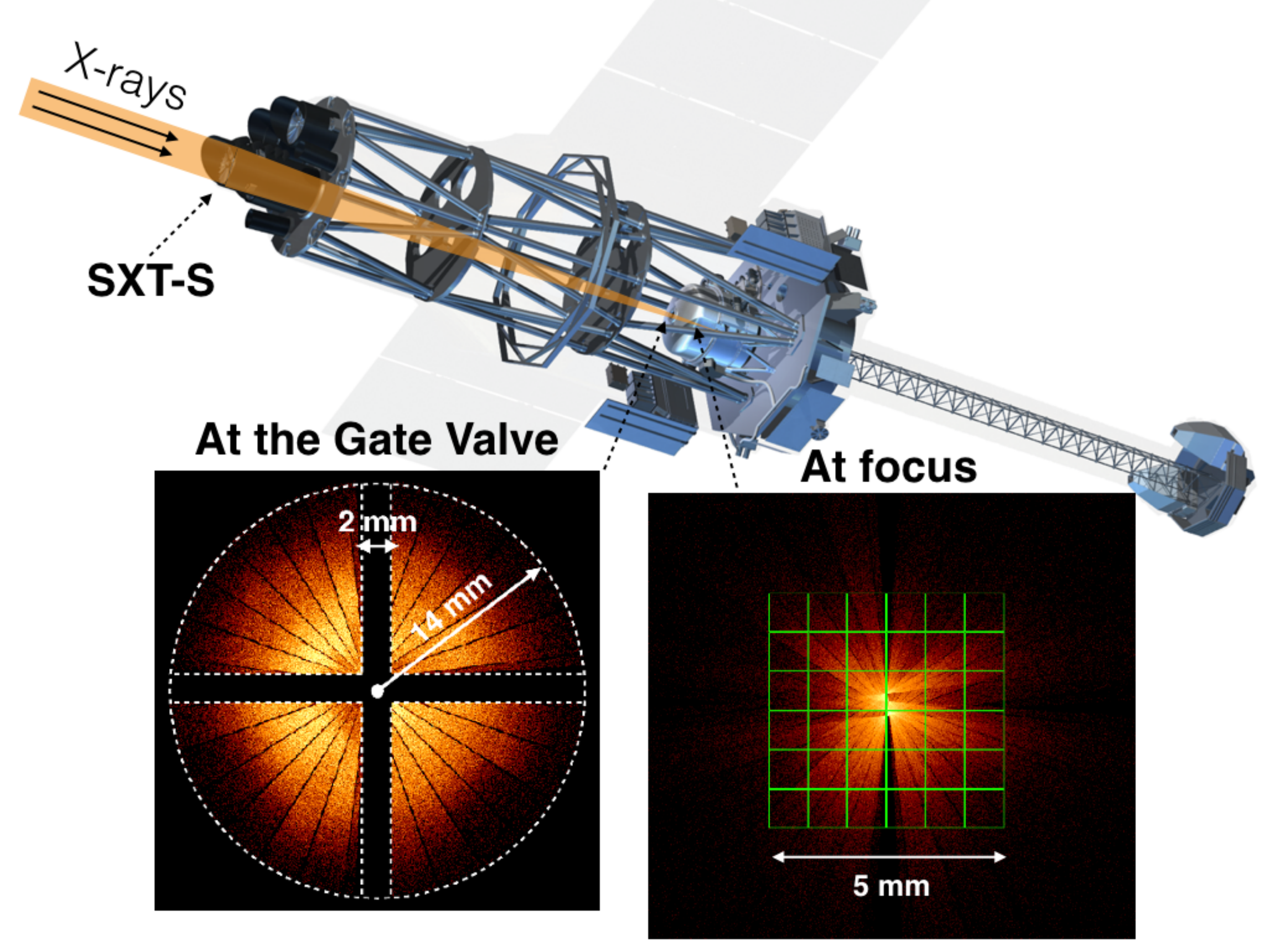}
 \end{center}
 %
\caption{Simulated SXT-S images at the Gate valve and the focal position. Green boxes at the focus show the SXS pixels. The structural model of the Hitomi observatory is overlaid. The incident X-rays are collimated by the SXT-S mirror assembly and are partially blocked at the gate valve on the off-focus position before they arrive at the detector array on the focal plane. 
The cross-bar obstruction structure of the gate valve is aligned with the orientation of the quadrant boundaries of the SXT-S. 
The obstruction structure of the gate valve including the cross bar is hardly discernable in the under-sampled $6\times6$ image at the focal plane. }
\label{f06}
\end{figure}

Second is the modeling of the obstruction structure between the SXT-S and the pixel array at the focal plane. 
The dominant structure is the gate valve of the detector entrance (figure~\ref{f06}).


In the figure, we assumed that the center of the gate valve structure is just aligned with the
telescope symmetry axis and the center of the SXS detector.\footnote{This assumption is not exactly the same as 
that made in the standard software (\cite{2016SPIE.9905E..14A}).  In the software, the structure is offset in DETX and DETY from the detector center $=$ $-$0.051mm and $-$0.221 mm. The effect to the image by the $\sim$0.2 mm shift is discussed later. } 
This assumption also applies to the ground measurements, namely that the structure was aligned within $\sim$0.2~mm. 
At the focus, little spatial information about the obstruction structure remains. 

Third is the focus position at the pixel array.  In~figure~\ref{f06}, we show the focal-plane image of the photons.  The pixel size of the array is about 0\farcm5. It is seen that the pixel size is not fine enough to resolve the sharp PSF core of the focused image. 
In order to find the focused position on a sub-pixel scale, we make a cross correlation of the in-orbit image with the raytracing image. The raytracing image is shifted on a sub-pixel scale of 10$\mu$m or 0\farcs37 and is binned to the pixel scale, but eliminating the pixel gaps.  We adopted the best-correlated position of the two images, which appears at the array center within 0\farcm02. 

\begin{figure}
 \begin{center}
\ \ \ \ \ \ \ \ (a) {\Large $\frac{{\rm The\ Crab\ pulsar\ image}}{{\rm The\ raytracing\ image\ with\ GV}}$} \hspace{1.5cm} (b) {\Large $\frac{{\rm The\ ground\ measurements}}{{\rm The\ raytracing\ image\ without\ GV}}$} \  \ \ \ \ \ \ \ \  \\
\includegraphics[width =80mm, bb=0 0 418 426]{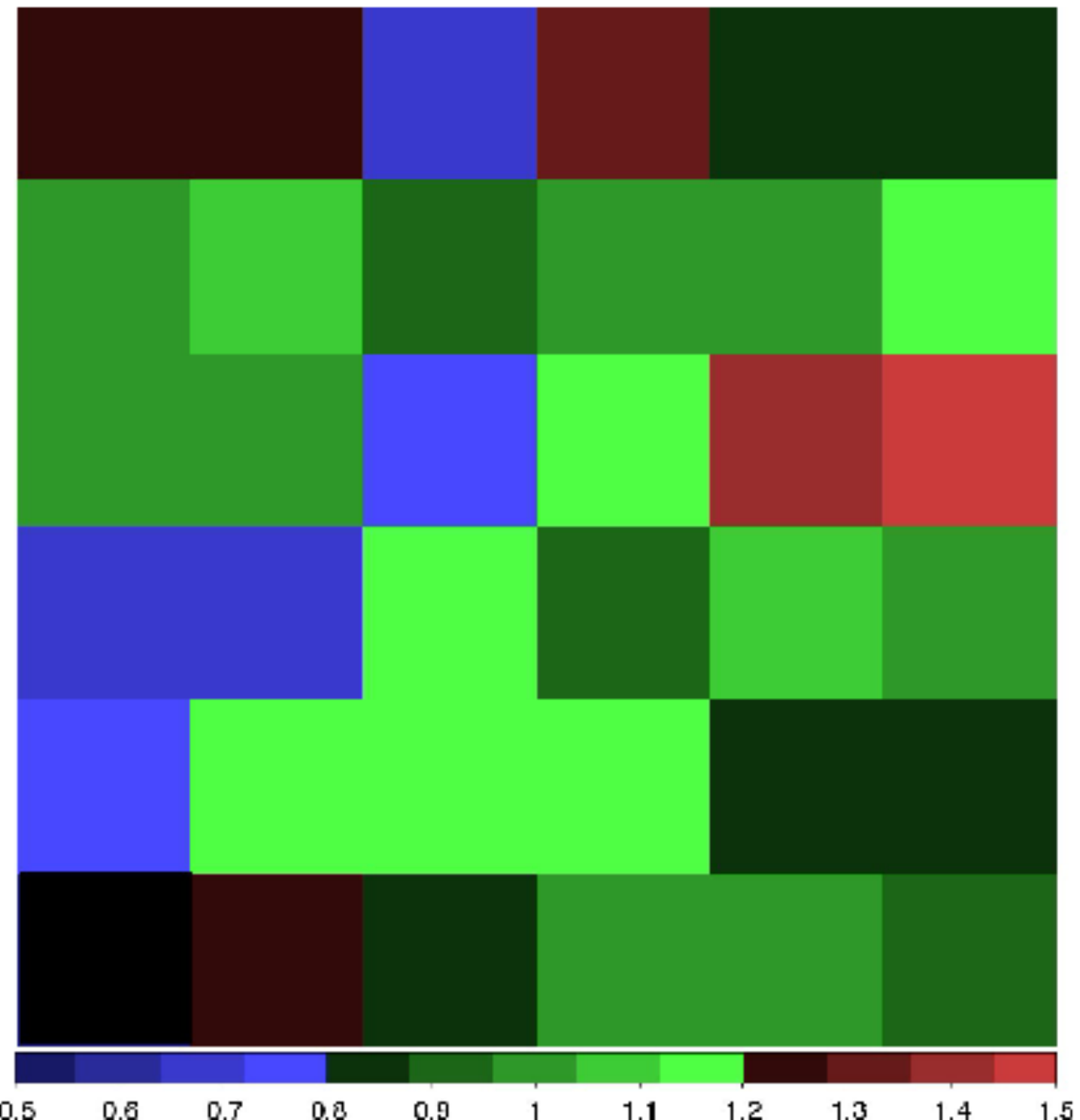} \
\includegraphics[width =80mm, bb=0 0 418 426]{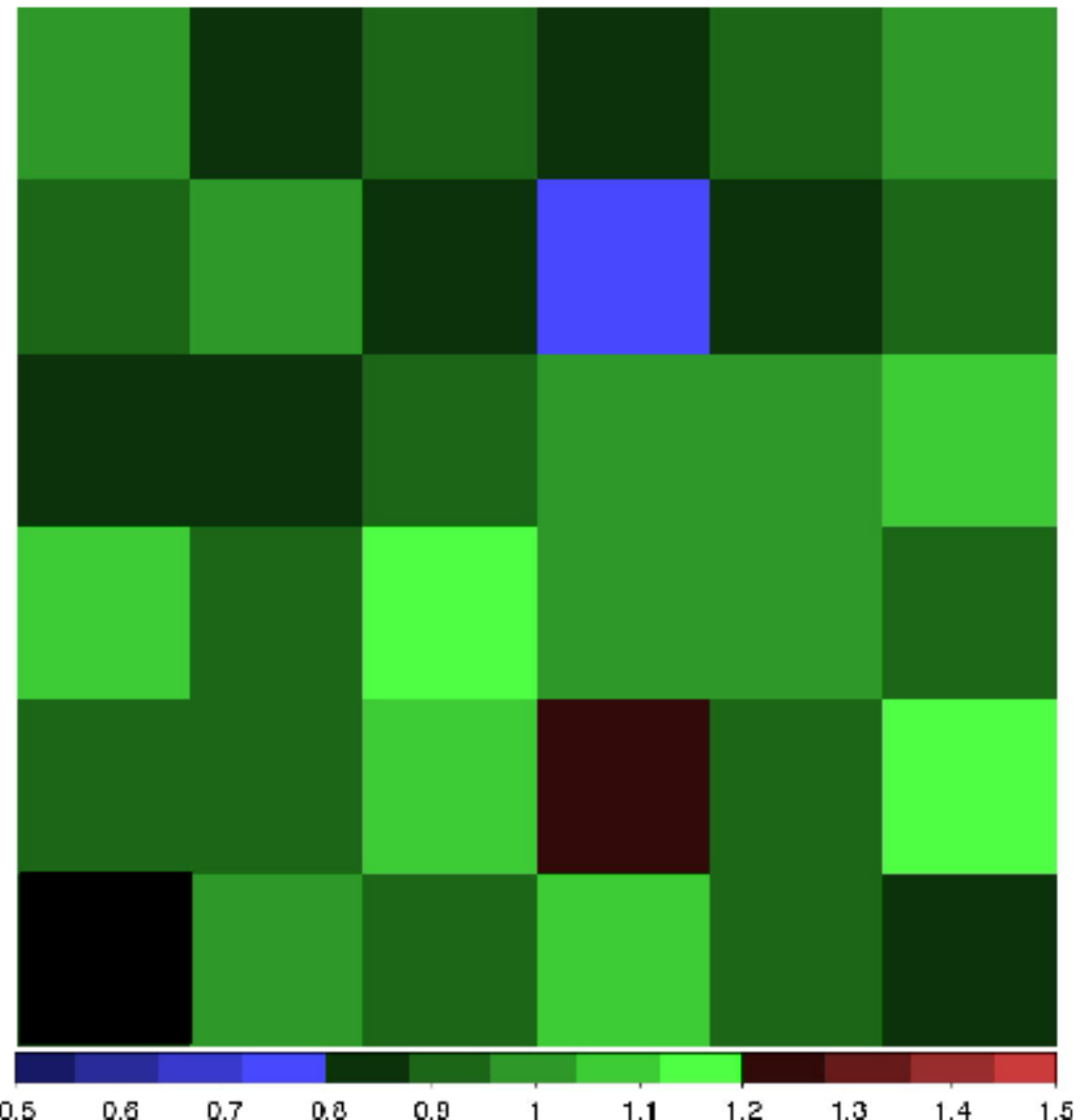} 
 \end{center}
 \caption{(a) Ratio between the Crab pulsar image (figure \ref{f07}a) and the raytracing image with the gate valve structure  (figure \ref{f07}c.).
(b) Ratio between the ground measurements (figure \ref{f07}b.) and the raytracing image without the gate valve structure (figure \ref{f07}d.)
}
\label{f08}
\end{figure}

\begin{figure*}
 \begin{center}
  \includegraphics[width =50mm, bb=0 0 489 504]{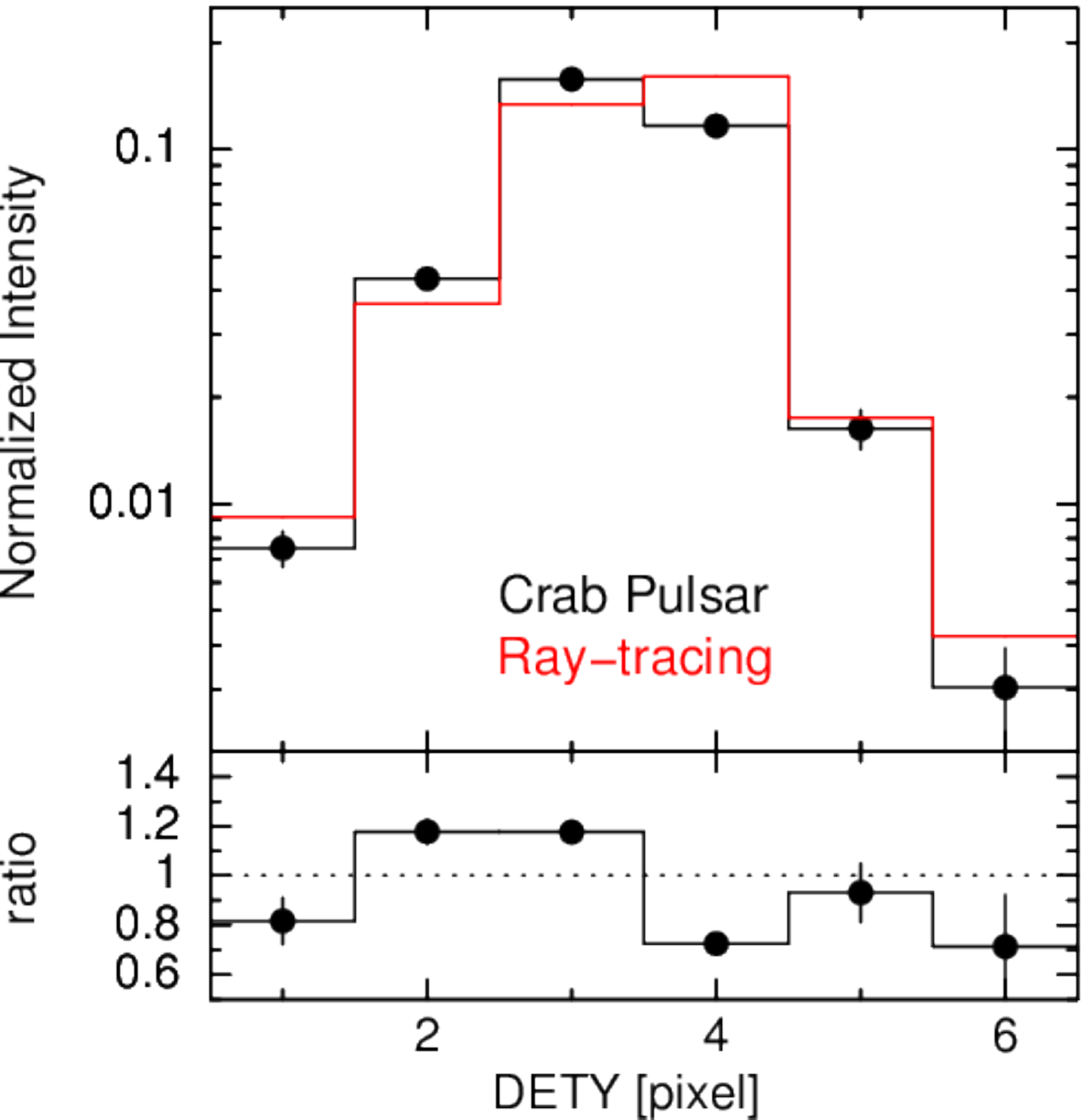}
  \includegraphics[width =50mm, bb=0 0 489 504]{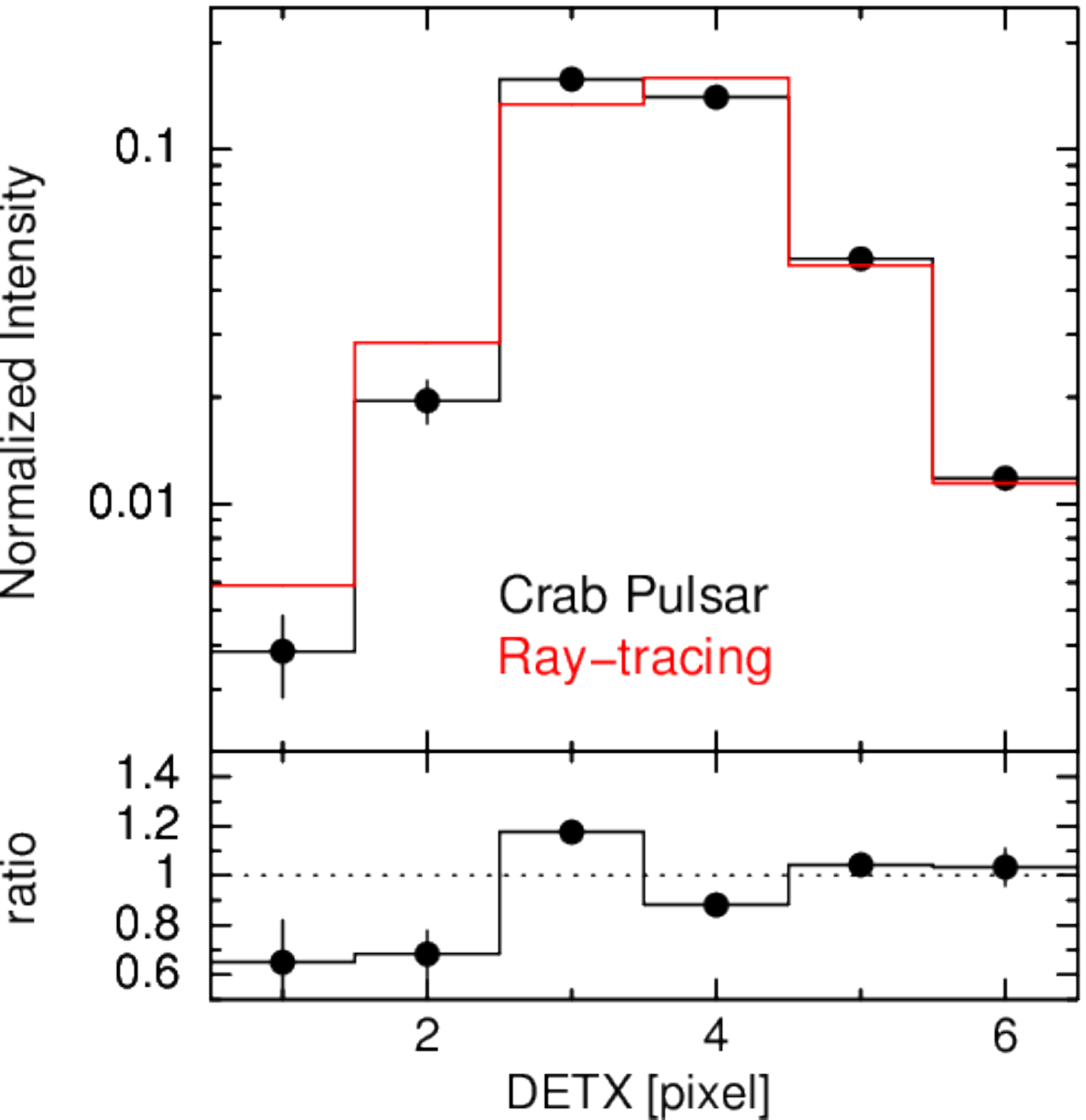}
  \includegraphics[width =50mm, bb=0 0 489 504]{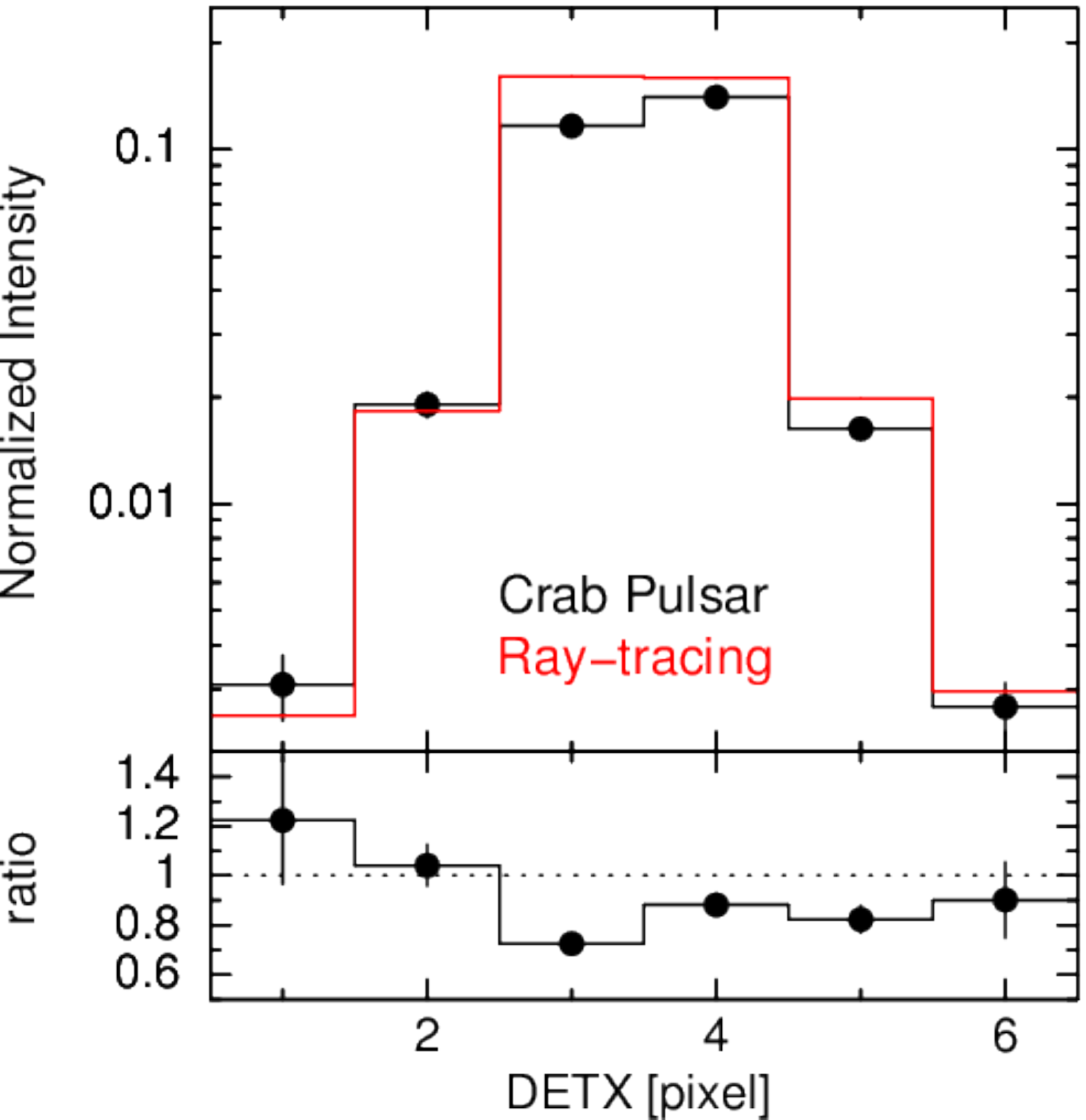}
 \end{center}
 
 \caption{Left: comparison of slice profiles of figure~\ref{f07}(a,c) at pixel $x = 3$, which corresponds to the third line counting from the bottom. Middle: comparison of slice profiles at pixel $y = 3$, which corresponds to the third row counting from the left. 
 Right: comparison of slice profiles ($x + y = 7$ pixel). Black, and red profiles show the Crab pulsar, and the raytracing image
 with the gate valve structure, respectively. Bottom panel shows the ratio between the raytracing image
 and the Crab pulsar image (slices of figure~\ref{f08}a). The error bars correspond to a 1~$\sigma$ statistical level. }
\label{f09}
\end{figure*}

In~figure~\ref{f07}a and c, we show the Crab-pulsar image and the image of the raytracing output respectively.
The detailed numbers of the PSF are also given in table~\ref{tab:psf}.
We calculate the raytracing image using a 1 keV pitch, weighting the result
at each energy with the Crab-pulsar's counting statistics for the respective energy band, and thus make the PSF. 
The overall pattern of the event distribution of the Crab-pulsar image is similar to that generated by the raytracing.  
Figure~\ref{f08}a shows the ratio of the raytracing image to the Crab-pulsar image.
In figure~\ref{f09}, we also show one-dimensional slices of the ratio distributions along particular lines of pixels, 
as indicated in its caption.

The ratio has a pixel-to-pixel scatter, but is 40\% or less at any pixel except for the one with $(x,y)=(6,4)$. 
The pixel-to-pixel ratio is scattered by 20\% on average for the 15 edge pixels, with an averaged statistical error of 17\% (1~$\sigma$). 
At the central 16 pixels, the ratio is 15\% with an error of 6\%. 

Although the image ratio shows scatter
amongst pixels according to the above numbers, the differences in normalised intensity (or probability) are not large. 
The differences are listed in table \ref{tab:psf}. The difference is largest at the central pixel (x,y)=(3,3),
and is about 2\% of the total counts in the 6x6 pixels.
The difference is much smaller than $1$\% at the outer pixels since the normalised intensity is not high. 

The pixel-to-pixel scattering between the Crab pulsar and the raytracing results cannot be explained 
by statistical fluctuations.
We briefly discuss here the systematic error that possibly explains the discrepancy. 
In figure~\ref{f07}b, we also show the image measured on ground. 
The gate valve was not inserted for the ground measurements. The raytracing image is calculated again without the gate-valve structure, and is shown in figure~\ref{f07}d. 
Figure~\ref{f08}b shows the ratio of the raytracing image to the ground-based image. The pixel-to-pixel ratios for the ground measurement (figure~\ref{f08}b) have less scatter than those of the inflight data (figure~\ref{f08}a). This indicates that the pixel-to-pixel scatter in the ratio of the inflight Crab-pulsar image to the raytracing image is not mainly due to the tuning error of the raytracing to the ground measurements. 

 We suspect that the inflight Crab-pulsar image is distorted by the in-orbit pointing error. 
The pointing direction has a scatter of 4\farcs7 in the Crab observations. 
Since the image of the SXT-S has a sharp core,  about 50\% of photons are included in the central $2\times2$ pixels and make the count rate at the core sensitive to the pointing error. 
The shift of 4\farcs7 causes a change in the count rate of 30\% in the central 4 pixels.
The pointing error may be the origin of the 10\% scatter in the central pixels of the image ratio. 


 The largest image ratios can be identified in the pixels at $(x,y)= (1,3) , (2,3) , (3,4),  (3,6),  (4,6), (5,4)$ and $(6,4)$. These pixels detect light that was reflected from the reflectors near the quadrant boundaries and slipped through the gate valve criss-cross structure (see figure~\ref{f06} ).  The PSF properties of the reflectors near the quadrant boundaries are found to be more complex than those from the other reflectors (\cite{2016JATIS...2d4001S}).  These facts may contribute to the large scatter between the Crab pulsar image and the raytracing output. 

 In our analysis, we assumed an ideal alignement of the gate valve center with the detector center. If we apply the 0.2$\mu$m shift that was measured on ground, the pixel-to-pixel count rate changed by 10\% r.m.s., and 20\% at most.  The alignment of the gate valve cross structure gives an uncertainty of up to 20\% in the flux map.  
 Taking into account the ratio of the pixel-by-pixel distributions of the data to raytracing events ($\leq30$\% due to pointing error, $\leq10$\% due to deadtime correction error, and $\leq20$\% due to the gate valve obstruction uncertainty),  the in-orbit image is consistent with that predicted by the raytracing code.  


In the PSF calculation with the gate valve, 
about 17\% of the total photons collected by the SXT-S are expected to arrive outside of the pixel array 
for the raytracing or ground measurements of the PSF. 
If the PSF tail is more extended in orbit than we expected, more photons will be lost, resulting in
a significantly reduced effective area. The Crab nebula is also a rare example of an
X-ray standard candle with known spectral parameters and flux.  
The consistency of the Crab flux with the canonical value (\cite{EA2017a}) endorses the 
view that the tail of the PSF is not significantly different in orbit compared to that found from ground 
measurements.

\begin{table}[h]
\caption{The pixel-by-pixel probability distribution (or normalized intensity) $P$ of an on-axis point-source}
\begin{center}
\begin{footnotesize}
\begin{tabular}{cccccc}
\hline \hline
Pixel ID & ($x$,$y$) & \multicolumn{2}{c}{Normalised intensity} & Ratio & Difference \\ 
 & & $P_{\rm \ The\ Crab\ pulsar}$ & $P_{\rm \ Raytracing}$ & $\frac{P_{\rm \ The\ Crab\ pulsar}}{P_{\rm \ Raytracing}}$ & $P_{\rm \ The\ Crab\ pulsar} - P_{\rm \ Raytracing}$\\ \hline 
11 & (1,2) & 0.0040 $\pm$ 0.0009 & 0.0051 & 0.78 $\pm$ 0.18 & $-$0.0011 $\pm$ 0.0009 \\
9 & (1,3) & 0.0038 $\pm$ 0.0010 & 0.0059 & 0.65 $\pm$ 0.17 & $-$0.0021 $\pm$ 0.0010 \\
19 & (1,4) & 0.0036 $\pm$ 0.0011 & 0.0036 & 1.00 $\pm$ 0.30 & $-$0.0000 $\pm$ 0.0011 \\
21 & (1,5) & 0.0047 $\pm$ 0.0009 & 0.0048 & 0.98 $\pm$ 0.19 & $-$0.0001 $\pm$ 0.0009 \\
23 & (1,6) & 0.0031 $\pm$ 0.0007 & 0.0025 & 1.22 $\pm$ 0.26 & $+$0.0006 $\pm$ 0.0007 \\
14 & (2,1) & 0.0103 $\pm$ 0.0008 & 0.0082 & 1.26 $\pm$ 0.10 & $+$0.0022 $\pm$ 0.0008 \\
13 & (2,2) & 0.0258 $\pm$ 0.0016 & 0.0220 & 1.17 $\pm$ 0.07 & $+$0.0038 $\pm$ 0.0016 \\
10 & (2,3) & 0.0196 $\pm$ 0.0027 & 0.0286 & 0.68 $\pm$ 0.09 & $-$0.0090 $\pm$ 0.0027 \\
20 & (2,4) & 0.0264 $\pm$ 0.0025 & 0.0269 & 0.98 $\pm$ 0.09 & $-$0.0005 $\pm$ 0.0025 \\
22 & (2,5) & 0.0191 $\pm$ 0.0015 & 0.0183 & 1.04 $\pm$ 0.08 & $+$0.0008 $\pm$ 0.0015 \\
24 & (2,6) & 0.0047 $\pm$ 0.0009 & 0.0038 & 1.25 $\pm$ 0.23 & $+$0.0010 $\pm$ 0.0009 \\
16 & (3,1) & 0.0075 $\pm$ 0.0009 & 0.0092 & 0.82 $\pm$ 0.09 & $-$0.0017 $\pm$ 0.0009 \\
15 & (3,2) & 0.0432 $\pm$ 0.0019 & 0.0366 & 1.18 $\pm$ 0.05 & $+$0.0065 $\pm$ 0.0019 \\
17 & (3,3) & 0.1577 $\pm$ 0.0047 & 0.1341 & 1.18 $\pm$ 0.04 & $+$0.0236 $\pm$ 0.0047 \\
18 & (3,4) & 0.1165 $\pm$ 0.0053 & 0.1605 & 0.73 $\pm$ 0.03 & $-$0.0440 $\pm$ 0.0053 \\
25 & (3,5) & 0.0163 $\pm$ 0.0021 & 0.0175 & 0.93 $\pm$ 0.12 & $-$0.0012 $\pm$ 0.0021 \\
26 & (3,6) & 0.0030 $\pm$ 0.0009 & 0.0043 & 0.71 $\pm$ 0.21 & $-$0.0012 $\pm$ 0.0009 \\
8 & (4,1) & 0.0092 $\pm$ 0.0008 & 0.0094 & 0.99 $\pm$ 0.09 & $-$0.0001 $\pm$ 0.0008 \\
7 & (4,2) & 0.0414 $\pm$ 0.0015 & 0.0354 & 1.17 $\pm$ 0.04 & $+$0.0061 $\pm$ 0.0015 \\
0 & (4,3) & 0.1404 $\pm$ 0.0025 & 0.1591 & 0.88 $\pm$ 0.02 & $-$0.0187 $\pm$ 0.0025 \\
35 & (4,4) & 0.1382 $\pm$ 0.0033 & 0.1171 & 1.18 $\pm$ 0.03 & $+$0.0211 $\pm$ 0.0033 \\
33 & (4,5) & 0.0163 $\pm$ 0.0018 & 0.0166 & 0.99 $\pm$ 0.11 & $-$0.0002 $\pm$ 0.0018 \\
34 & (4,6) & 0.0050 $\pm$ 0.0008 & 0.0038 & 1.34 $\pm$ 0.20 & $+$0.0013 $\pm$ 0.0008 \\
6 & (5,1) & 0.0088 $\pm$ 0.0007 & 0.0085 & 1.03 $\pm$ 0.08 & $+$0.0003 $\pm$ 0.0007 \\
4 & (5,2) & 0.0163 $\pm$ 0.0011 & 0.0198 & 0.82 $\pm$ 0.05 & $-$0.0035 $\pm$ 0.0011 \\
2 & (5,3) & 0.0492 $\pm$ 0.0015 & 0.0471 & 1.04 $\pm$ 0.03 & $+$0.0021 $\pm$ 0.0015 \\
28 & (5,4) & 0.0454 $\pm$ 0.0016 & 0.0320 & 1.42 $\pm$ 0.05 & $+$0.0134 $\pm$ 0.0016 \\
31 & (5,5) & 0.0147 $\pm$ 0.0013 & 0.0153 & 0.96 $\pm$ 0.08 & $-$0.0006 $\pm$ 0.0013 \\
32 & (5,6) & 0.0025 $\pm$ 0.0007 & 0.0029 & 0.85 $\pm$ 0.23 & $-$0.0004 $\pm$ 0.0007 \\
5 & (6,1) & 0.0027 $\pm$ 0.0005 & 0.0030 & 0.90 $\pm$ 0.15 & $-$0.0003 $\pm$ 0.0005 \\
3 & (6,2) & 0.0066 $\pm$ 0.0007 & 0.0080 & 0.82 $\pm$ 0.08 & $-$0.0014 $\pm$ 0.0007 \\
1 & (6,3) & 0.0118 $\pm$ 0.0009 & 0.0114 & 1.04 $\pm$ 0.08 & $+$0.0004 $\pm$ 0.0009 \\
27 & (6,4) & 0.0119 $\pm$ 0.0008 & 0.0065 & 1.83 $\pm$ 0.13 & $+$0.0054 $\pm$ 0.0008 \\
29 & (6,5) & 0.0082 $\pm$ 0.0007 & 0.0073 & 1.12 $\pm$ 0.10 & $+$0.0009 $\pm$ 0.0007 \\
30 & (6,6) & 0.0019 $\pm$ 0.0005 & 0.0021 & 0.87 $\pm$ 0.23 & $-$0.0003 $\pm$ 0.0005 \\ \hline
Total & & 1 & 1 & 35 & 0 \\ \hline
\end{tabular}
\end{footnotesize}
\label{tab:psf}
\end{center}
Only the 1$\sigma$ statistical error is written.  The error is rather dominated by systematics. See text for details.  
\end{table}

\clearpage

\section{Summary}


We presented the inflight calibration status for the angular 
resolution of the SXT-S that focuses X-rays at the SXS detector array. In order to
examine the performance, a point-like source, ``the Crab pulsar,''  was extracted from the Crab nebula 
emission and  was compared with raytracing image data.
The raytracing calibration files were tuned to reproduce the ground measurements. 
Within the limited statistics afforded by an exposure time of only 6.9~ksec,
we found that the Crab pulsar image is consistent with the raytracing results, within uncertainties. 
The ratio between the Crab pulsar image and the corresponding raytracing image
shows scatter from pixel to pixel, but this is 40\% or less in all pixels except one.  
The pixel-to-pixel ratio has a scatter of 20\%, on average, for the 15 edge
 pixels, with an averaged statistical error of 17\% ($1\sigma$). At the central 
16 pixels, the ratio is 15\% with an error of 6\%. 


\section*{Appendix: Off-axis PSF calibration using the ground measurements}

Not only is the on-axis PSF important, but the off-axis PSF is also crucially important 
for studying extended sources. 
However, due to the limited life of the Hitomi satellite, the inflight PSF calibration was not made for off-axis
positions. 
We instead summarize the reproducibility of the
ground measurements by the raytracing software {\texttt xrtraytrace} as a calibration of the off-axis PSF, but without the 
gate valve obstruction. 
Assuming that the off-axis PSF measured on ground is exactly the same as that in orbit, 
the reproducibility is equivalent to the calibration error of the raytracing outputs for the in-orbit off-axis PSFs. 
The ground measurements of the off-axis PSF at 4.5 keV are made at 3\farcm0, 4\farcm5 and 8\farcm6 
off-axis and are presented by \citet{2014SPIE.9144E..58I}, \citet{2016SPIE.9905E..3XS} and \citet{2016SPIE.9905E..5DH}. 
The energy of 4.5 keV corresponds to the K-shell transition line of titanium, which is the target material of the X-ray source in the ground calibration facility. It is located near the Fe-K lines that are of importance for Hitomi science.  

Because of the moderate angular resolution of the SXT-S, 
a significant fraction of the collected photons arrives at the pixel array
even if the target source is focused outside of the array.
The majority of these light paths are identified as normal double reflections in the PSF tail, in which the incoming X-rays reflect first at a primary reflector and next at a secondary reflector (\cite{2014SPIE.9144E..58I}). 

Since the off-axis light suffers from the vignetting effect, the effective area (EA) is reduced at off-axis angles. We summarize the PSF tail in units of the effective area for the integration area of the 3\farcm05$\times$3\farcm05 area that roughly corresponds to the pixel array. 
The FWHM of the vignetting is 15\farcm8. The EA for the whole focal plane at 8$'$ off-axis is about
half of that at the on-axis. 

\begin{figure}[h]
 (On-axis)\\
 \begin{center}
\includegraphics[width =40mm, bb=0 0 742 743]{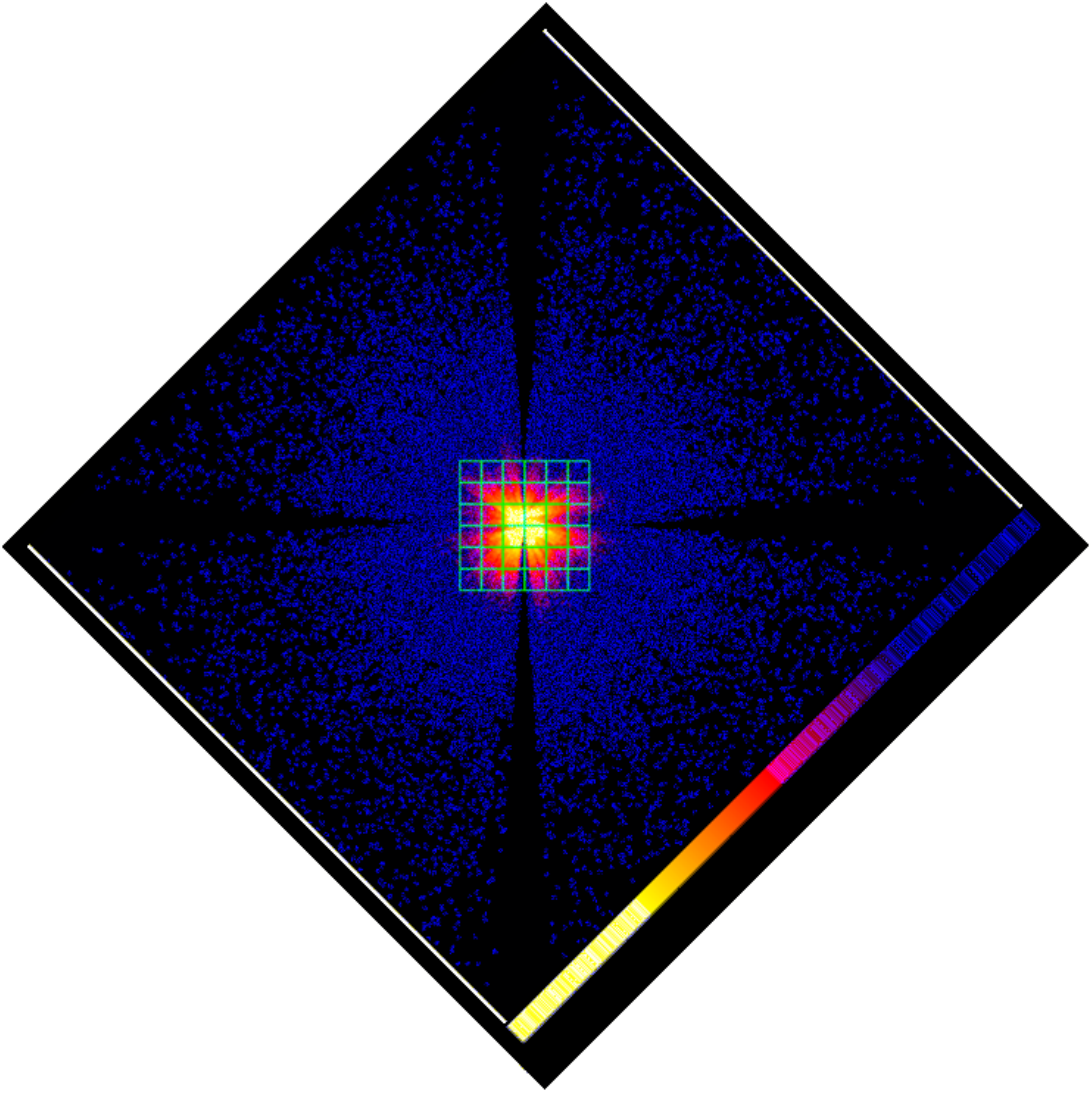}
\includegraphics[width =40mm, bb=0 0 742 743]{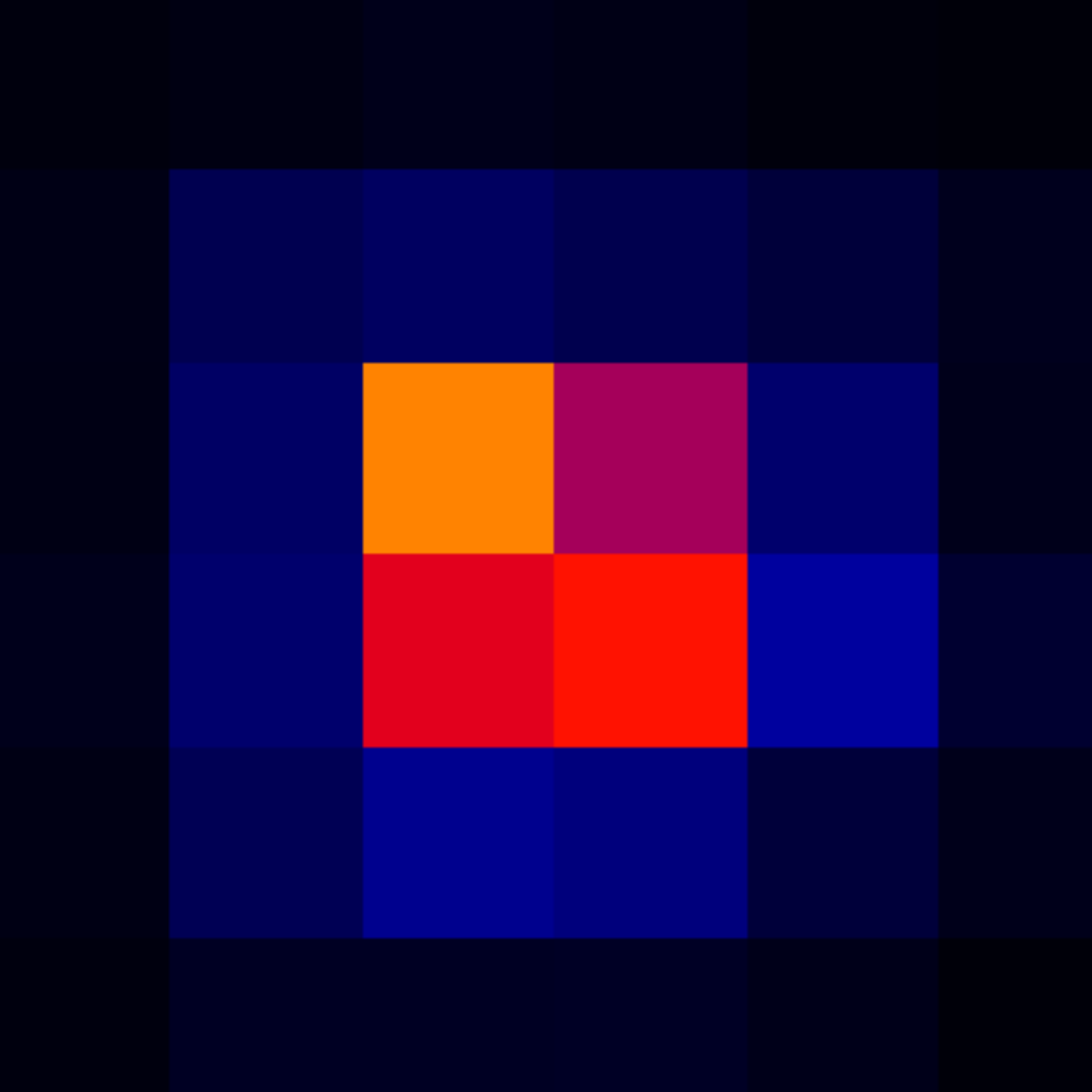}
\includegraphics[width =40mm, bb=0 0 742 743]{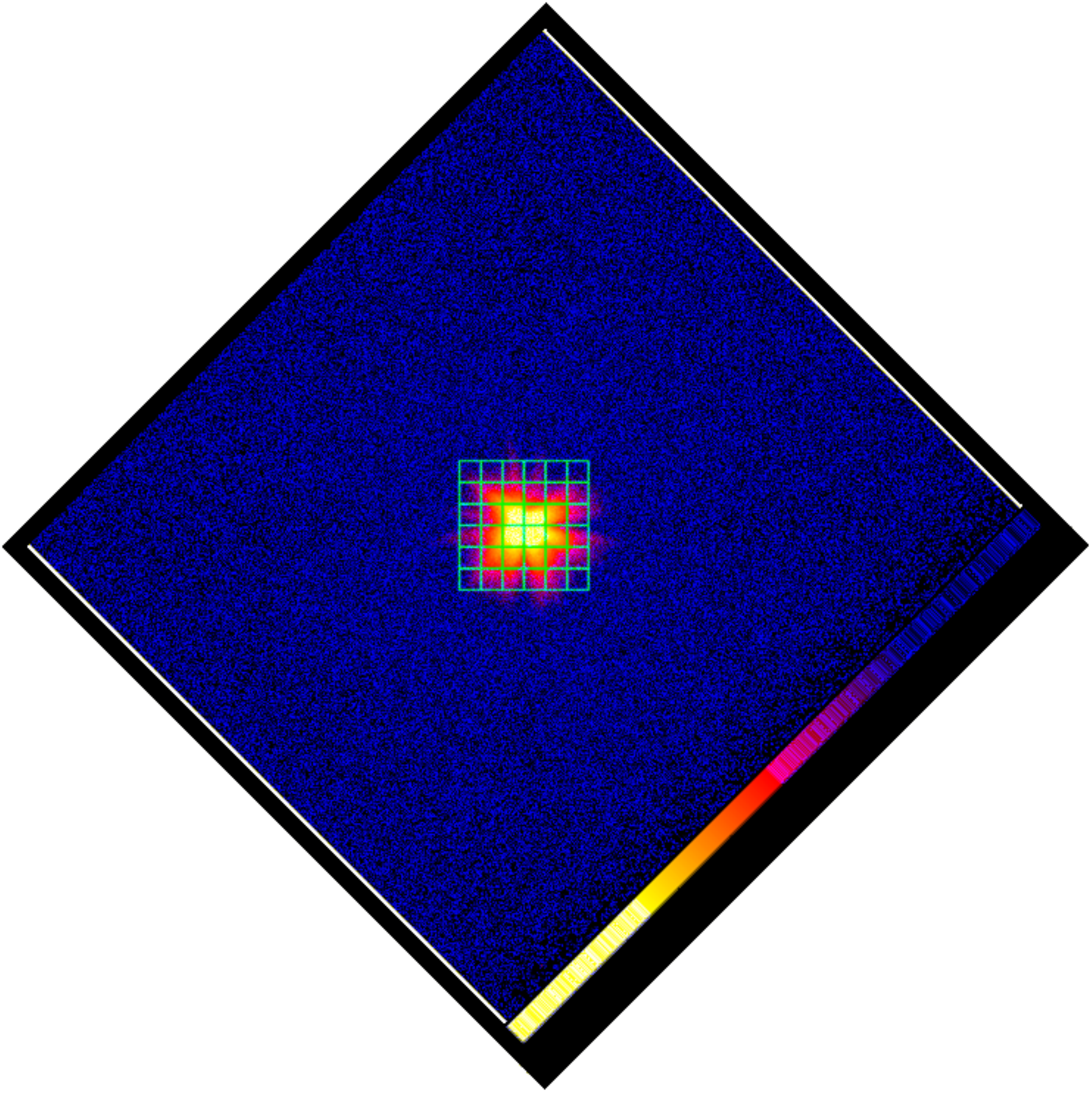}
\includegraphics[width =40mm, bb=0 0 742 743]{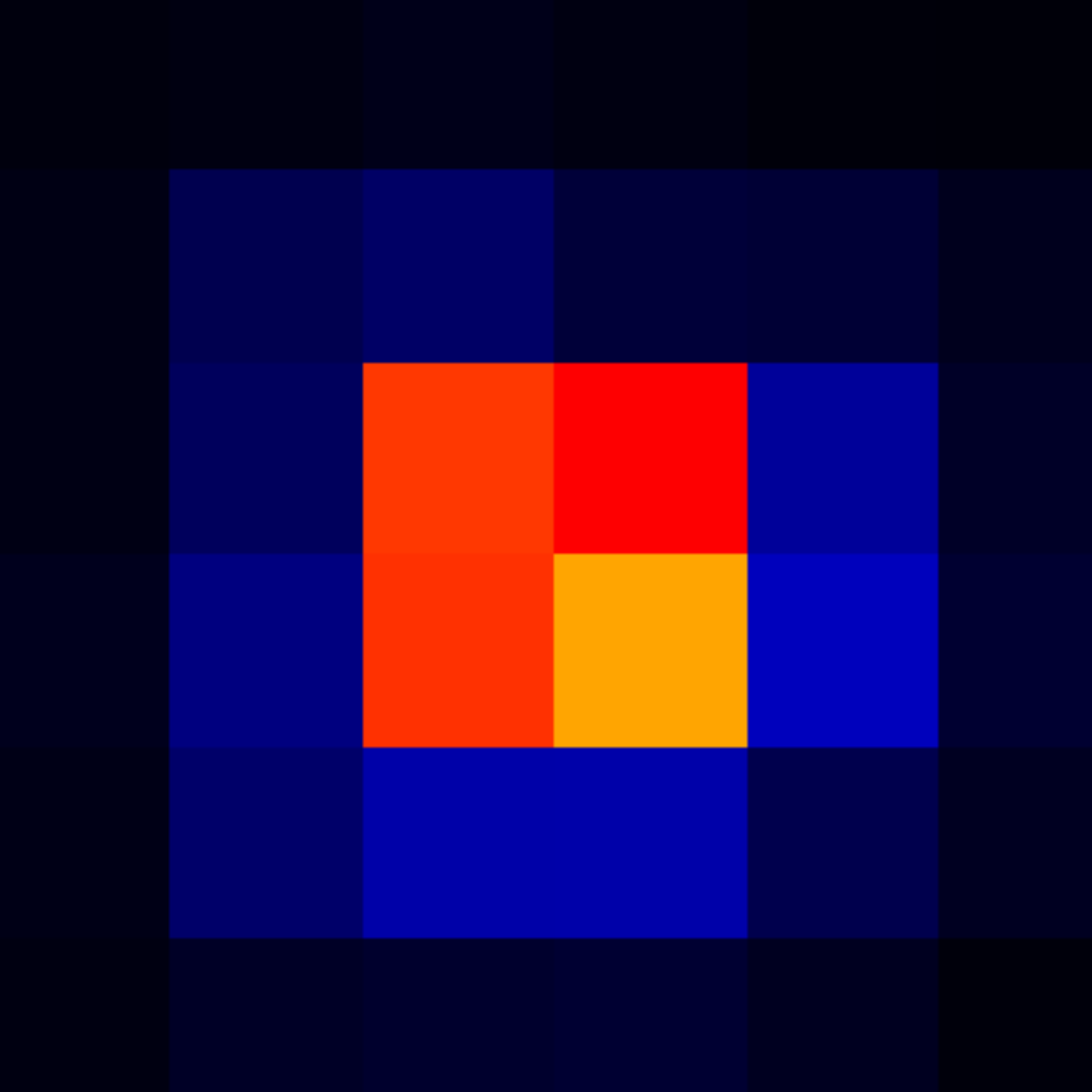}\\
\hspace*{-1.3cm}
\includegraphics[width =40mm, bb=0 0 525 85]{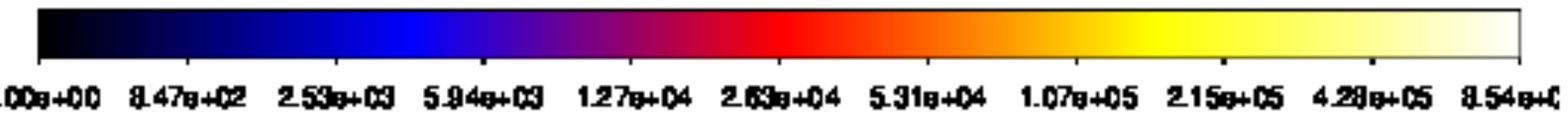}
\hspace{4cm}
\includegraphics[width =40mm, bb=0 0 525 85]{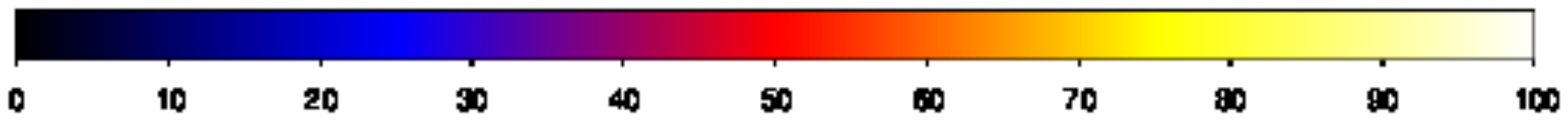}\\
-- Raytracing outputs -- \hspace{4cm} -- Ground measurements --
 \end{center}
\caption{Event distribution of the point-like source at on-axis. The position of the pixel array is displayed in the green grid. The center of the pixel array is aligned along the on-axis direction.  The left two figures are made using the raytracing outputs, whereas the right two are made using ground measurements. }
\label{fig:hayashi00}
\end{figure}

In figures~\ref{fig:hayashi00}--\ref{fig:hayashi03}, we show the off-axis images at four offsets of 0\farcm0 (on-axis), 3\farcm0, 4\farcm5 and 8\farcm6 at 4.5 keV. The images are plotted by aligning them in DET coordinates.
The ground measurements were made with a CCD detector by rotating 45 degrees relative to the DET coordinate system. 
The raytracing output is trimmed with the same region of the CCD detector that we used on ground. 
In figures~\ref{fig:hayashi00}--\ref{fig:hayashi03}, the image of the $6\times6$ pixel array is extracted and is arranged next to the original image. 

In the ground measurements, the telescope is not fully illuminated by X-rays at the offset IDs  4\farcm5 (II$_{\rm 1,3,4}$) and 8\farcm6 (III). 
For example, in figures~\ref{fig:hayashi02} the 
tail spills over the pixel array have contributions from two of four quadrants of the SXT-S. 
Due to the tight schedule of the ground calibration before the launch,
 the other two quadrants were sometimes not measured.

\begin{figure}[h]
 \begin{center}
        \includegraphics[width=80mm,bb=0 0 960 540]{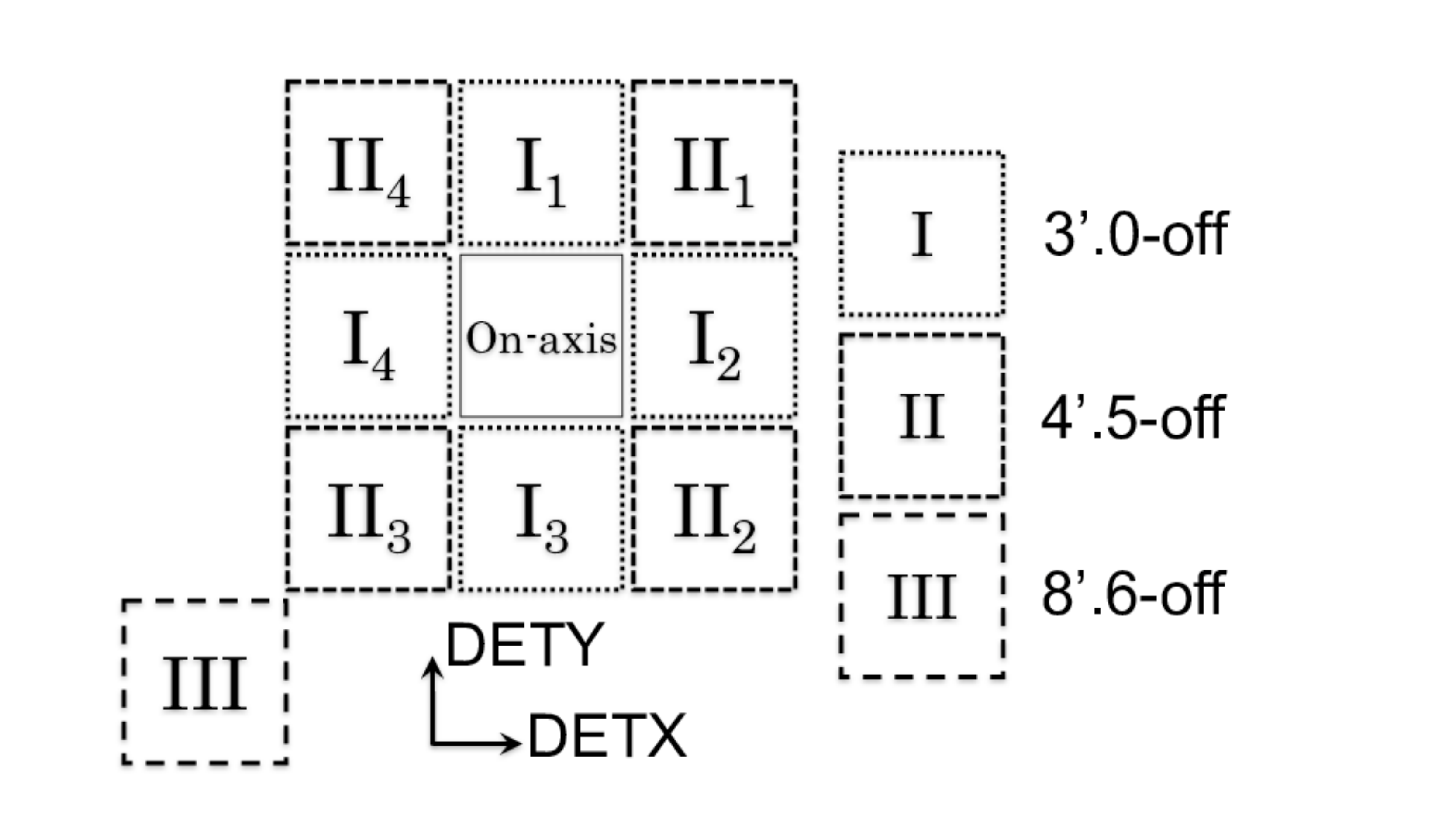}
 \end{center}
\caption{Offset positions to estimate the EA of the off-axis sources for the whole pixel array.}
\label{fig:fov_pos}
\end{figure}
\clearpage
\begin{table}[h]
\caption{EA for the off-axis sources in the whole pixel array at 4.5\,keV. 
}
\begin{center}
\begin{tabular}{ccccccc}
\hline
\multicolumn{2}{c}{Offset/Off-axis angle}& \multicolumn{2}{c}{Ground Measurements} & \multicolumn{2}{c}{Raytracing Outputs}  & PSF ratio \\ 
Region ID. & (DETX, DETY) & $EA$ [cm$^2$] & $R$($\frac{{\rm EA_{Off-axis}}}{{\rm EA_{On-axis}}}$) $^{\dagger}$  & $EA$& $R$($\frac{{\rm EA_{Off-axis}}}{{\rm EA_{On-axis}}}$)$^{\dagger}$   & $R$($\frac{{\rm EA_{Gnd.\  Meas.}}}{{\rm EA_{Raytracing}}}$)\\ 
 & &  [cm$^2$] &  [$\%$]  &  [cm$^2$] & [$\%$]   & [$\%$] \\ \hline\hline

On-axis                  & $(\pm0$\farcm0, $\pm0$\farcm0$)$ & 412 & -- & 380 & -- & 92 \\ 
3\farcm0 (I$_{\rm 1}$)$^{\ast}$ & $(\pm0$\farcm0, $+3$\farcm0$)$ & 7.4 & 1.8 & 8.7 & 2.3 & 118 \\ 
3\farcm0 (I$_{\rm 2}$)$^{\ast}$ &  $(+3$\farcm0, $\pm0$\farcm0$)$  & 10.5 & 2.6 & 11.8 & 3.1 & 112 \\
3\farcm0 (I$_{\rm 3}$)$^{\ast}$ &  $(\pm0$\farcm0, $-3$\farcm0$)$ & 9.2 & 2.2 & 11.4 & 3.0 & 124 \\ 
3\farcm0 (I$_{\rm 4}$)$^{\ast}$ &  $(-3$\farcm0, $\pm0$\farcm0$)$ &  7.4 & 1.8 & 8.7 & 2.3 & 118 \\ 
4\farcm5 (II$_{\rm 1}$)$^{\ast}$&  $(+3$\farcm0, $+3$\farcm0$)$ &  2.4 & 0.6 & 1.8 & 0.5 & 75 \\ 
4\farcm5 (II$_{\rm 2}$)$^{\ast}$& $(+3$\farcm0, $-3$\farcm0$)$ & 2.8 & 0.7 & 2.0 & 0.5 & 71 \\ 
4\farcm5 (II$_{\rm 3}$)$^{\ast}$& $(-3$\farcm0, $-3$\farcm0$)$ & 2.2 & 0.5 & 1.9 & 0.5 & 86 \\
4\farcm5 (II$_{\rm 4}$)$^{\ast}$& $(-3$\farcm0, $+3$\farcm0$)$ & 2.2 & 0.5 & 2.1 & 0.6 & 95 \\
8\farcm6 (III)$^{\ast}$& $(-6$\farcm0, $-6$\farcm0$)$ & 0.35 & 0.1 & 0.11 & 0.03 & 31 \\\hline\hline
\multicolumn{3}{l}{$^{\dagger}$ Ratio of EA to on-axis EA.}\\
\multicolumn{3}{l}{$^{\ast}$ Position label 
defined in figure\,\ref{fig:fov_pos}.}\\
\end{tabular}
\label{tab:ea_off}
\end{center}
The typical statistical error  is $\pm$0.2\% at on-axis,  $\pm$1.5\% at 3\farcm off-axis, $\pm$3\% at 4\farcm5 off-axis, and $\pm$10\% at 8\farcm6 off-axis and is reported in \citet{2016SPIE.9905E..5DH}. 
\end{table}
\clearpage
%

For the 3\farcm0 off-axis angle positions, the source is positioned
at offset points that are reached by shifts parallel to the array edges.
For the 4\farcm5 and 8\farcm6 off-axis angle positions, the source is positioned along
a diagonal across the array, as shown in figure \ref{fig:fov_pos}.
These positions reflect what would be useful in a tiled mapping of 
extended sources such as the Perseus Cluster. 
The EA values of the off-axis positions are tabulated in table\,\ref{tab:ea_off}.
The off-axis EAs from the ground calibration depend on the azimuthal direction and
for the range in azimuthal directions shown in table\,\ref{tab:ea_off} the EAs
have the range 
7.4--10.5 and 2.2-2.4 cm$^2$ for off-axis angles of 3\farcm0 and 4\farcm5, respectively.
These EAs correspond to $\sim$ 2\% and $\sim$ 0.6\% of the on-axis EA, respectively. 
The EA for the source at 8\farcm6 off-axis is $\sim$ 0.1\% of the on-axis EA. 

The EAs were also calculated using the raytracing software and are tabulated in table\,\ref{tab:ea_off}.
The ratios of the EA between the ground measurements and the corresponding raytracing 
results are dominated by the 
calibration error of the PSF. 
The ratio is smallest at the 8\farcm4 off-axis position, and has a value of about one third. 
At the other off-axes angles of 
3\farcm0 and 4\farcm5, the ratios are almost consistent with unity, with a scatter of up to 30\%. 

In the {\texttt aharfgen} software, the EA simulated by the raytracing is corrected for the gate valve obstruction.
However, the on-axis gate valve correction function pertinent for a 
point source is always used regardless of the actual off-axis angle and the spatial extent of the source.
For off-axis angles, the gate valve correction function also depends on the azimuthal angle.
The errors introduced by these assumptions have been quantified for different off-axis and azimuthal angles and
for extended sources, and are shown in the {\rm Hitomi} analysis guide which can 
be found at {\rm https://heasarc.gsfc.nasa.gov/docs/hitomi/analysis/}. 


\begin{figure}[h]
3\farcm0(I$_{\rm 1}$)
 \begin{center}
\includegraphics[width =40mm, bb=0 0 742 743]{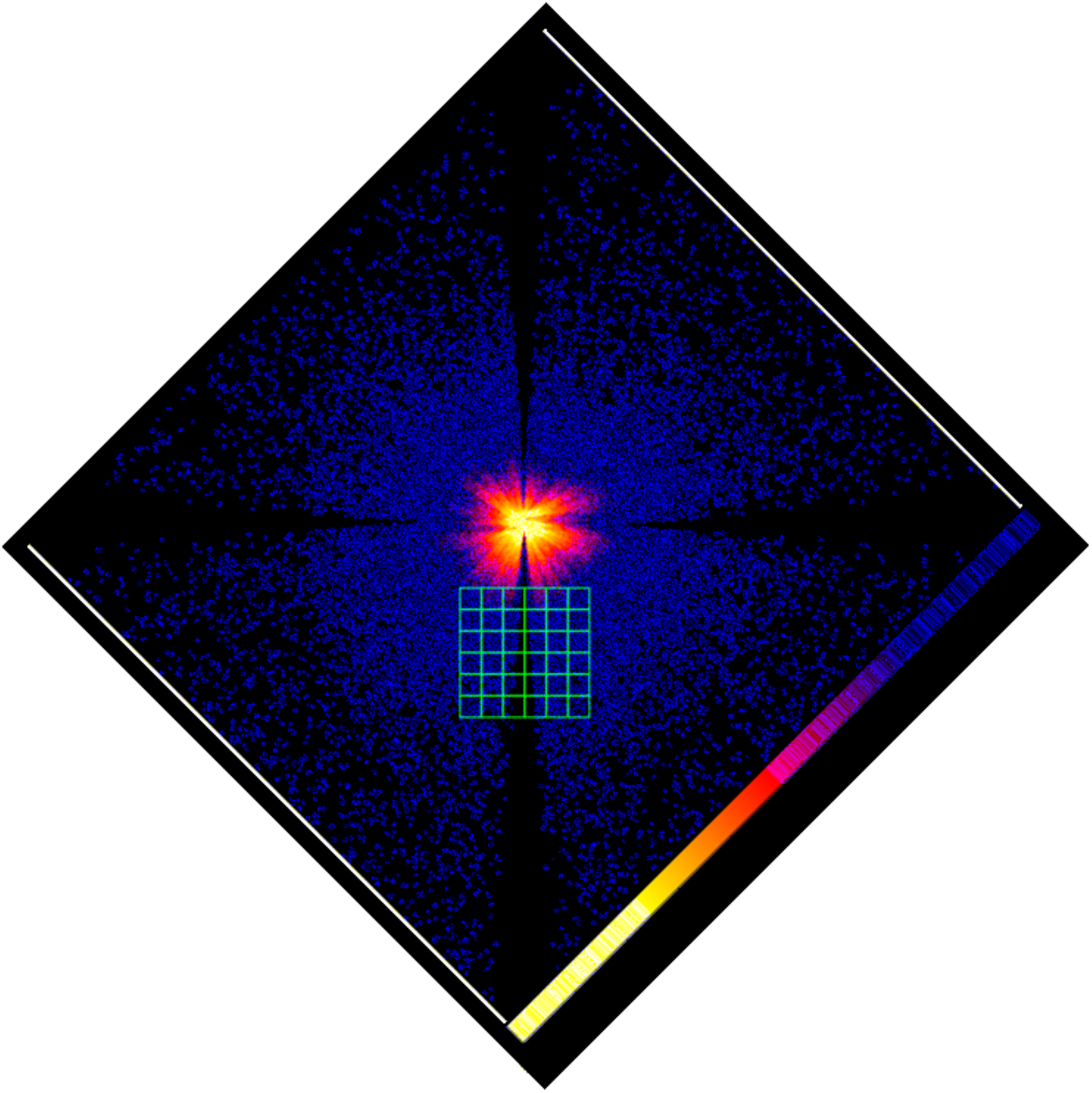}
\includegraphics[width =40mm, bb=0 0 742 743]{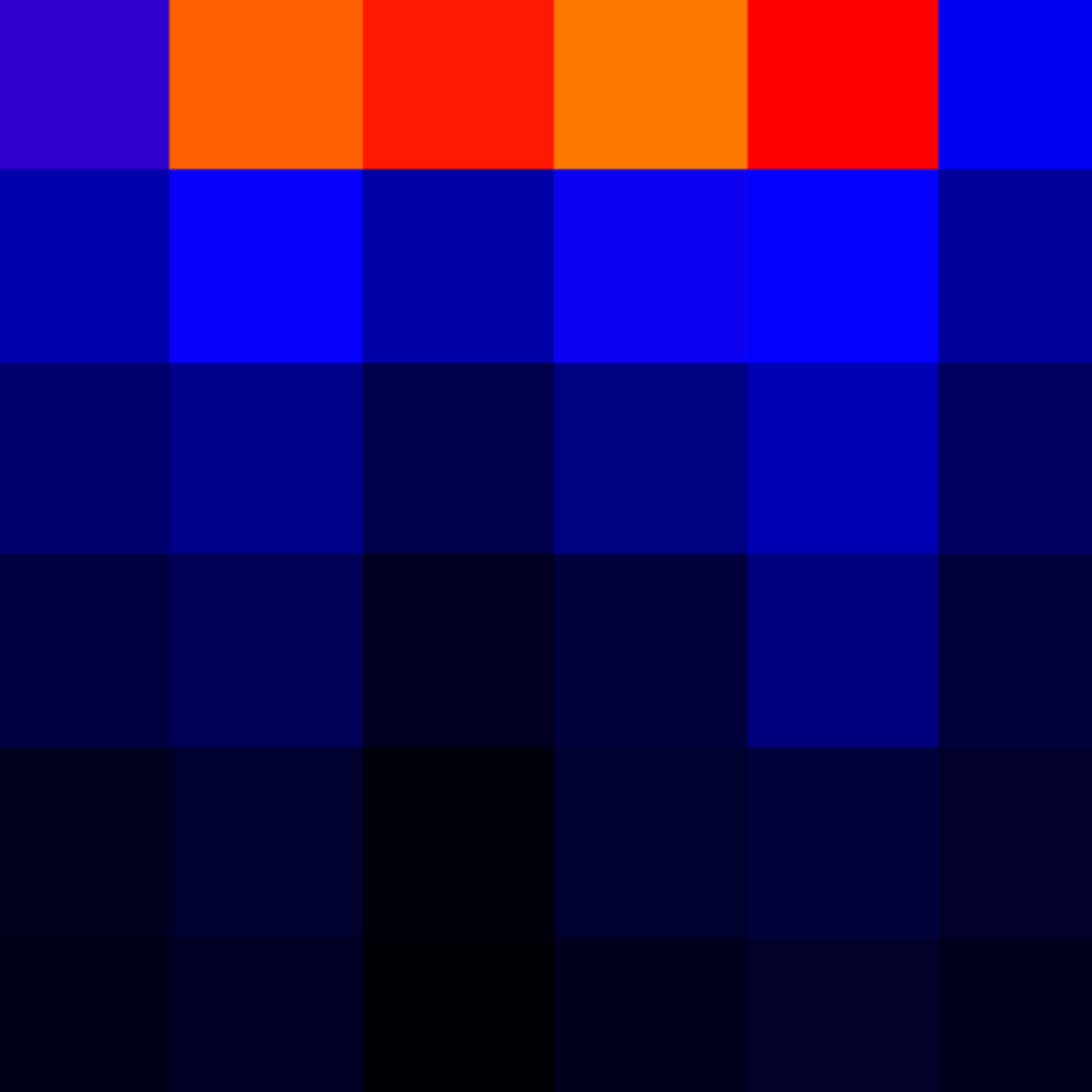}
\includegraphics[width =40mm, bb=0 0 742 743]{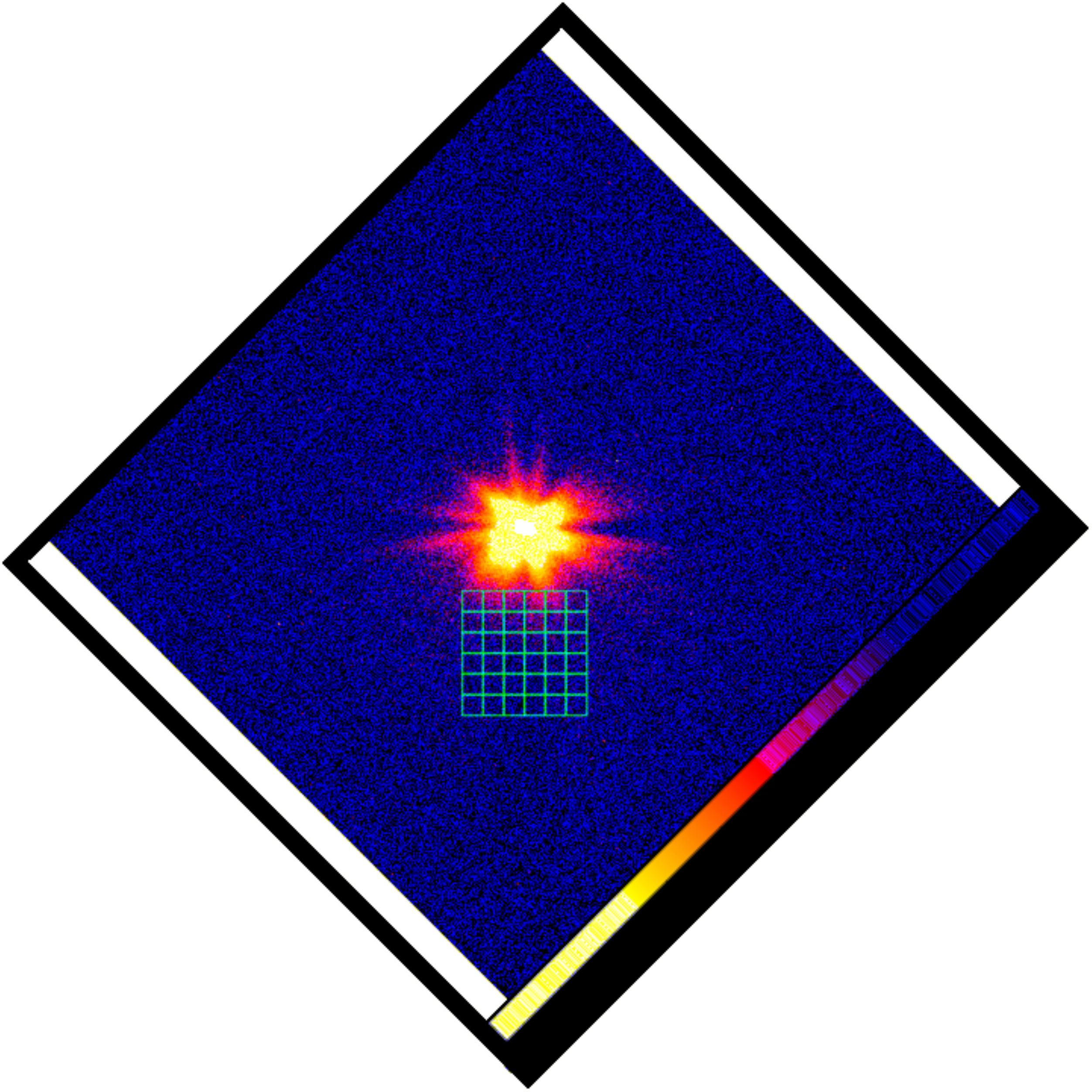}
\includegraphics[width =40mm, bb=0 0 742 743]{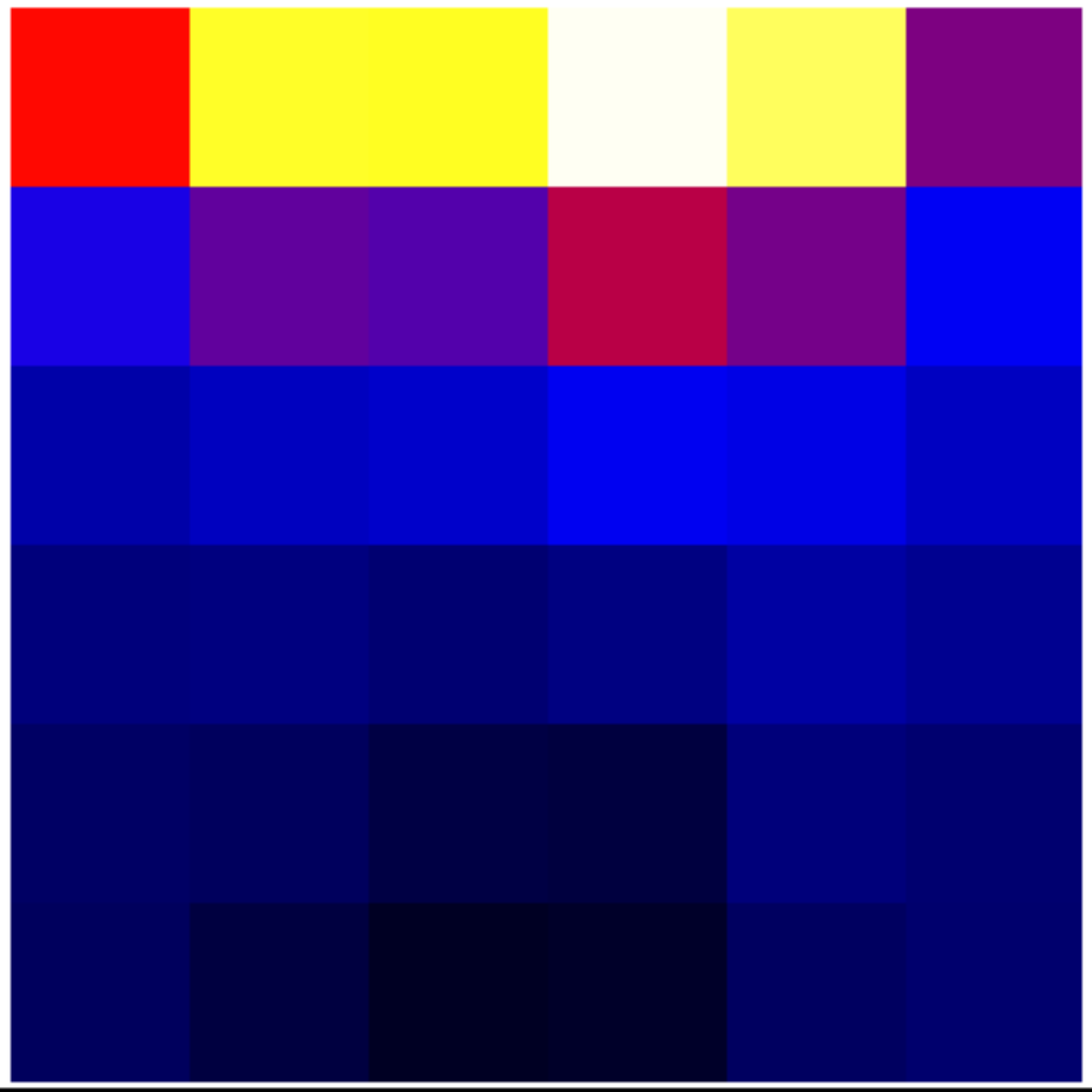}\\
\vspace*{0.3cm}
\hspace*{-1.3cm}
\includegraphics[width =40mm, bb=0 0 525 85]{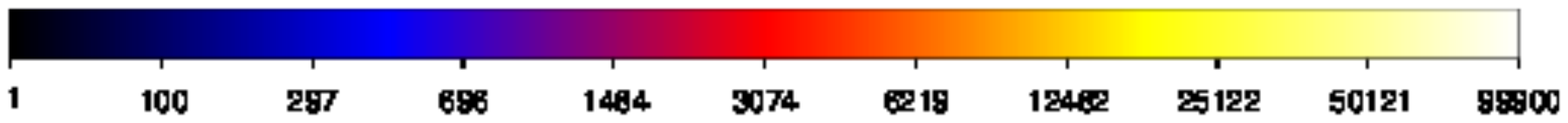}
\hspace{4cm}
\includegraphics[width =40mm, bb=0 0 525 85]{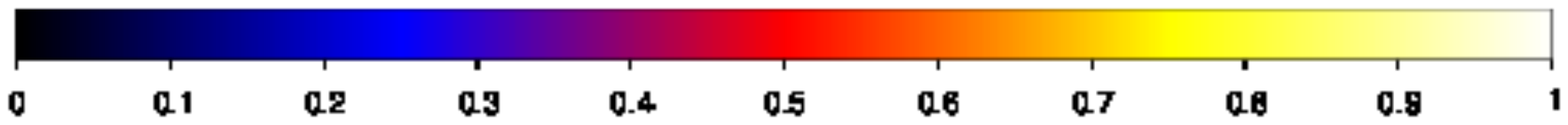}\\
 \end{center}

3\farcm0(I$_{\rm 2}$)
 \begin{center}
\includegraphics[width =40mm, bb=0 0 742 743]{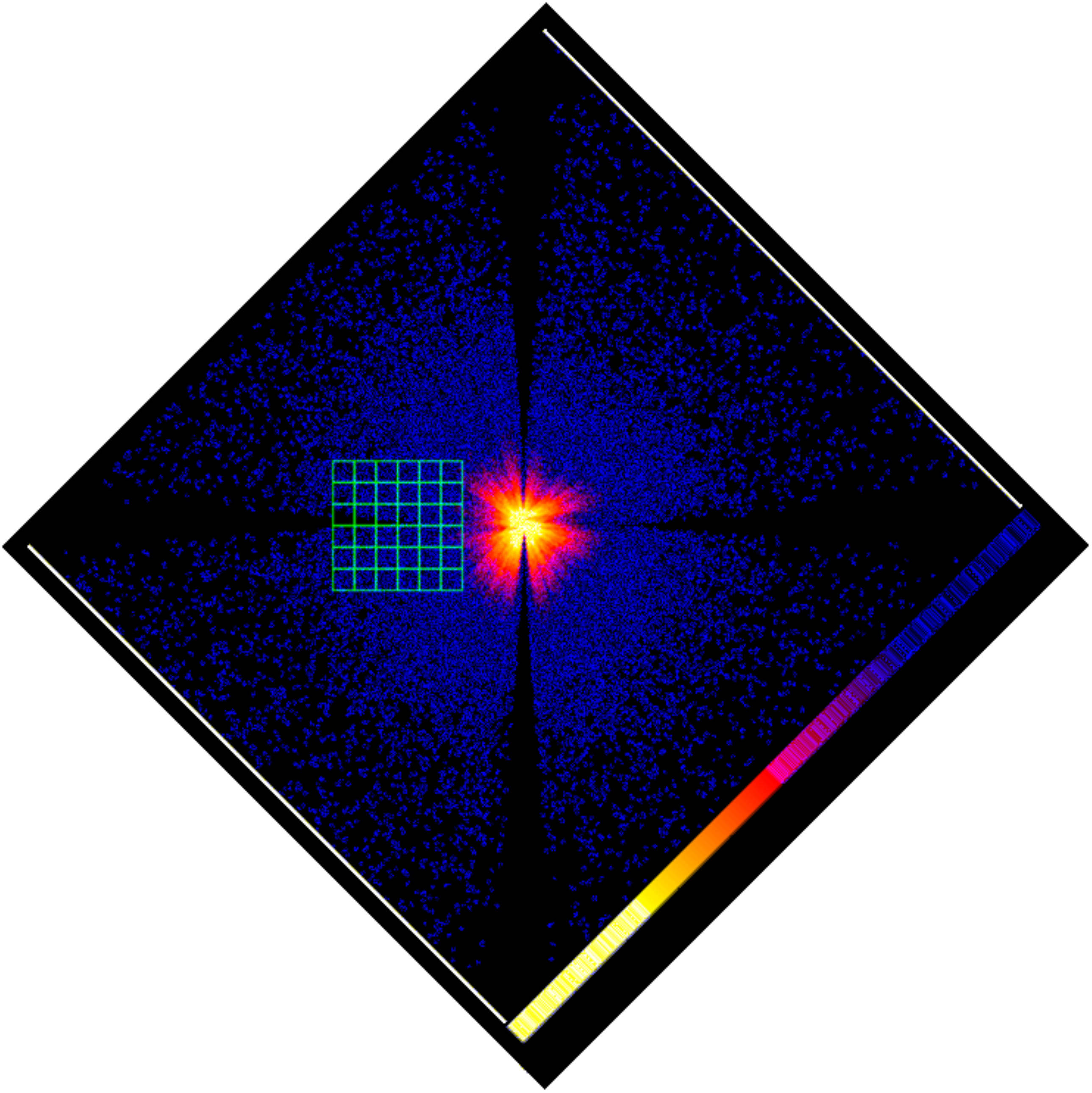}
\includegraphics[width =40mm, bb=0 0 742 743]{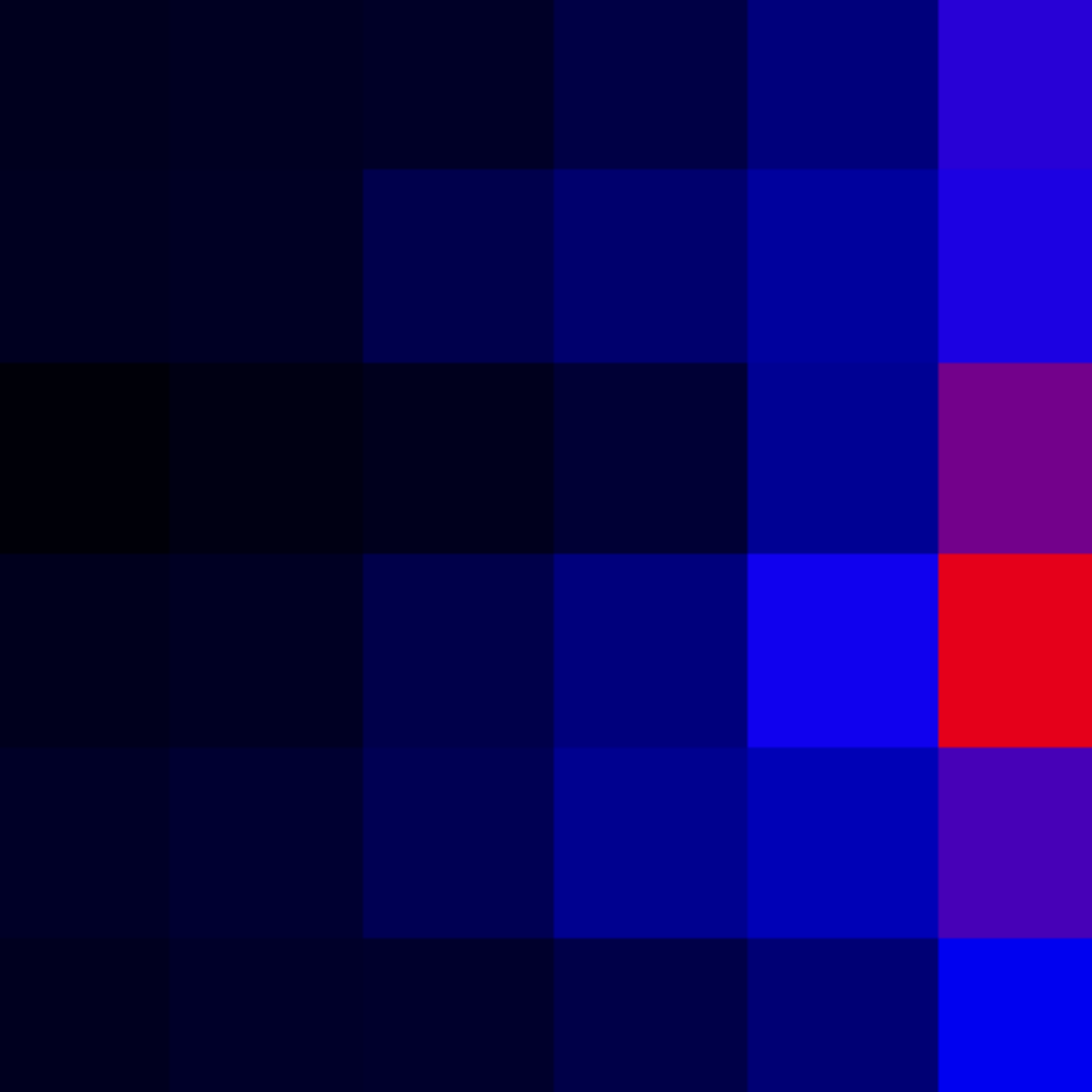}
\includegraphics[width =40mm, bb=0 0 742 743]{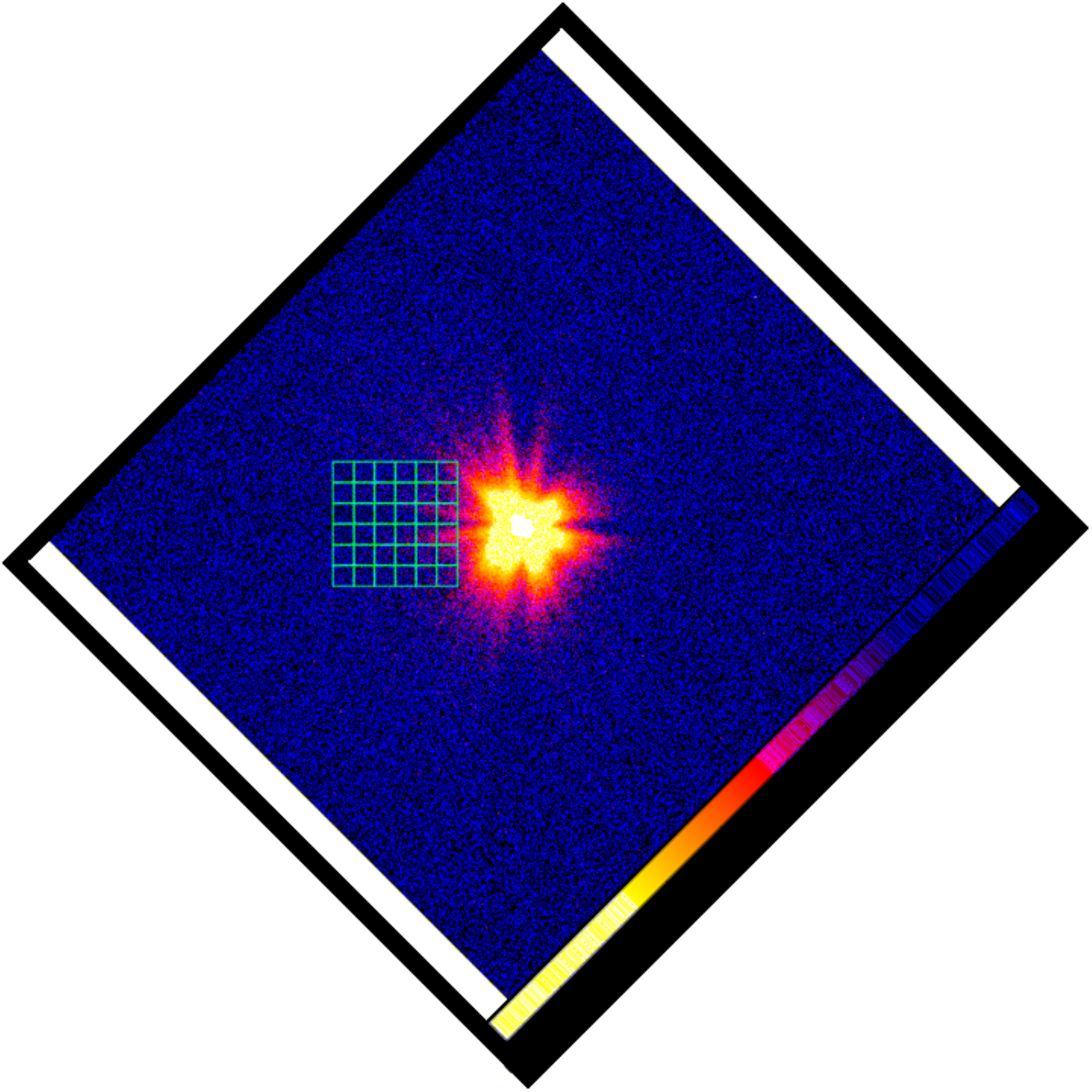}
\includegraphics[width =40mm, bb=0 0 742 743]{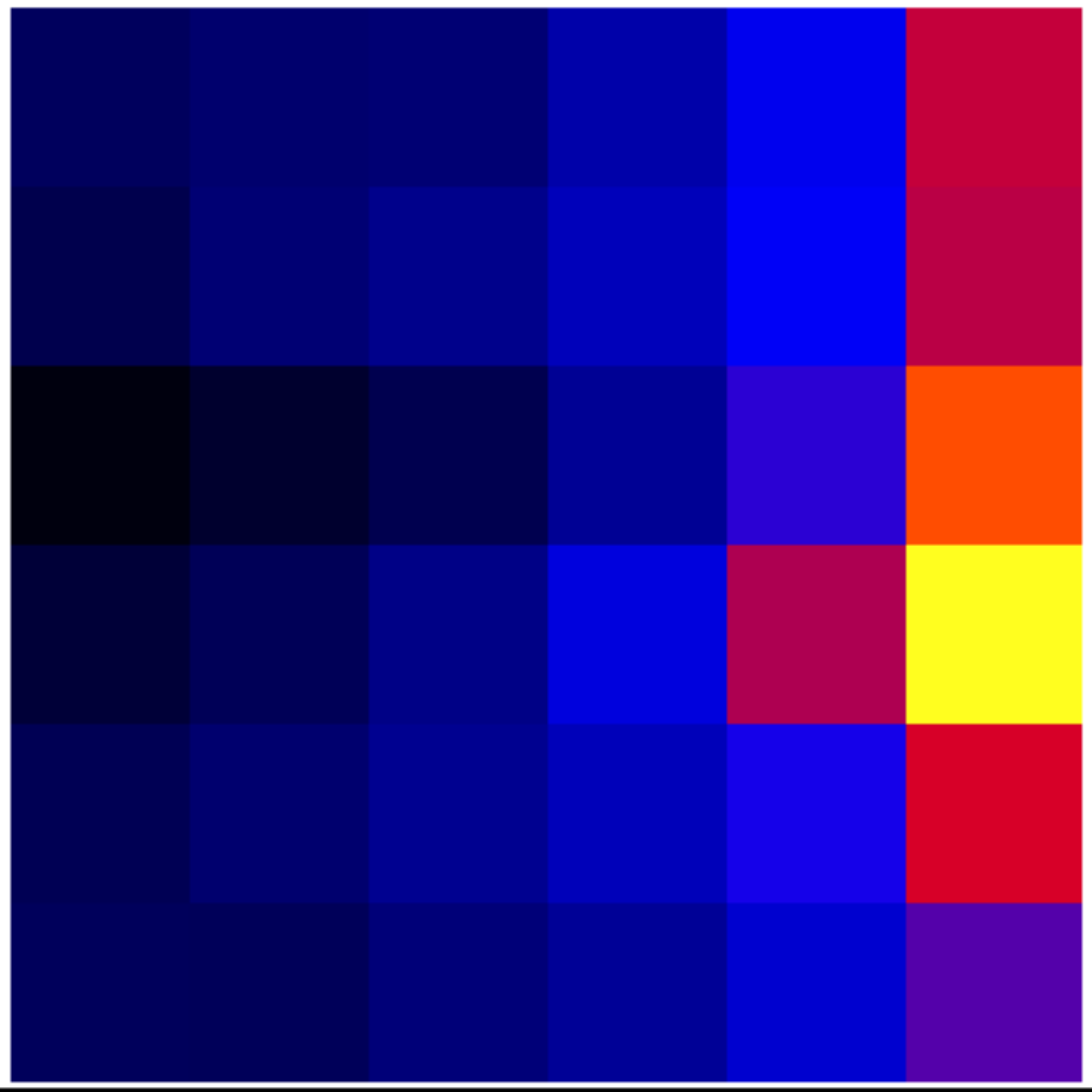}\\
\vspace*{0.3cm}
\hspace*{-1.3cm}
\includegraphics[width =40mm, bb=0 0 525 85]{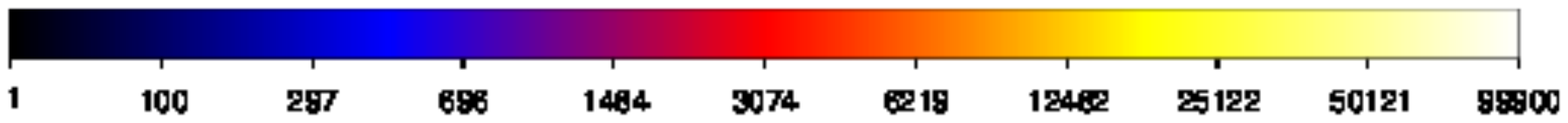}
\hspace{4cm}
\includegraphics[width =40mm, bb=0 0 525 85]{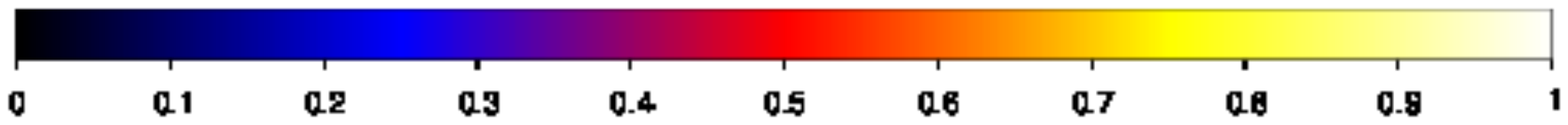}\\
 \end{center}

3\farcm0(I$_{\rm 3}$)
\begin{center}
\includegraphics[width =40mm, bb=0 0 742 743]{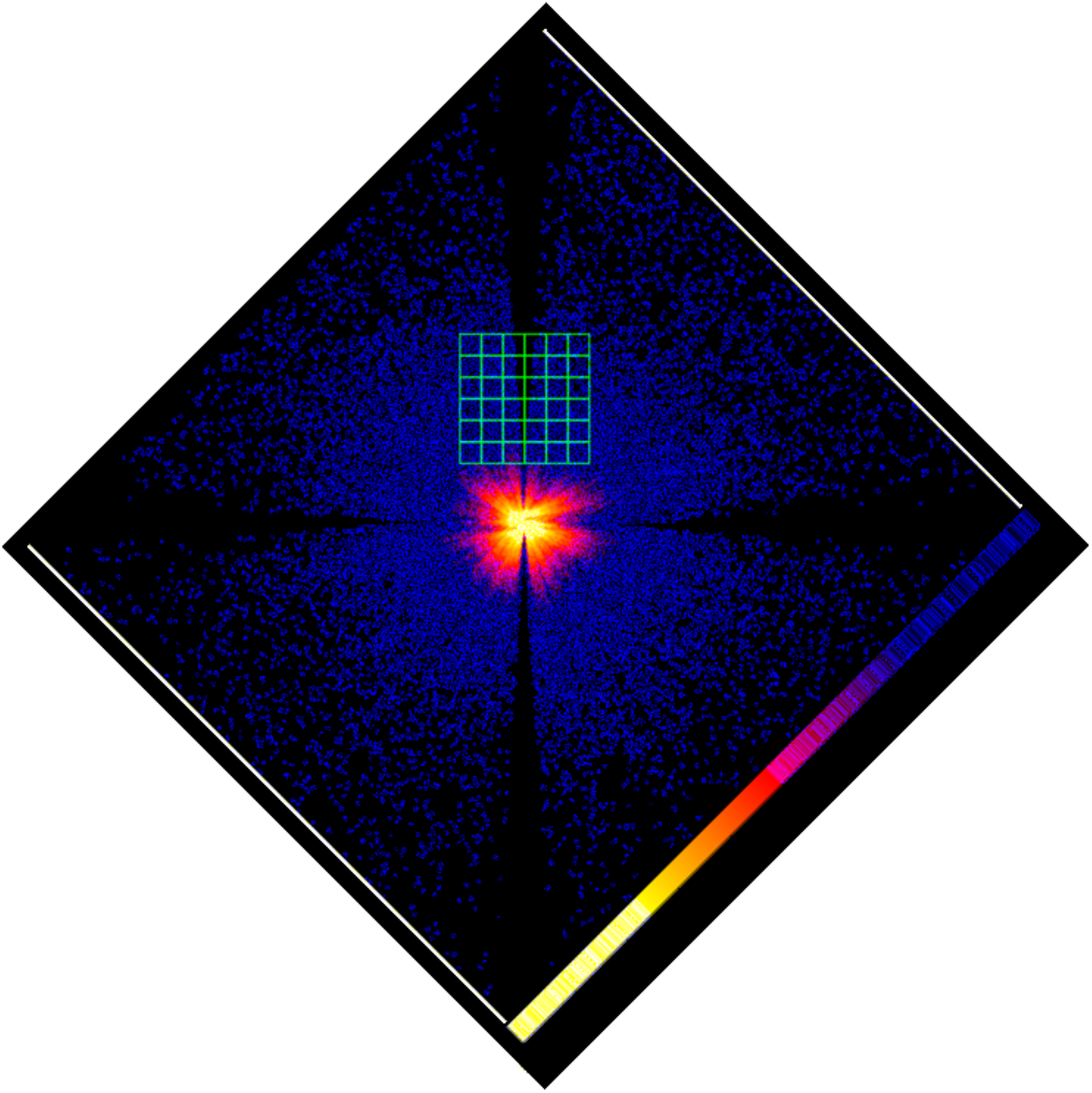}
\includegraphics[width =40mm, bb=0 0 742 743]{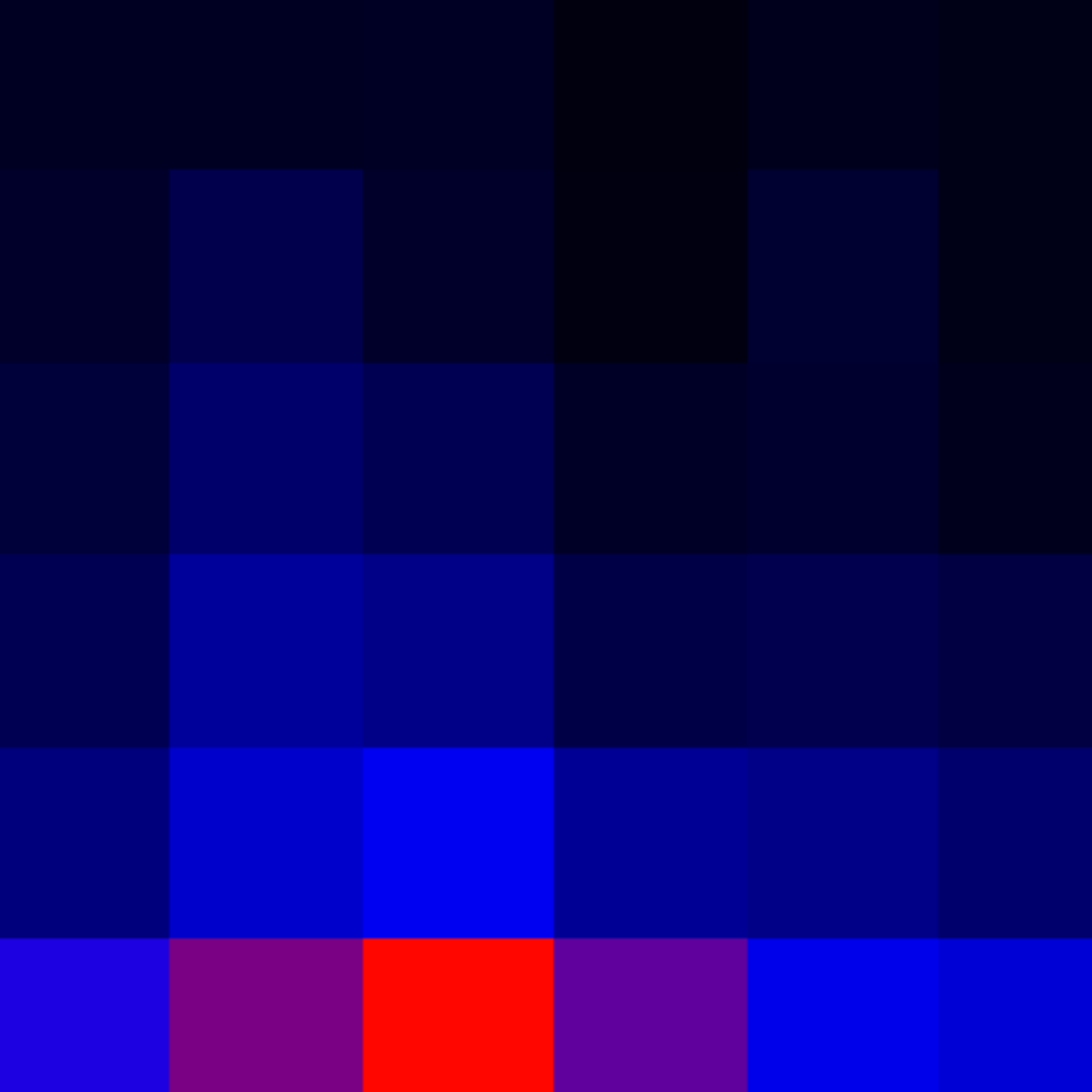}
\includegraphics[width =40mm, bb=0 0 742 743]{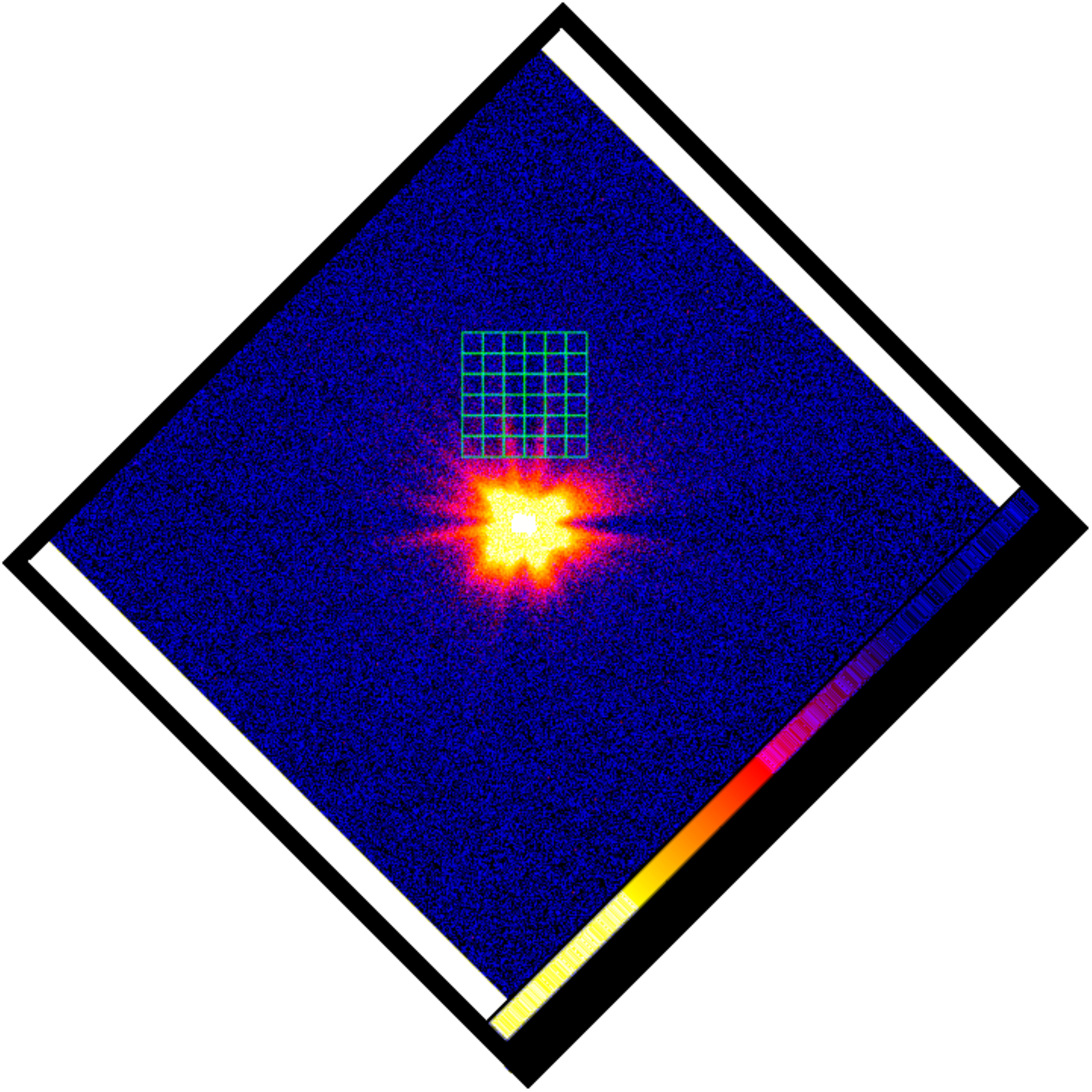}
\includegraphics[width =40mm, bb=0 0 742 743]{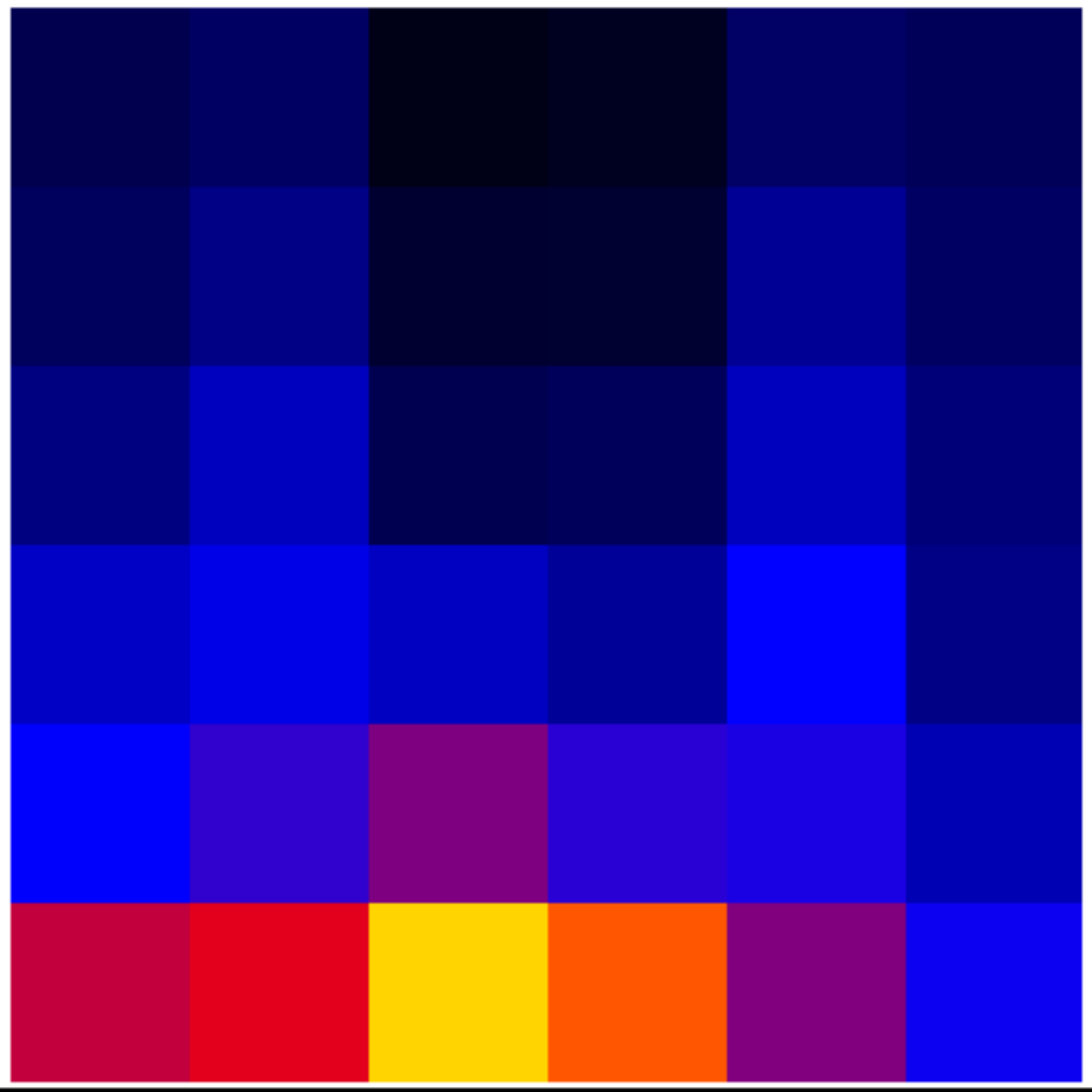}\\
\vspace*{0.3cm}
\hspace*{-1.3cm}
\includegraphics[width =40mm, bb=0 0 525 85]{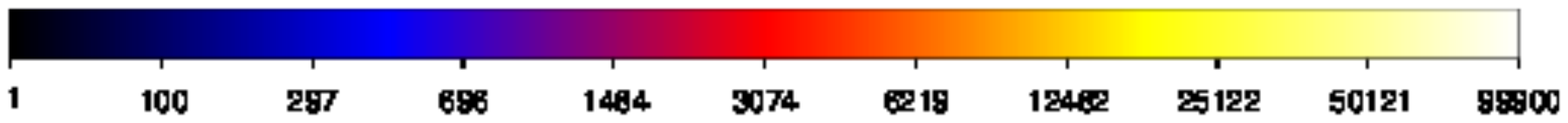}
\hspace{4cm}
\includegraphics[width =40mm, bb=0 0 525 85]{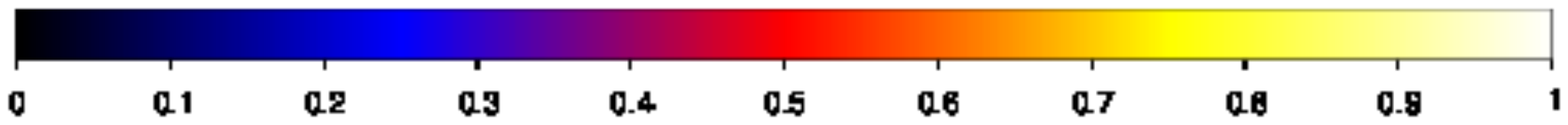}\\
 \end{center}
 
3\farcm0(I$_{\rm 4}$)
\begin{center}
\includegraphics[width =40mm, bb=0 0 742 743]{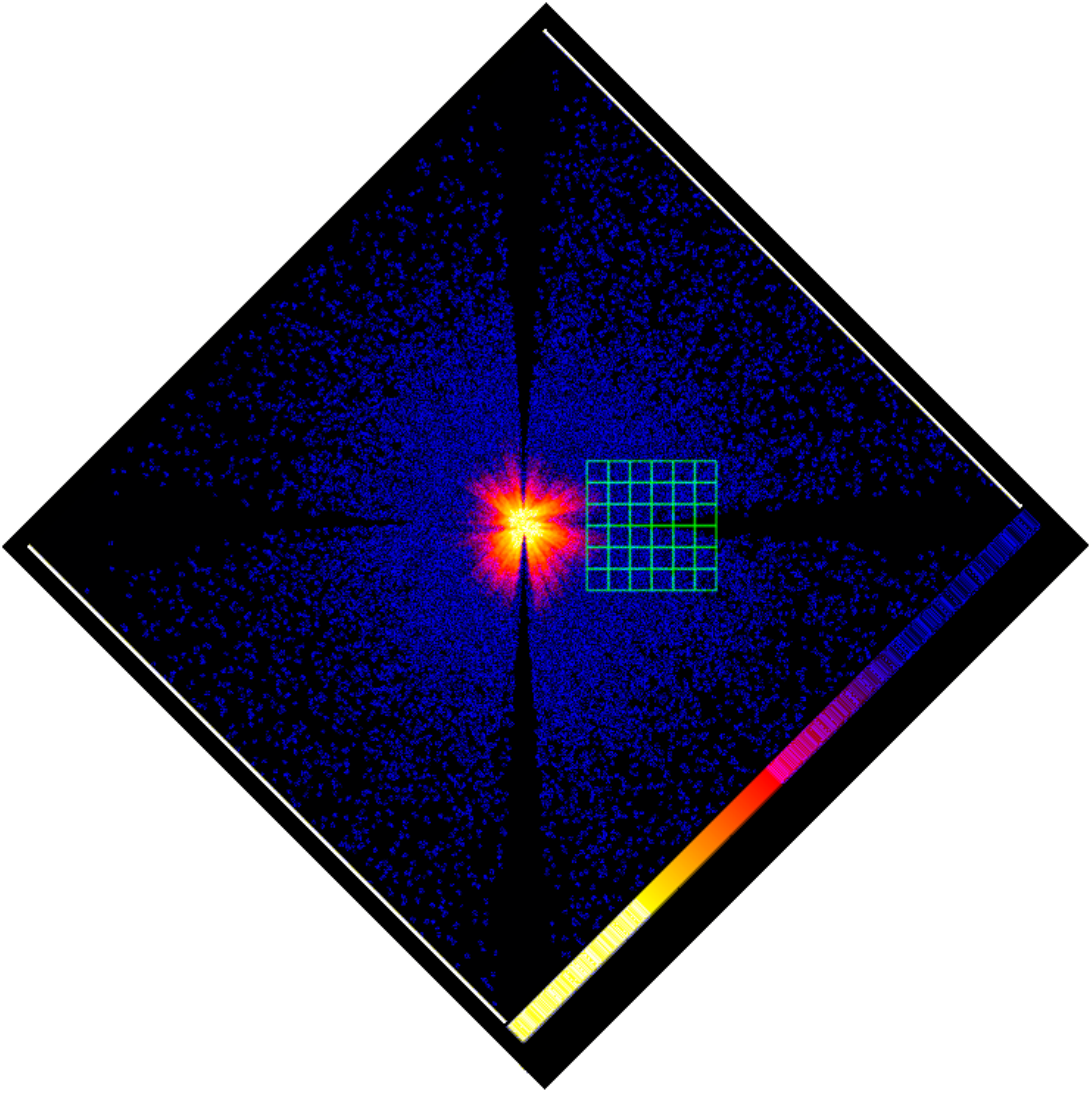}
\includegraphics[width =40mm, bb=0 0 742 743]{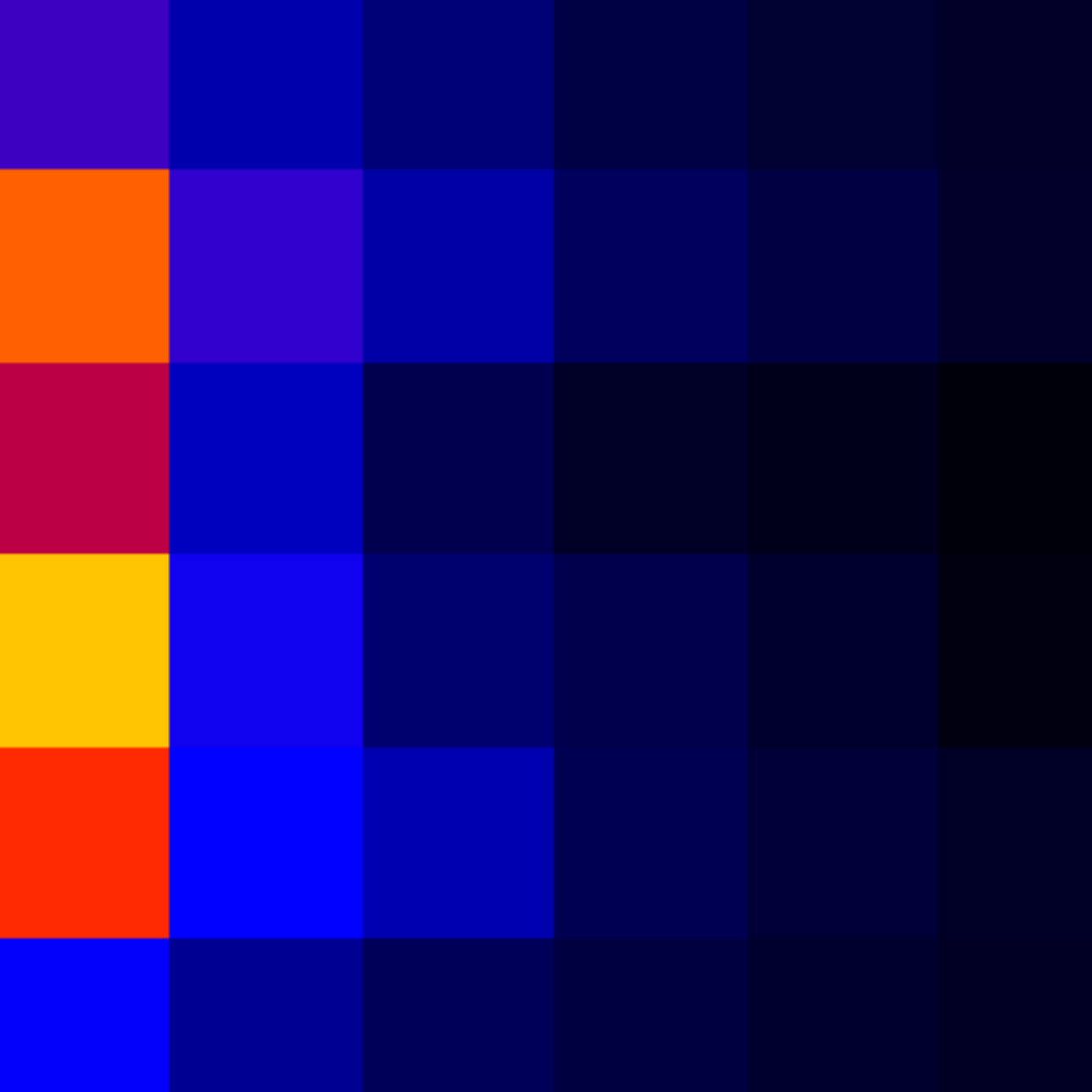}
\includegraphics[width =40mm, bb=0 0 742 743]{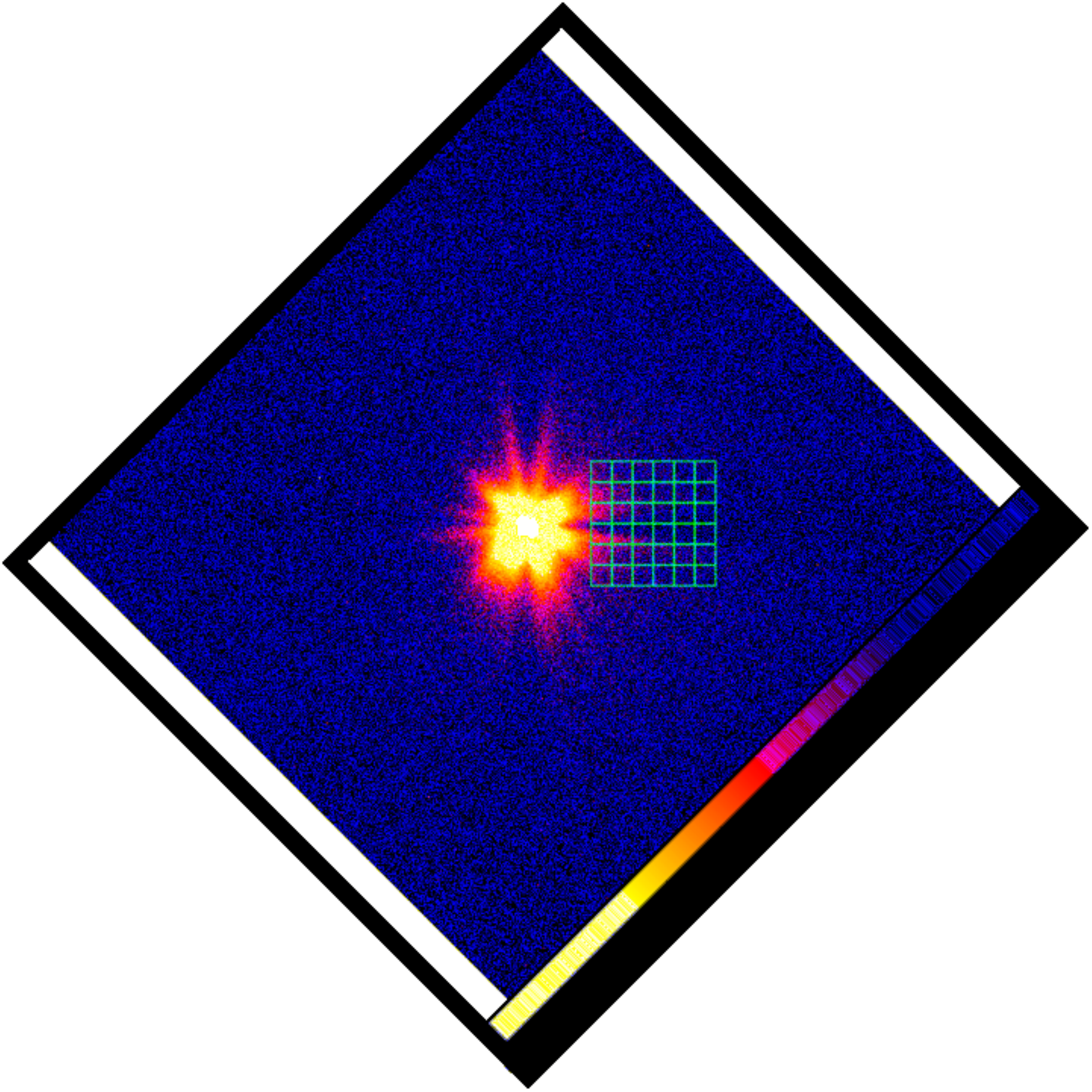}
\includegraphics[width =40mm, bb=0 0 742 743]{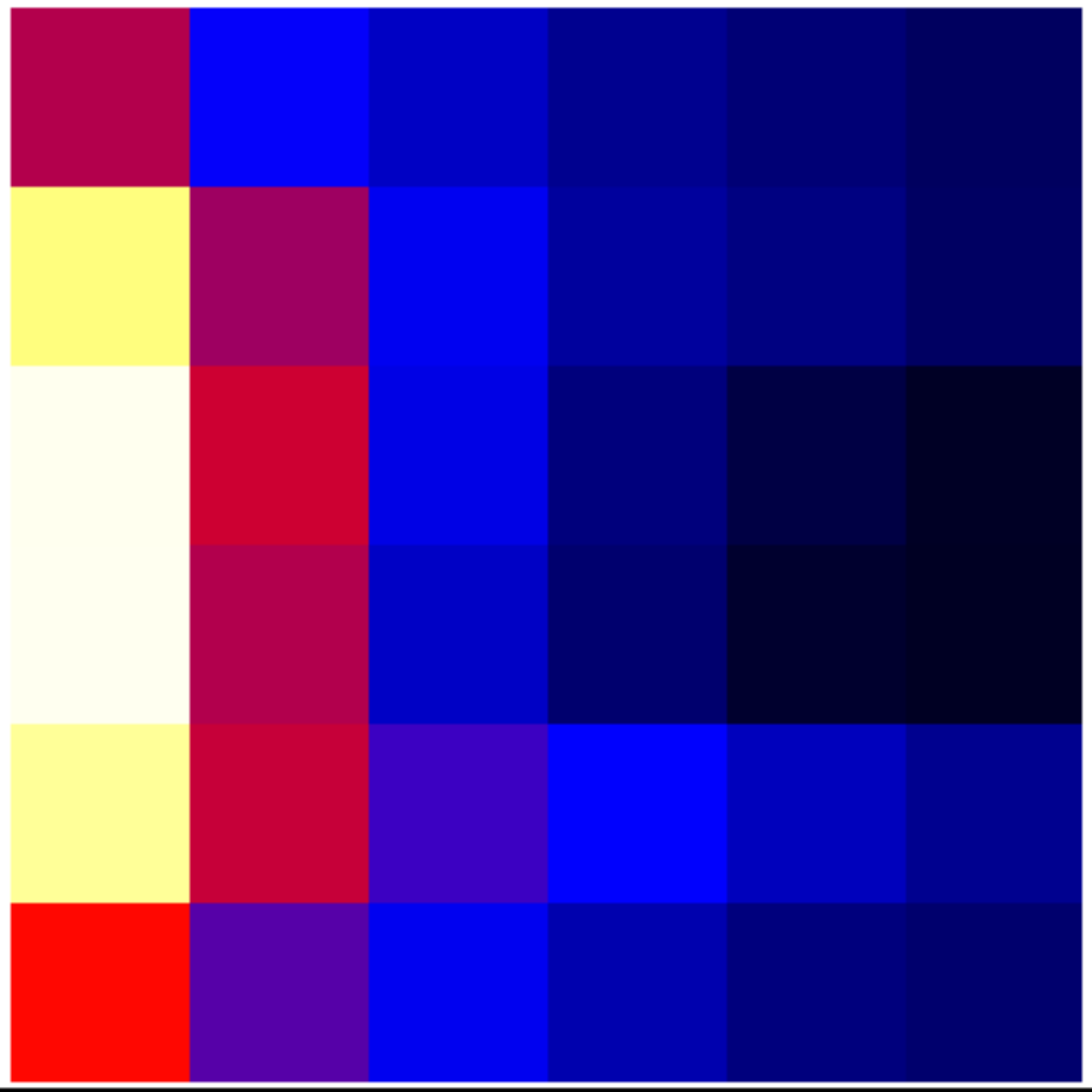}\\
\vspace*{0.3cm}
\hspace*{-1.3cm}
\includegraphics[width =40mm, bb=0 0 525 85]{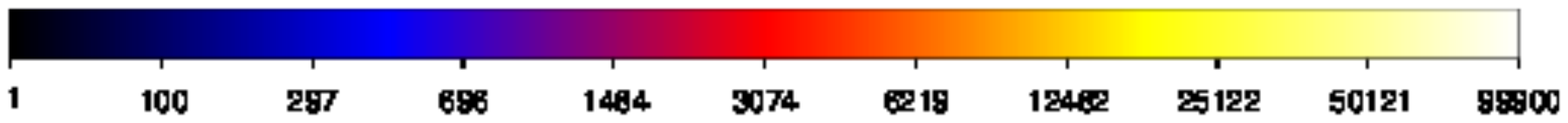}
\hspace{4cm}
\includegraphics[width =40mm, bb=0 0 525 85]{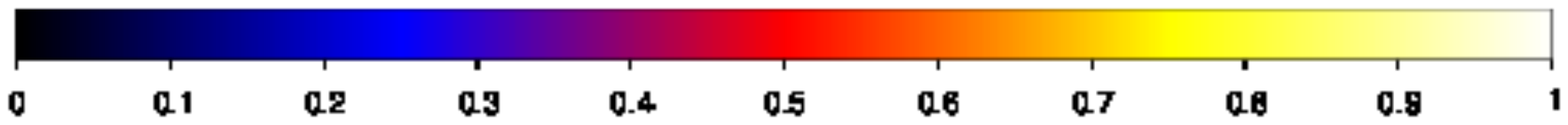}\\
-- Raytracing outputs -- \hspace{4cm} -- Ground measurements --
 \end{center}
\caption{Event distribution of a point-like source at 3\farcm0 off-axis positions. The region IDs for the
images correspond to those
defined in figure\,\ref{fig:fov_pos}. }
\label{fig:hayashi01}
\end{figure}

\begin{figure}[h]
4\farcm5(II$_{\rm 1}$)\\
 \begin{center}
\includegraphics[width =40mm, bb=0 0 742 743]{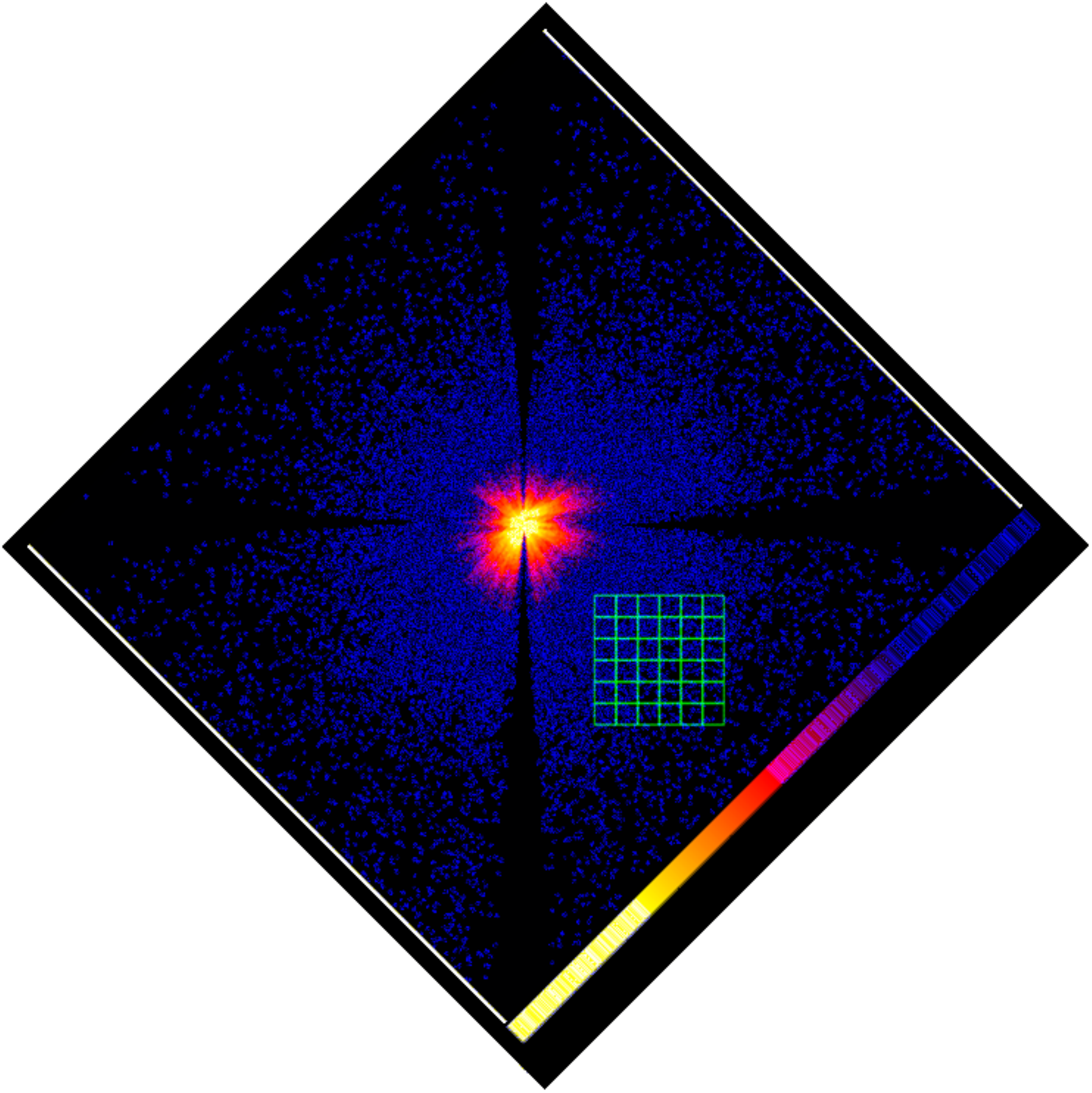}
\includegraphics[width =40mm, bb=0 0 742 743]{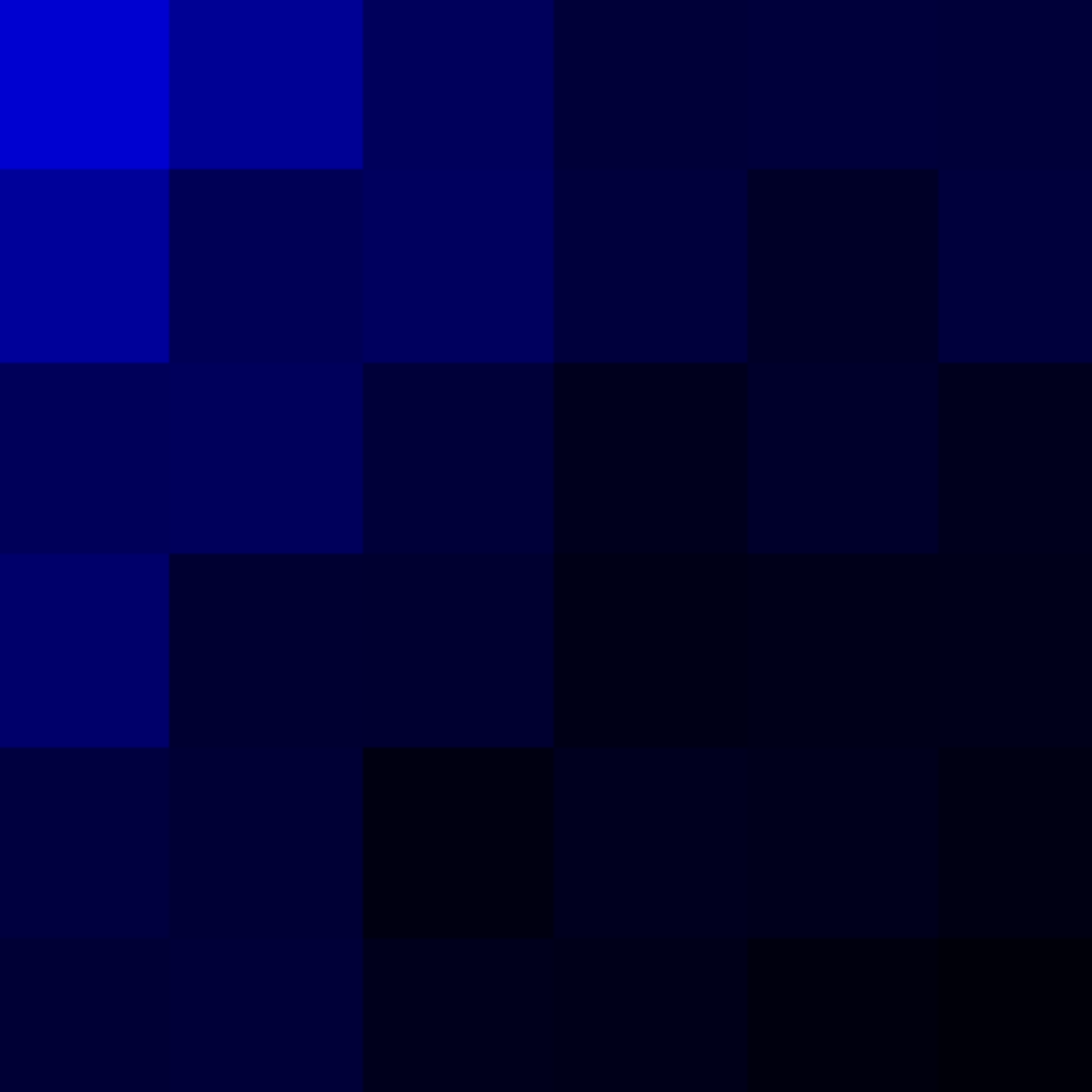}
\includegraphics[width =40mm, bb=0 0 742 743]{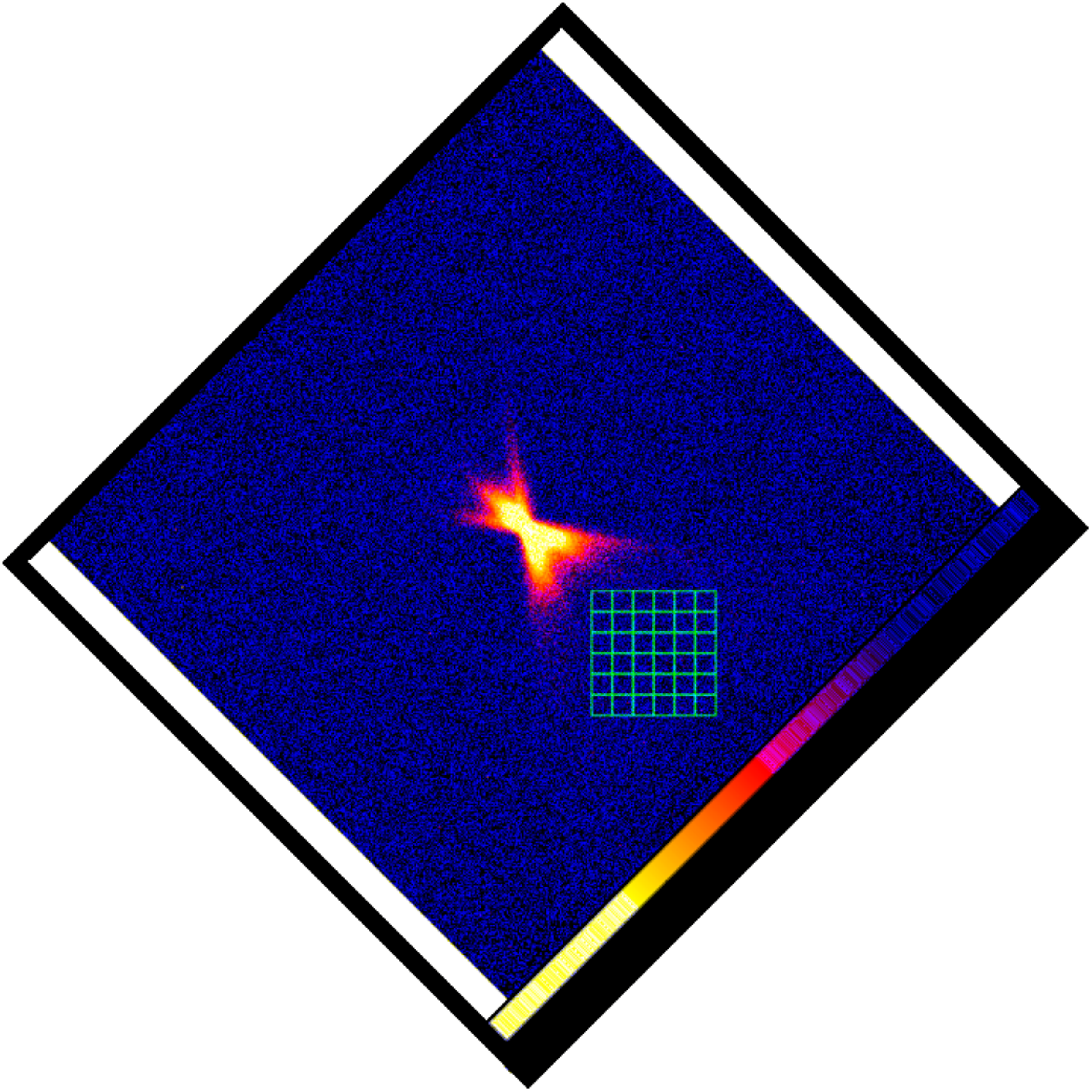}
\includegraphics[width =40mm, bb=0 0 742 743]{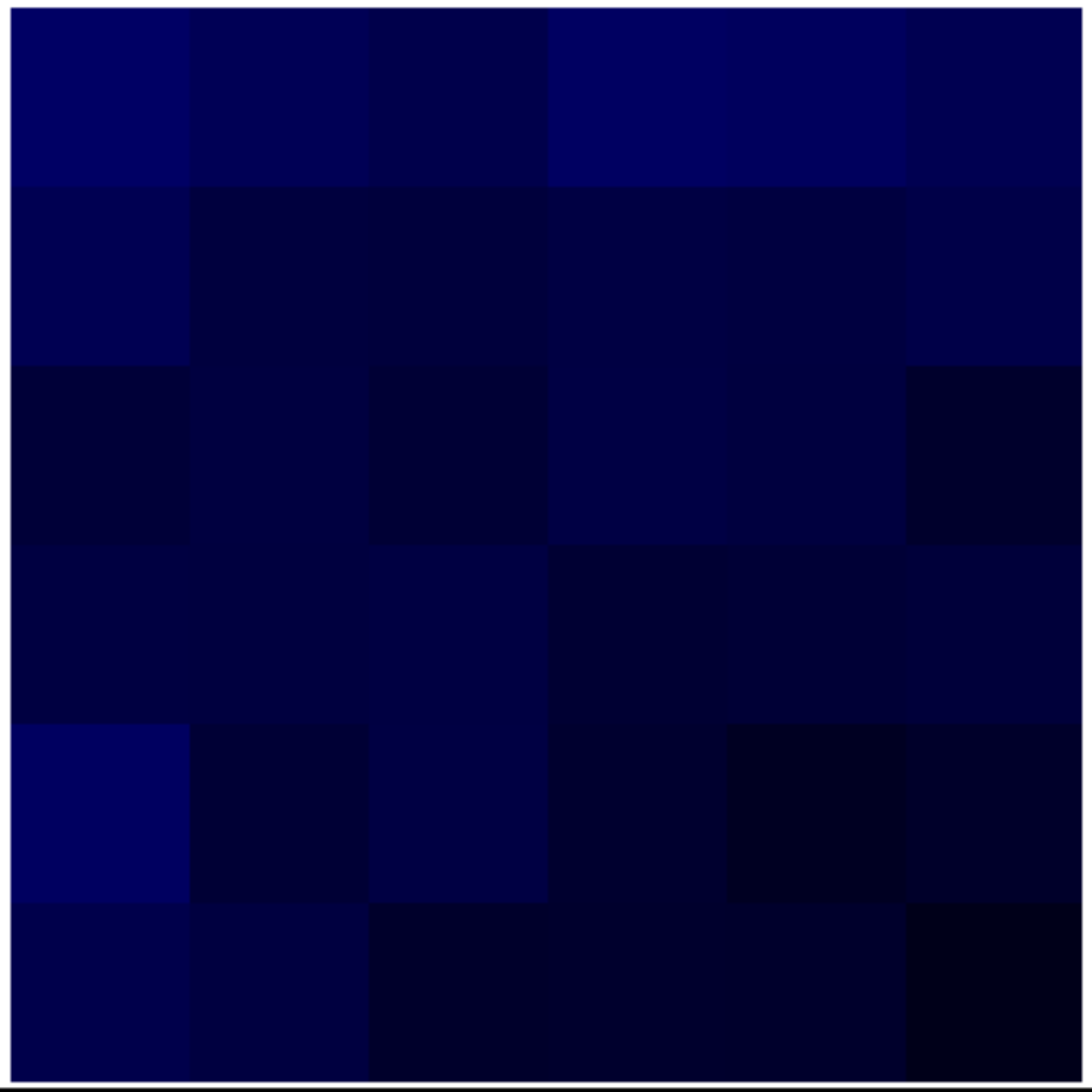}\\
\vspace*{0.3cm}
\hspace*{-1.3cm}
\includegraphics[width =40mm, bb=0 0 525 85]{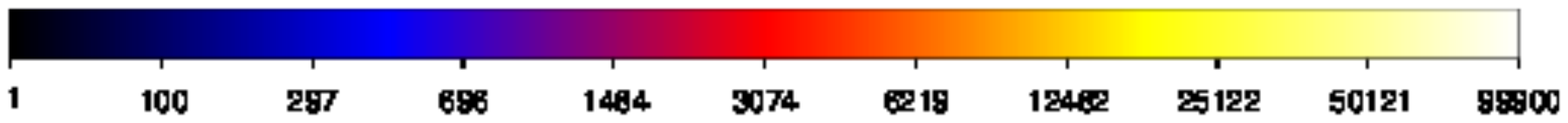}
\hspace{4cm}
\includegraphics[width =40mm, bb=0 0 525 85]{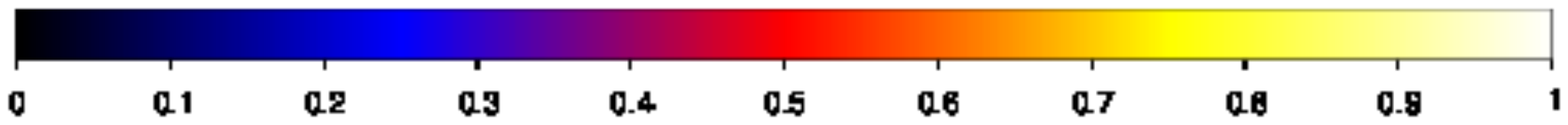}\\
 \end{center}
4\farcm5(II$_{\rm 2}$)\\
 \begin{center}
\includegraphics[width =40mm, bb=0 0 742 743]{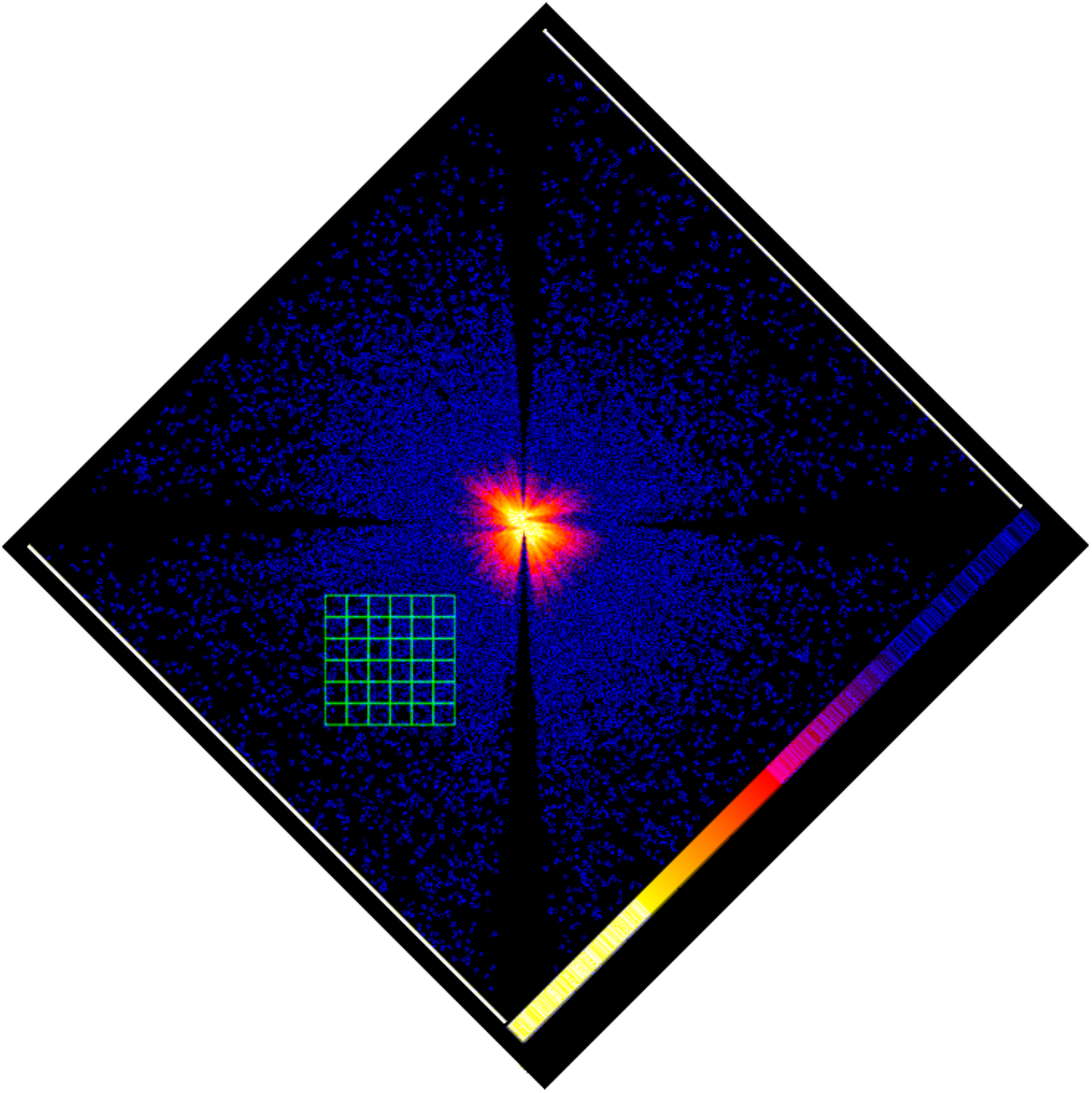}
\includegraphics[width =40mm, bb=0 0 742 743]{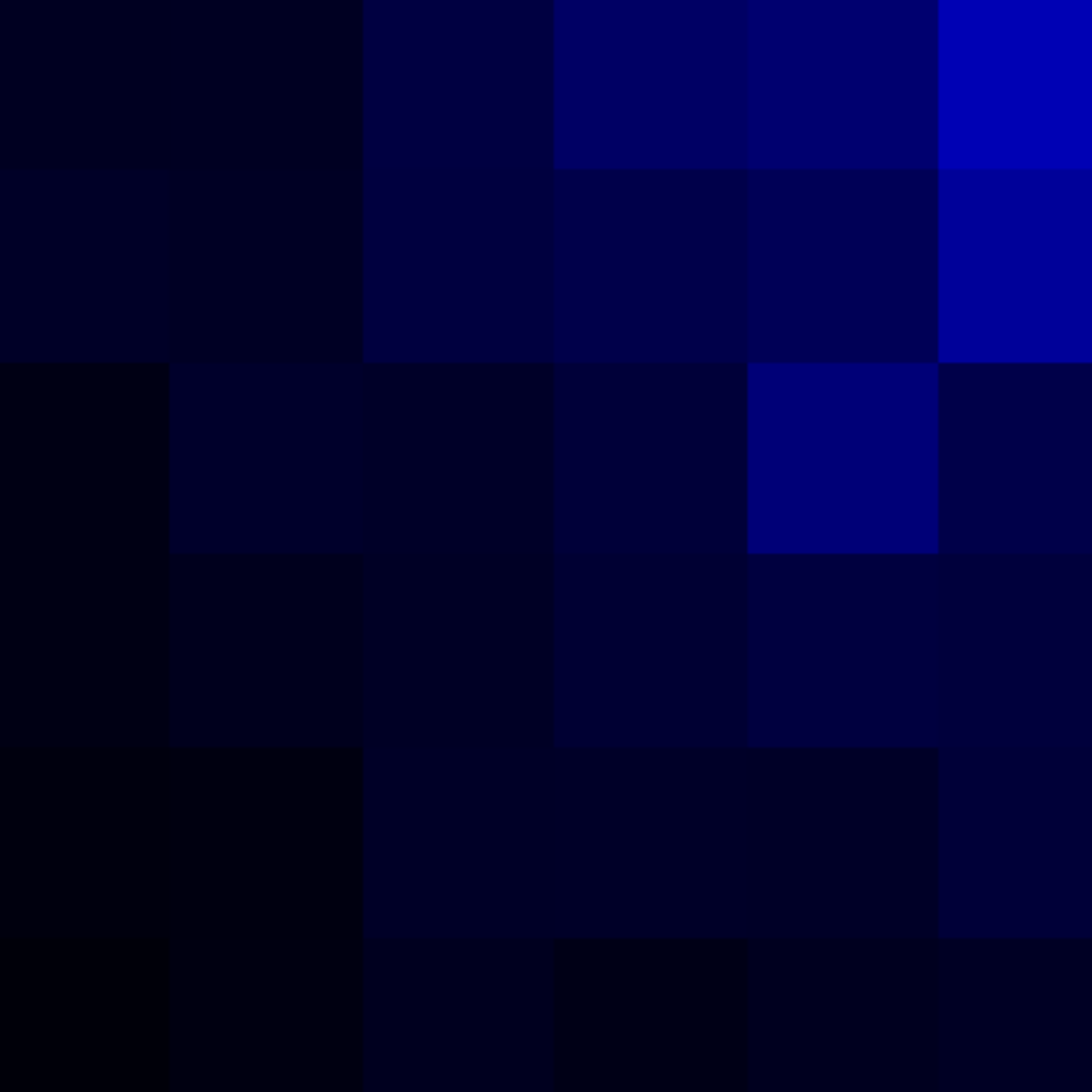}
\includegraphics[width =40mm, bb=0 0 742 743]{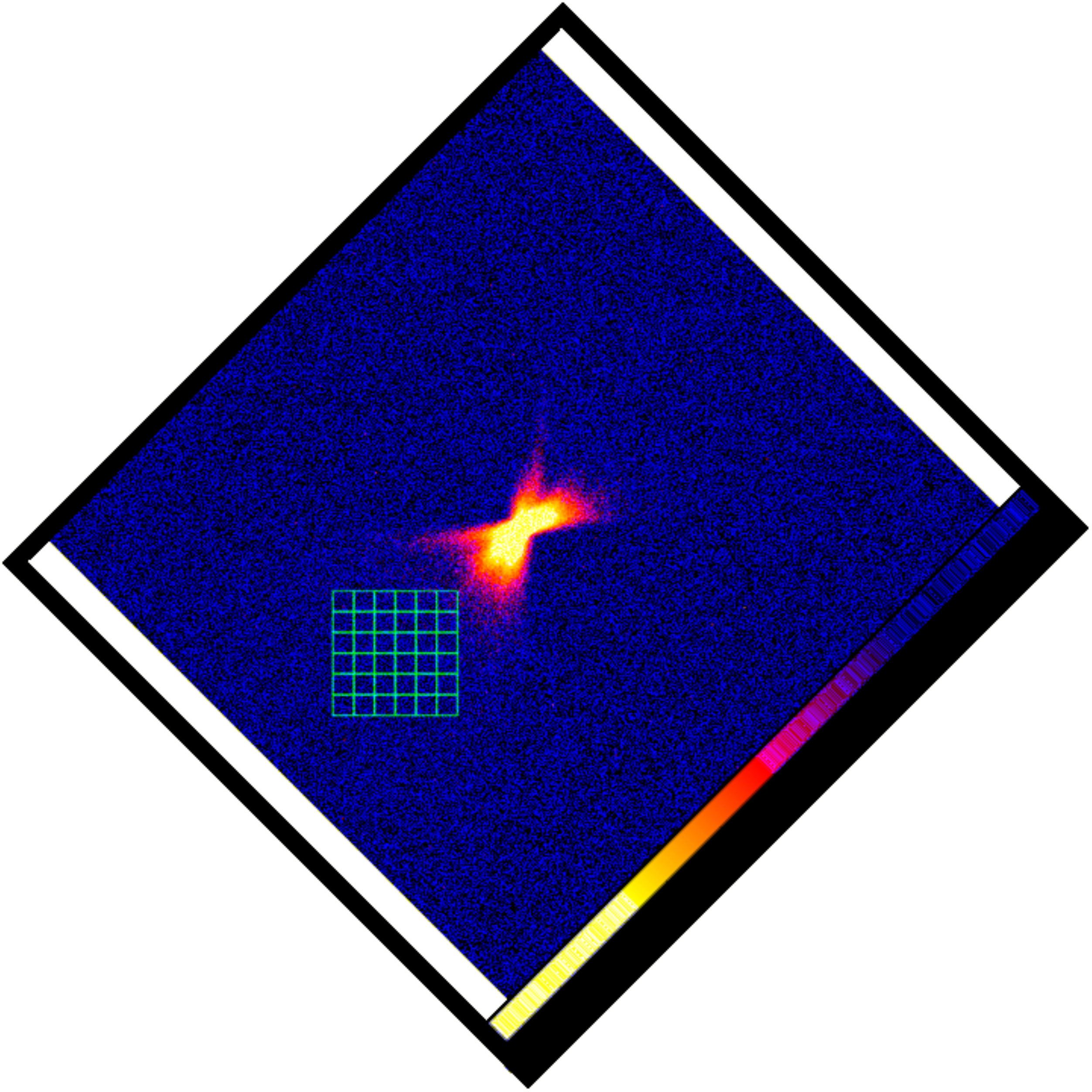}
\includegraphics[width =40mm, bb=0 0 742 743]{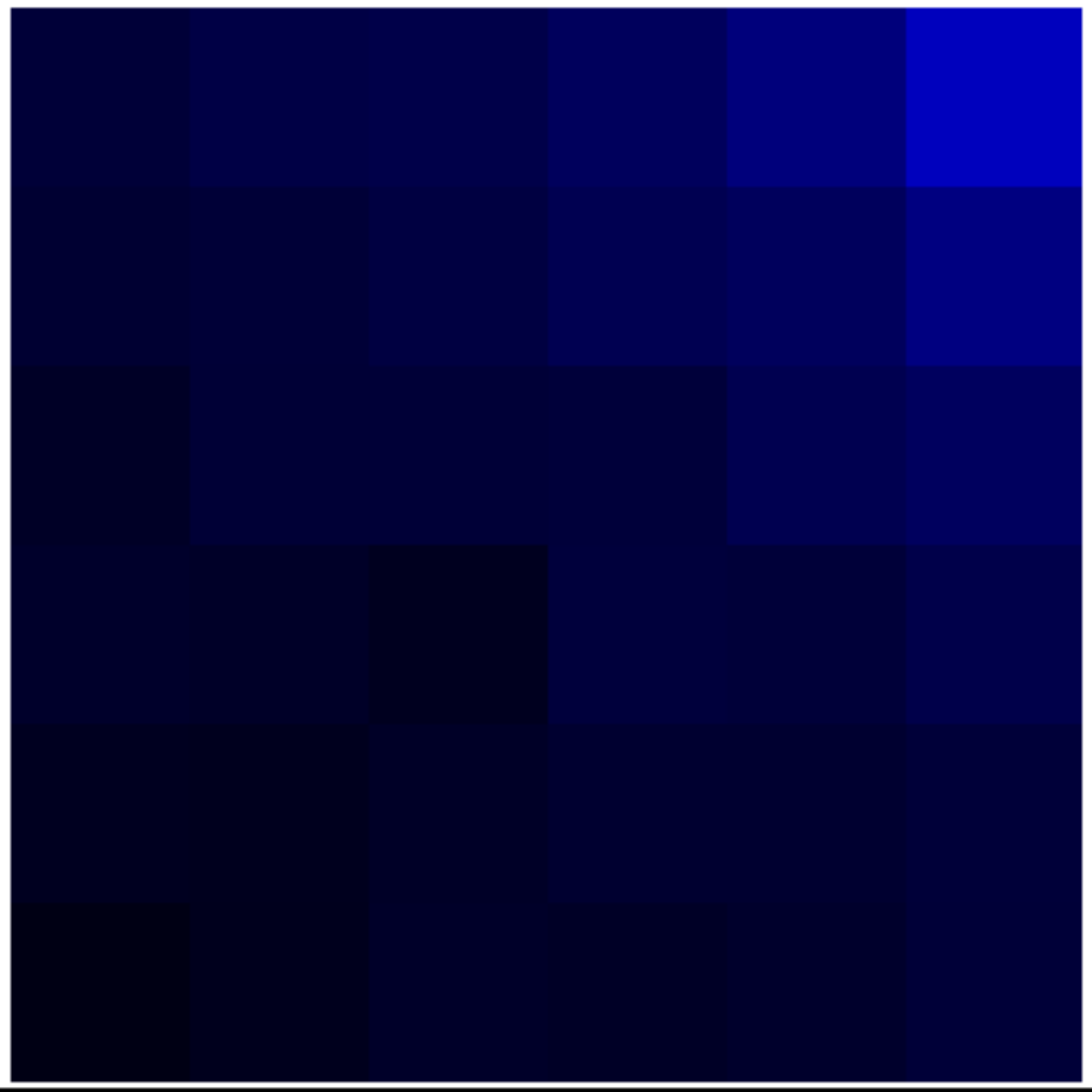}\\
\vspace*{0.3cm}
\hspace*{-1.3cm}
\includegraphics[width =40mm, bb=0 0 525 85]{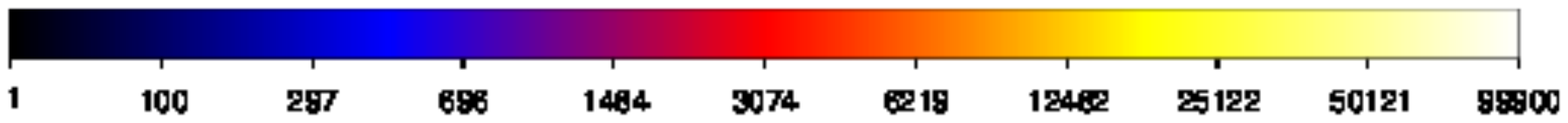}
\hspace{4cm}
\includegraphics[width =40mm, bb=0 0 525 85]{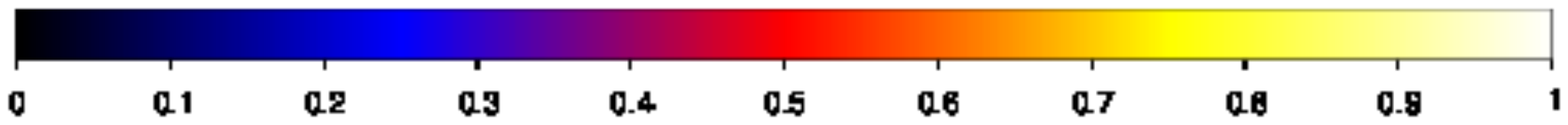}\\
 \end{center}
4\farcm5(II$_{\rm 3}$)\\
 \begin{center}
\includegraphics[width =40mm, bb=0 0 742 743]{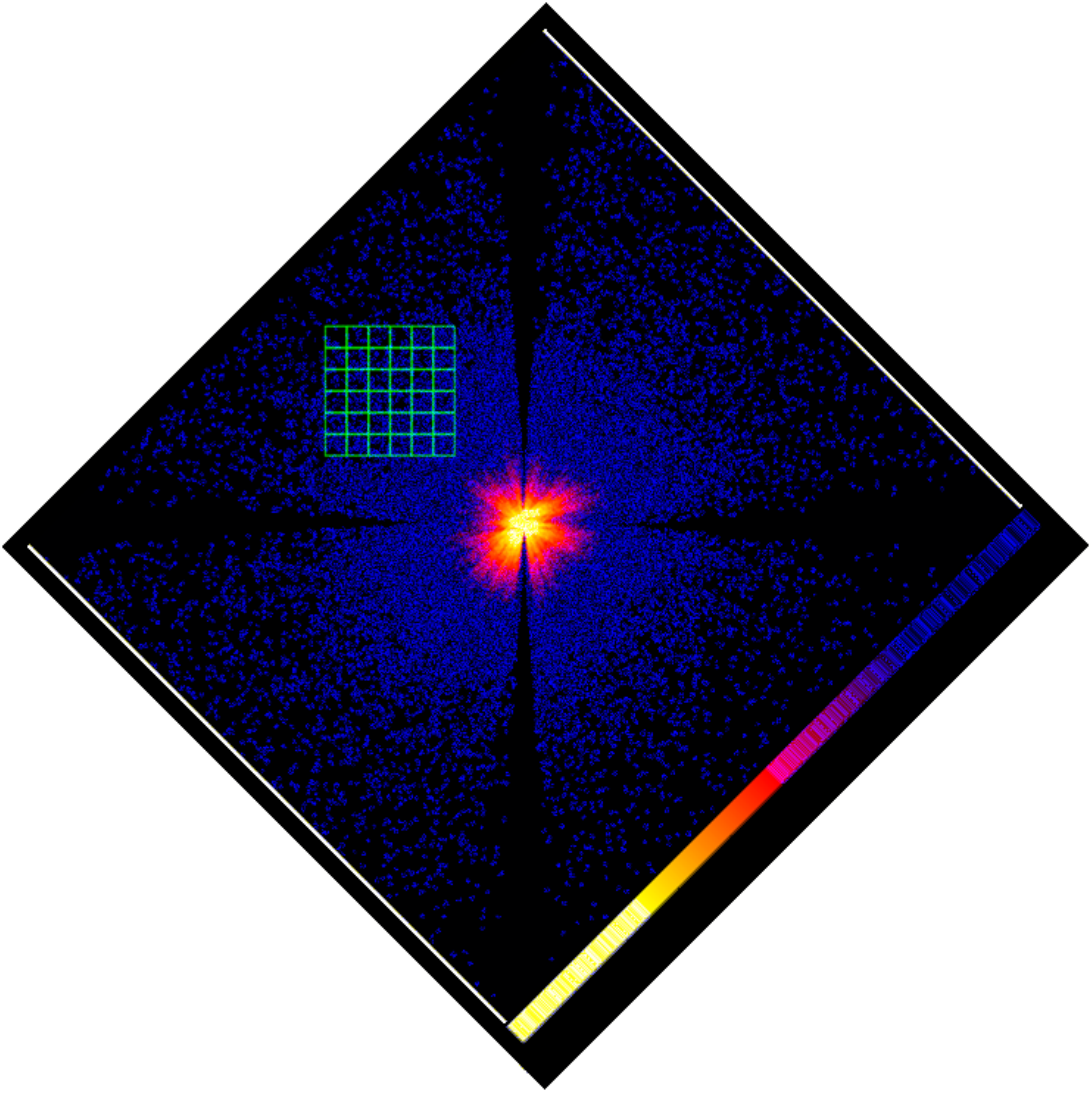}
\includegraphics[width =40mm, bb=0 0 742 743]{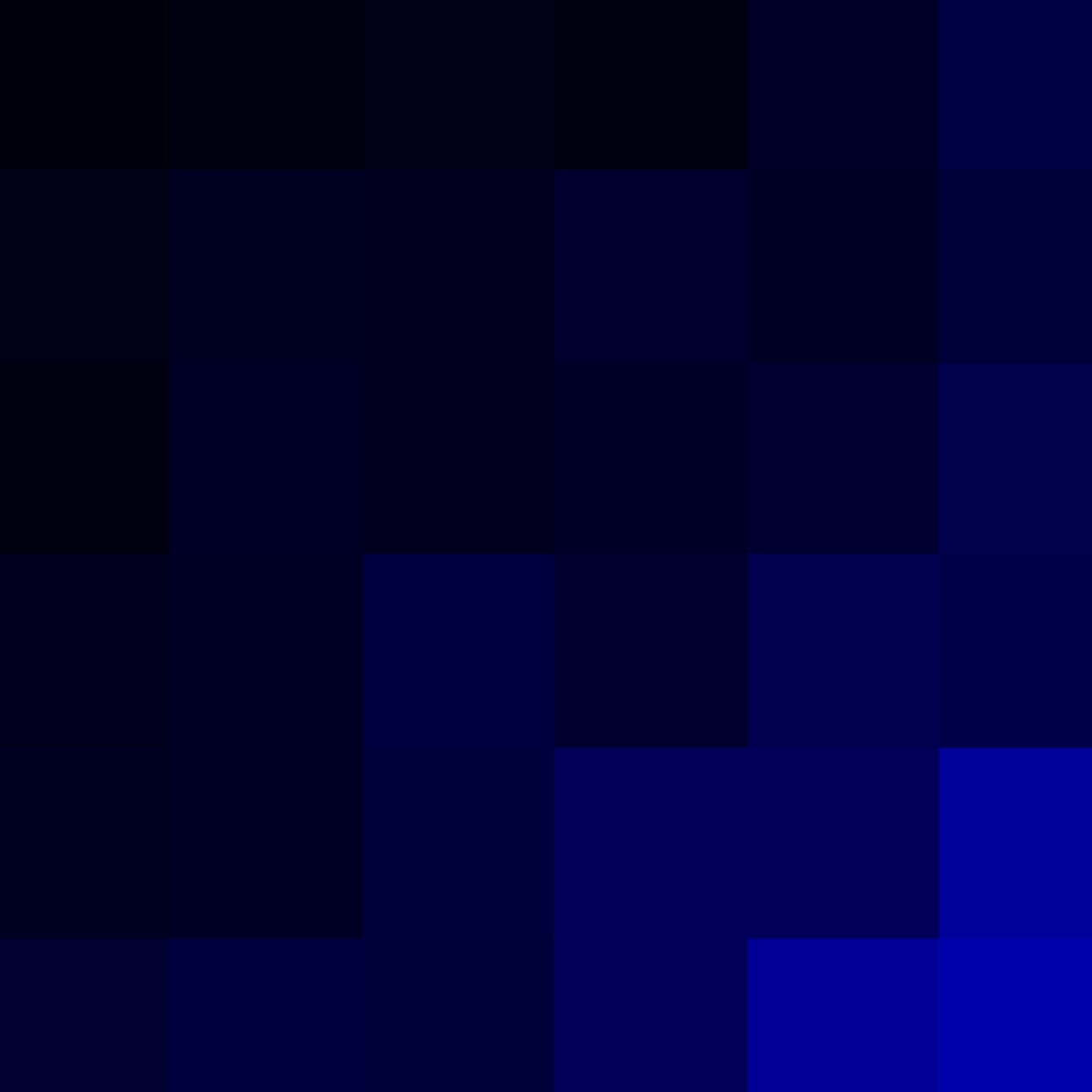}
\includegraphics[width =40mm, bb=0 0 742 743]{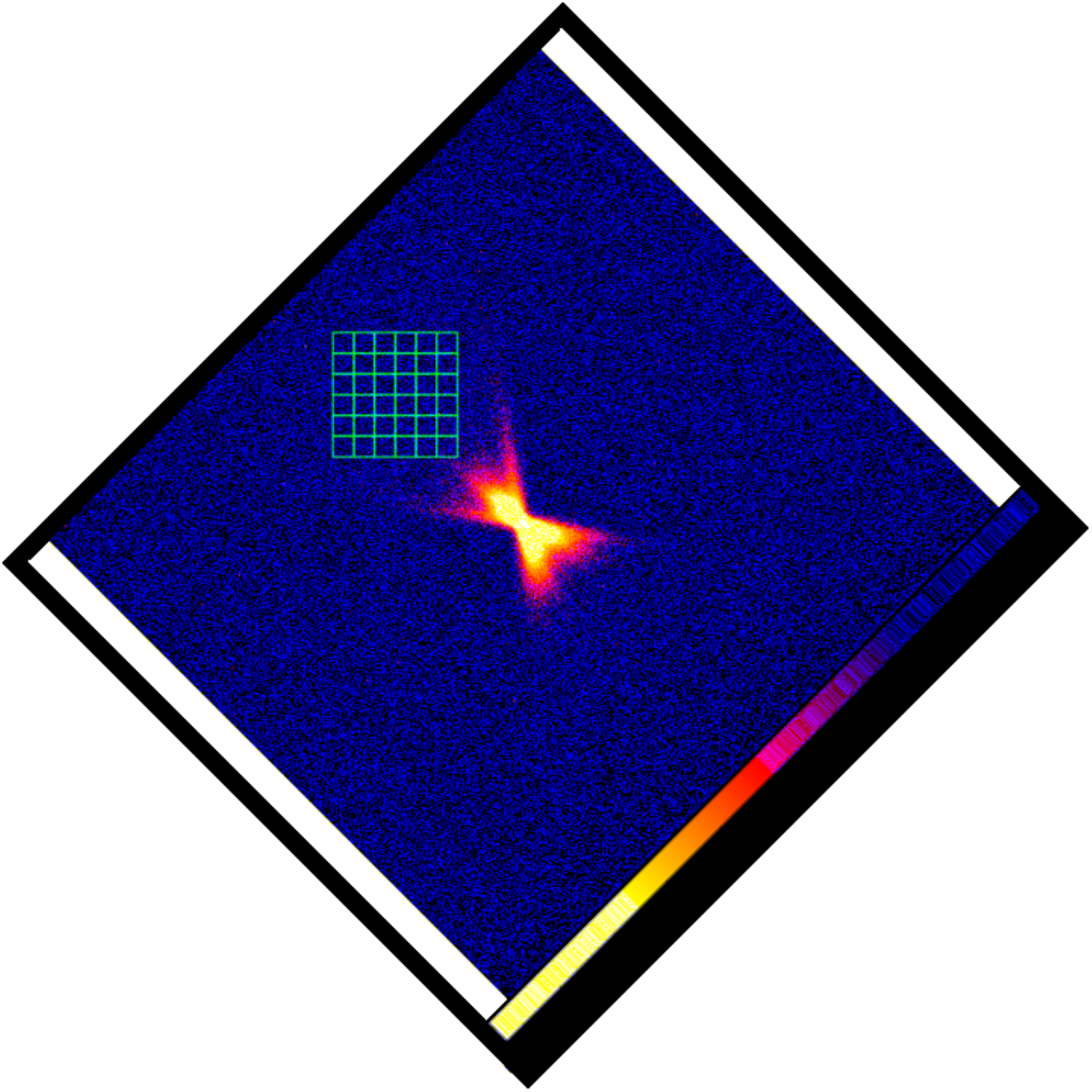}
\includegraphics[width =40mm, bb=0 0 742 743]{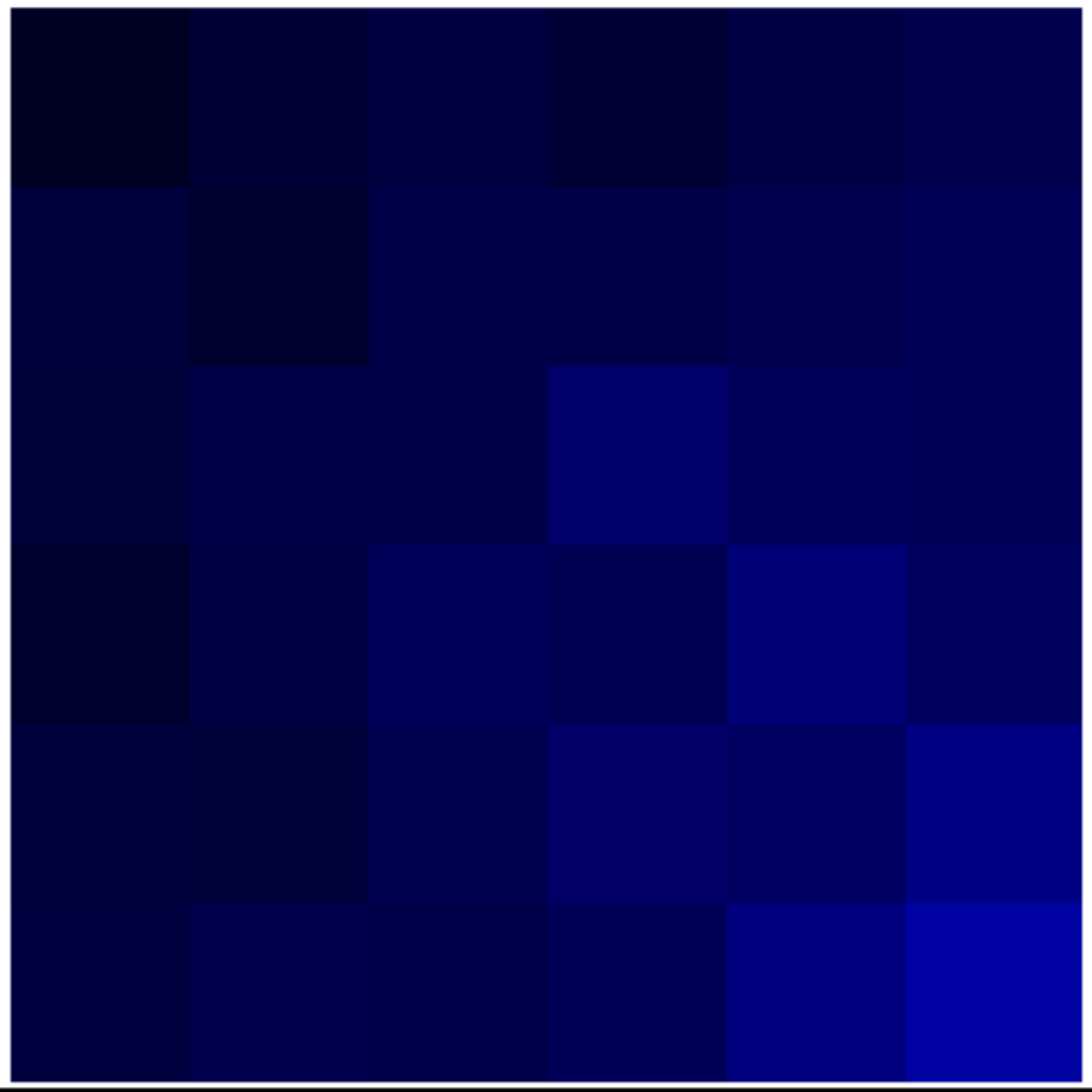}\\
\vspace*{0.3cm}
\hspace*{-1.3cm}
\includegraphics[width =40mm, bb=0 0 525 85]{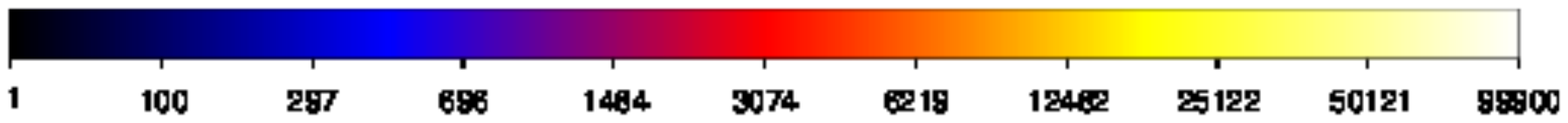}
\hspace{4cm}
\includegraphics[width =40mm, bb=0 0 525 85]{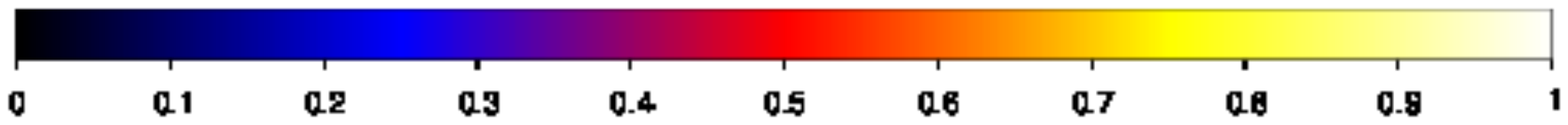}\\
 \end{center}
4\farcm5(II$_{\rm 4}$)\\
 \begin{center}
\includegraphics[width =40mm, bb=0 0 742 743]{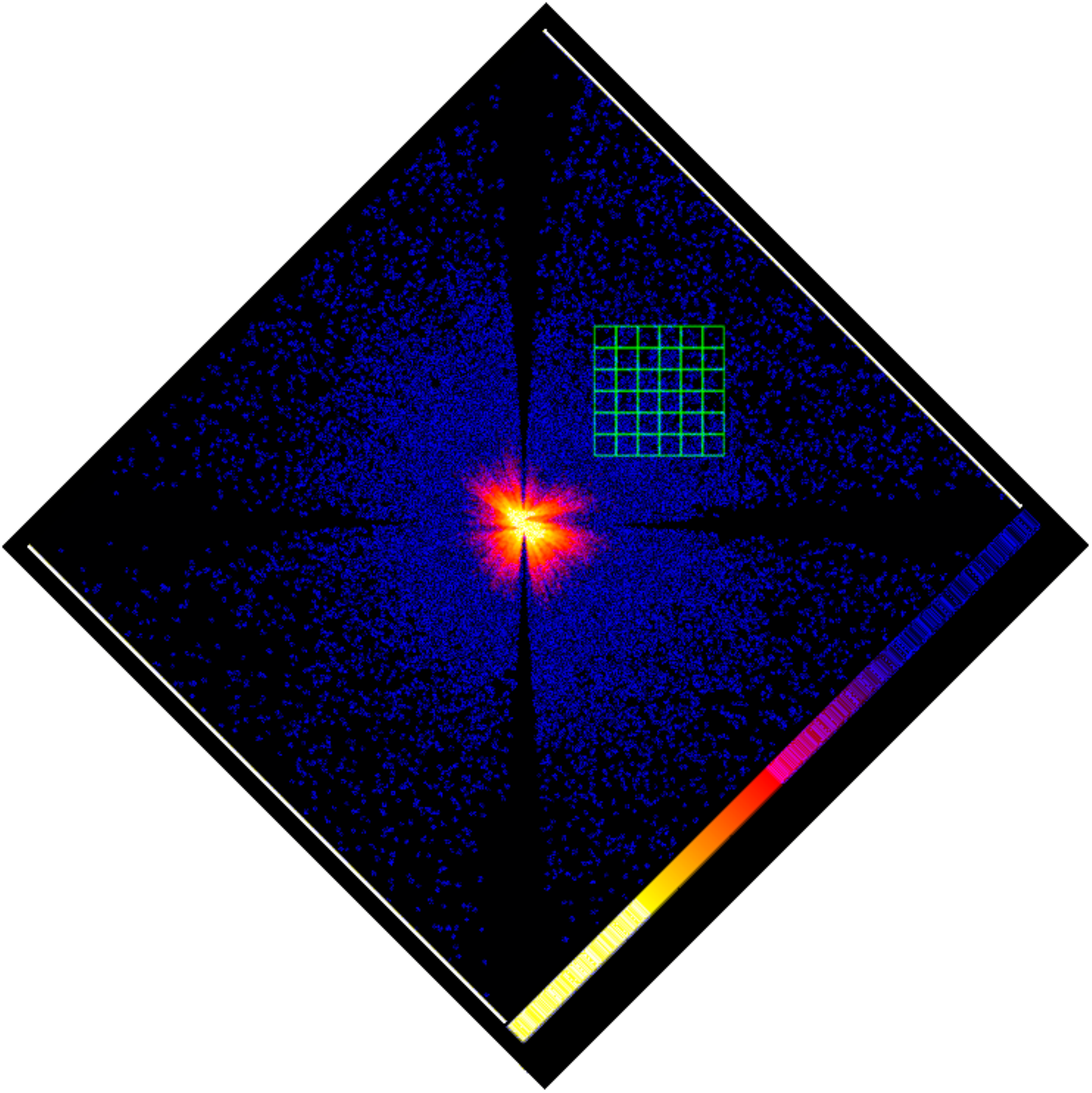}
\includegraphics[width =40mm, bb=0 0 742 743]{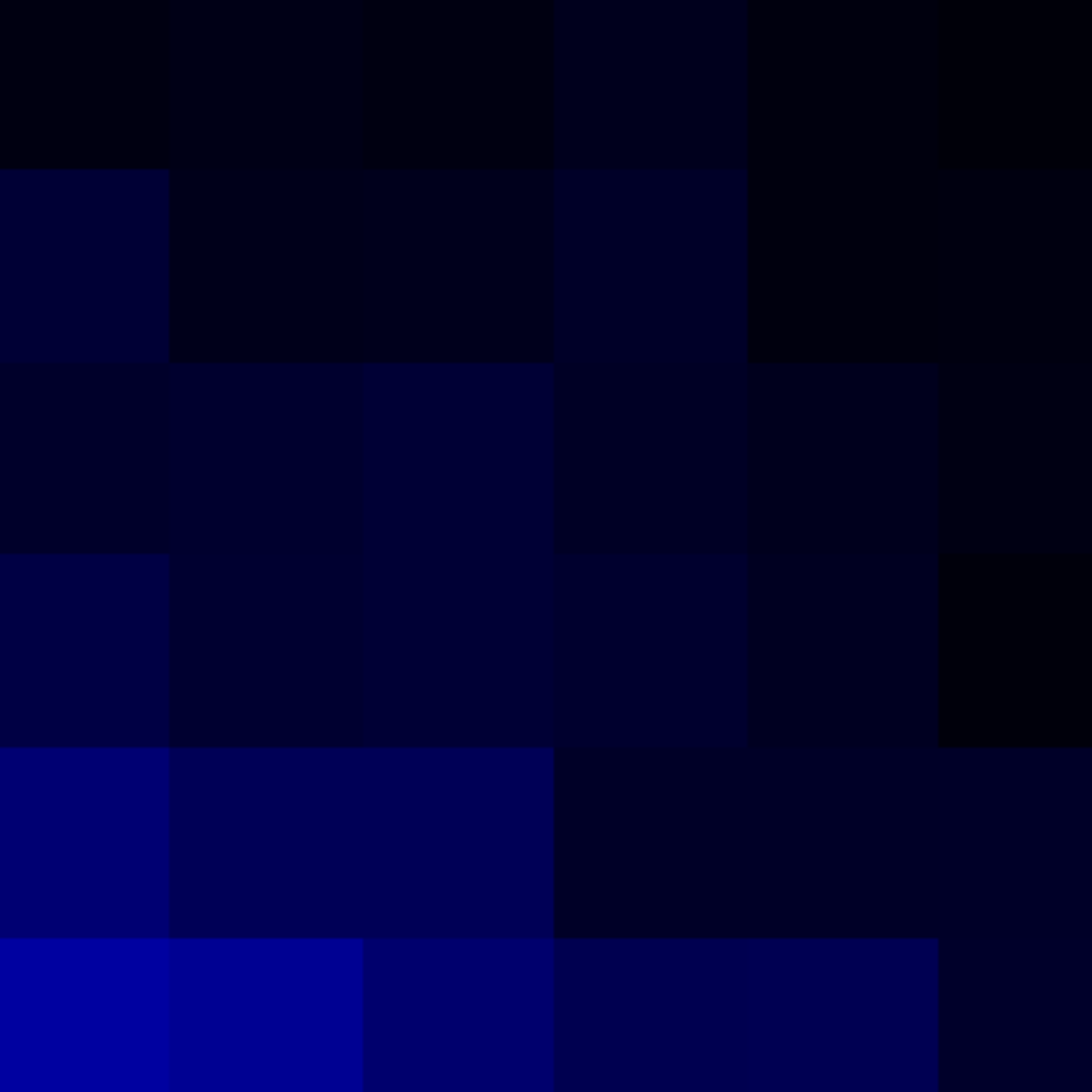}
\includegraphics[width =40mm, bb=0 0 742 743]{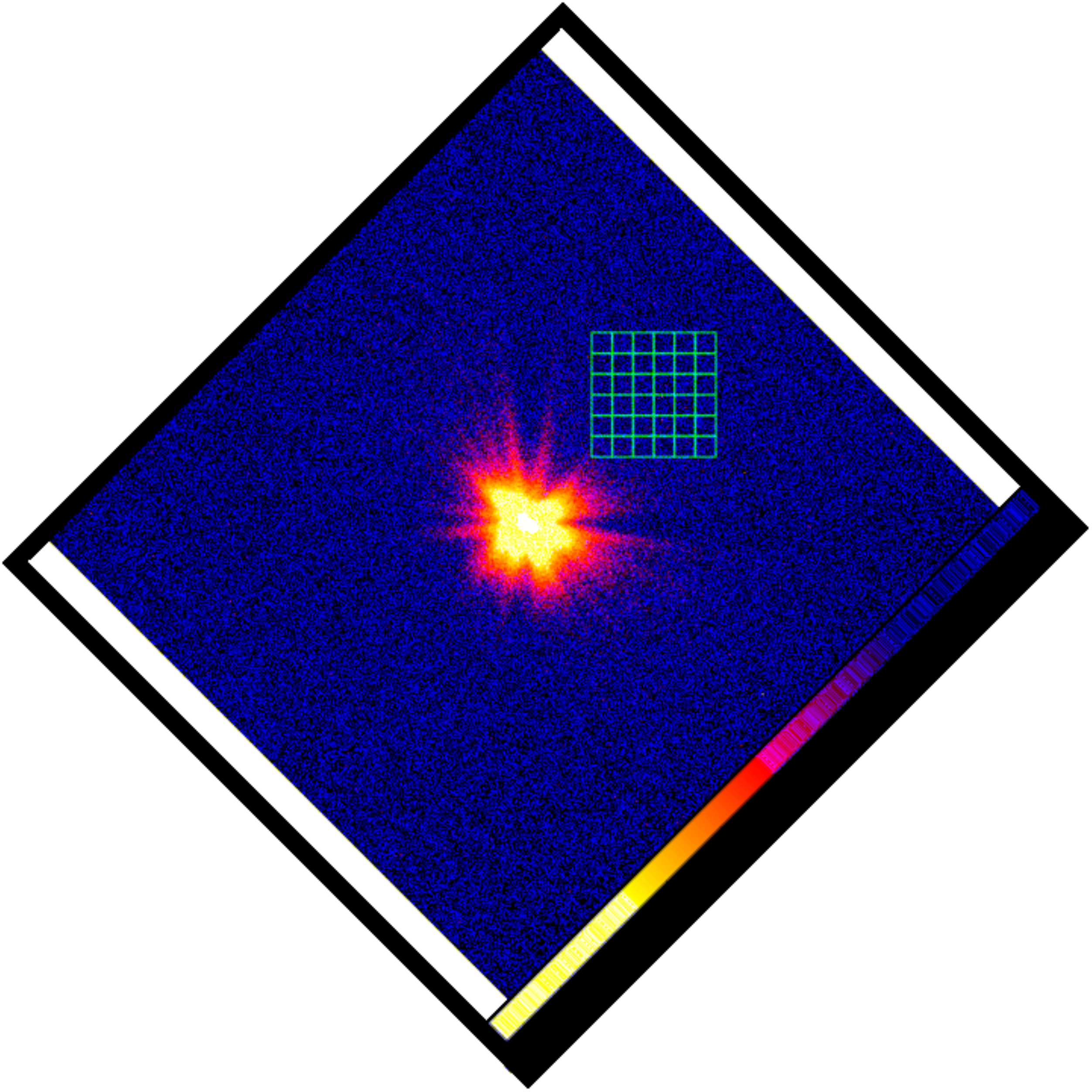}
\includegraphics[width =40mm, bb=0 0 742 743]{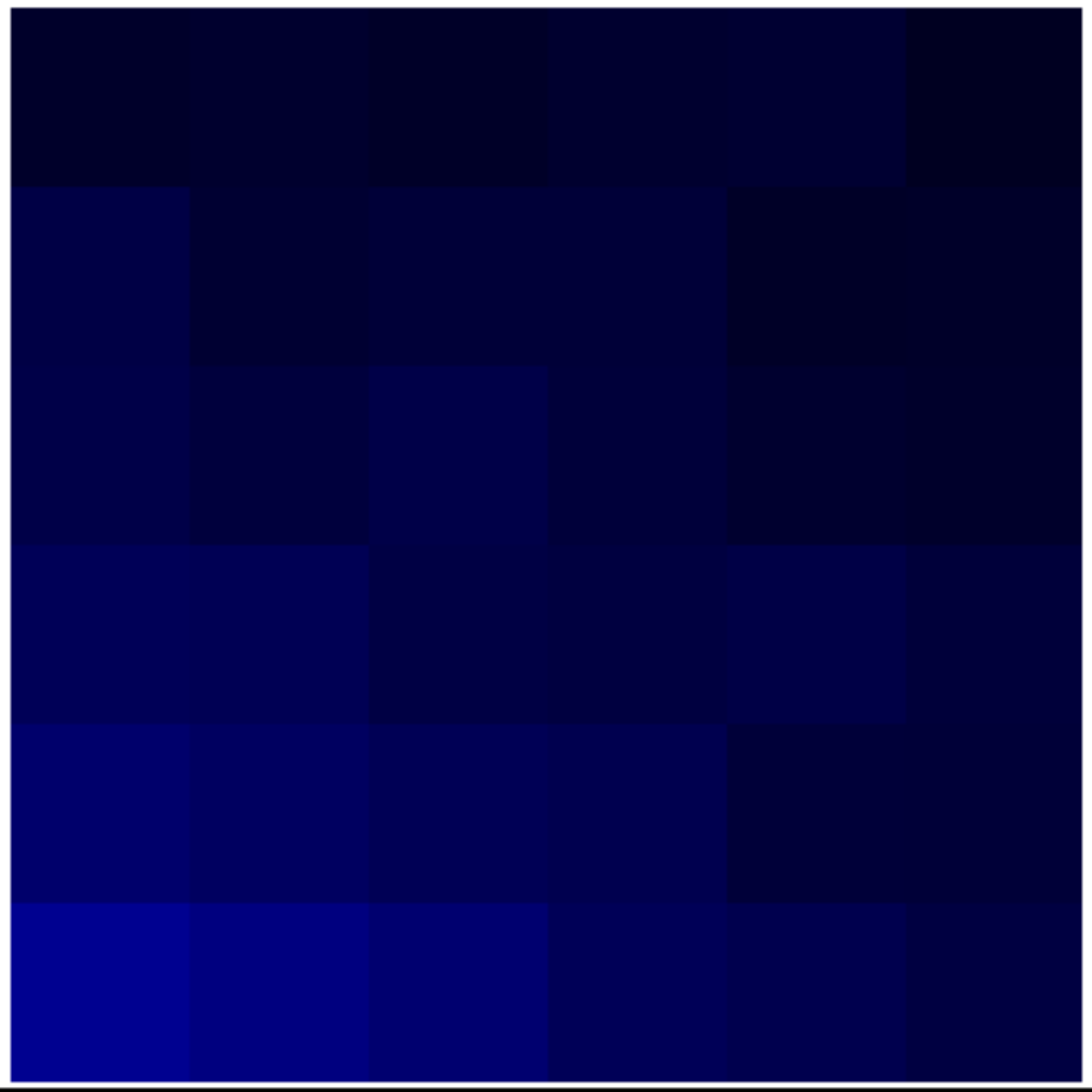}\\
\vspace*{0.3cm}
\hspace*{-1.3cm}
\includegraphics[width =40mm, bb=0 0 525 85]{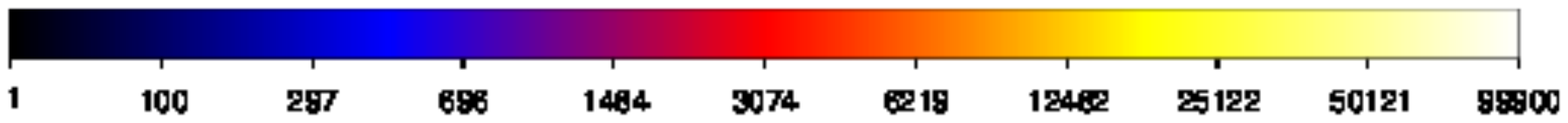}
\hspace{4cm}
\includegraphics[width =40mm, bb=0 0 525 85]{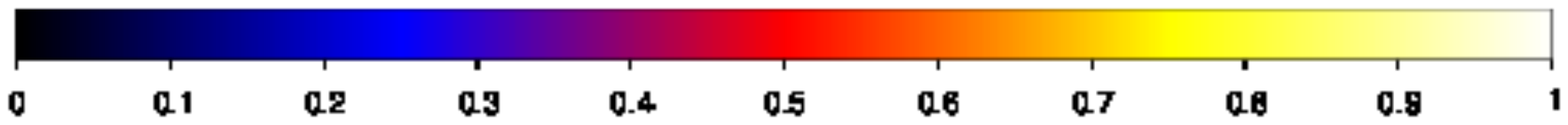}\\
-- Raytracing outputs -- \hspace{4cm} -- Ground measurements --
 \end{center}
\caption{Event distribution of the point-like source at 4\farcm5 off-axis positions. The region IDs for the
images correspond to those defined in figure\,\ref{fig:fov_pos}. }
\label{fig:hayashi02}
\end{figure}

\begin{figure}[h]
8\farcm6(III)\\
 \begin{center}
\includegraphics[width =40mm, bb=0 0 742 743]{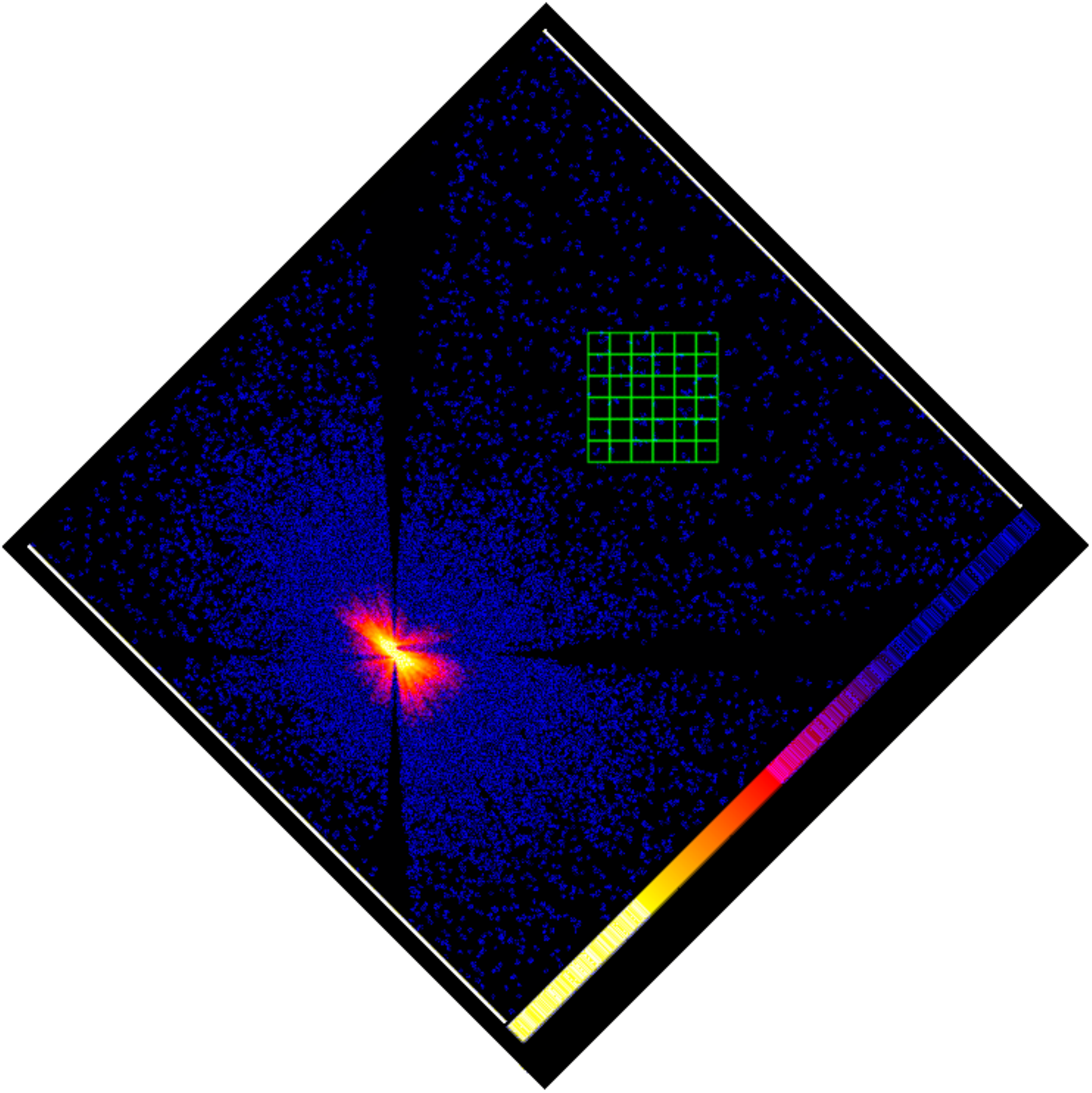}
\includegraphics[width =40mm, bb=0 0 742 743]{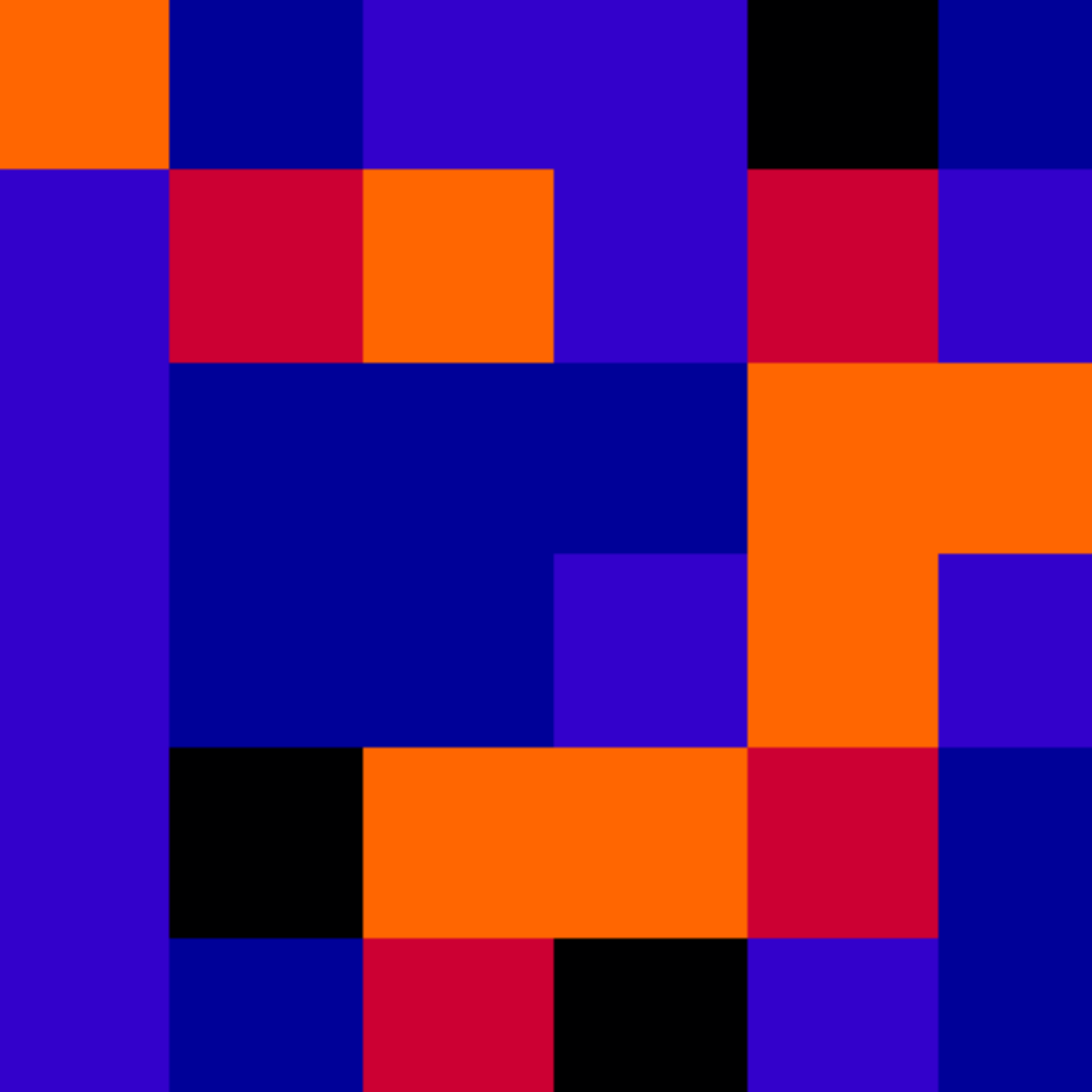}
\includegraphics[width =40mm, bb=0 0 742 743]{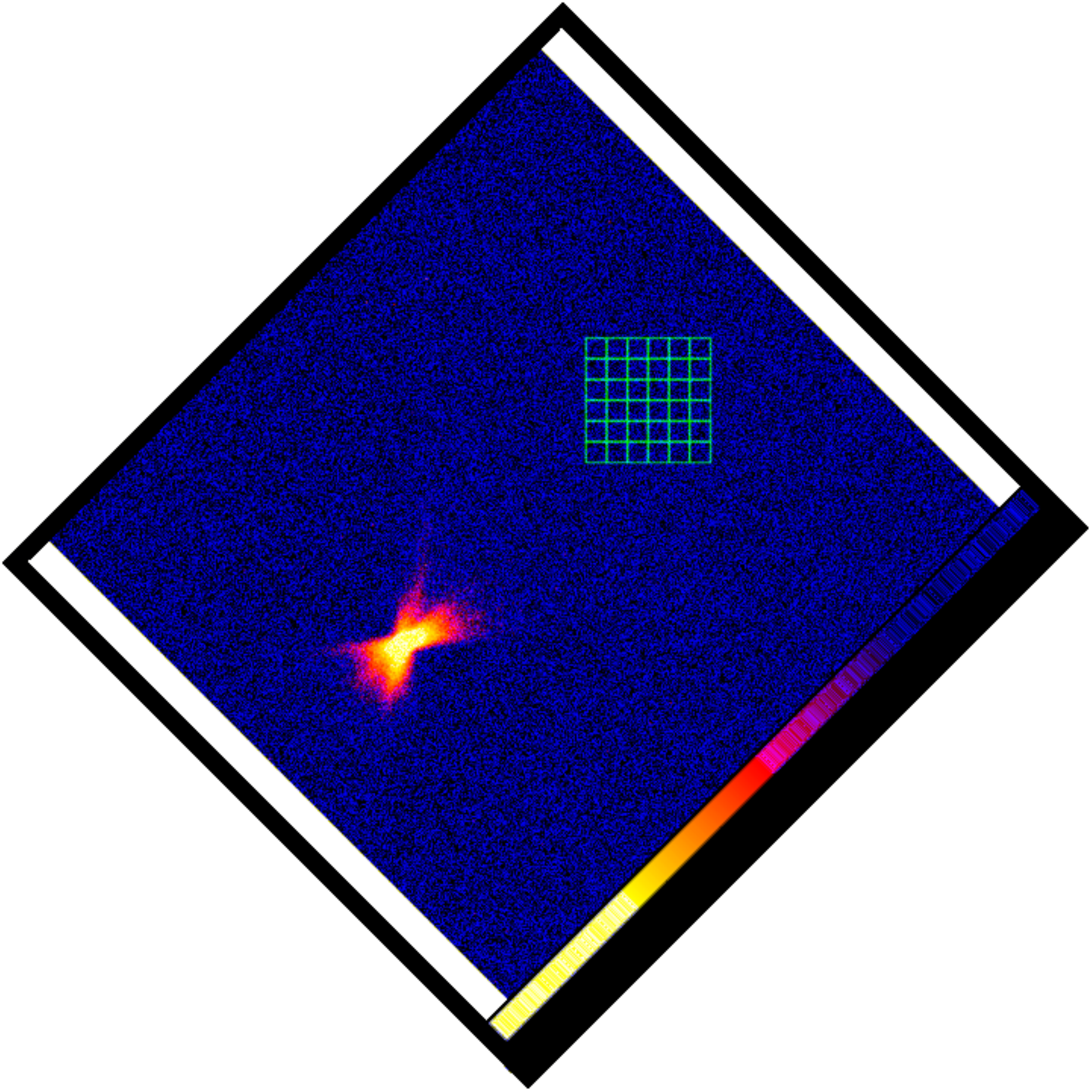}
\includegraphics[width =40mm, bb=0 0 742 743]{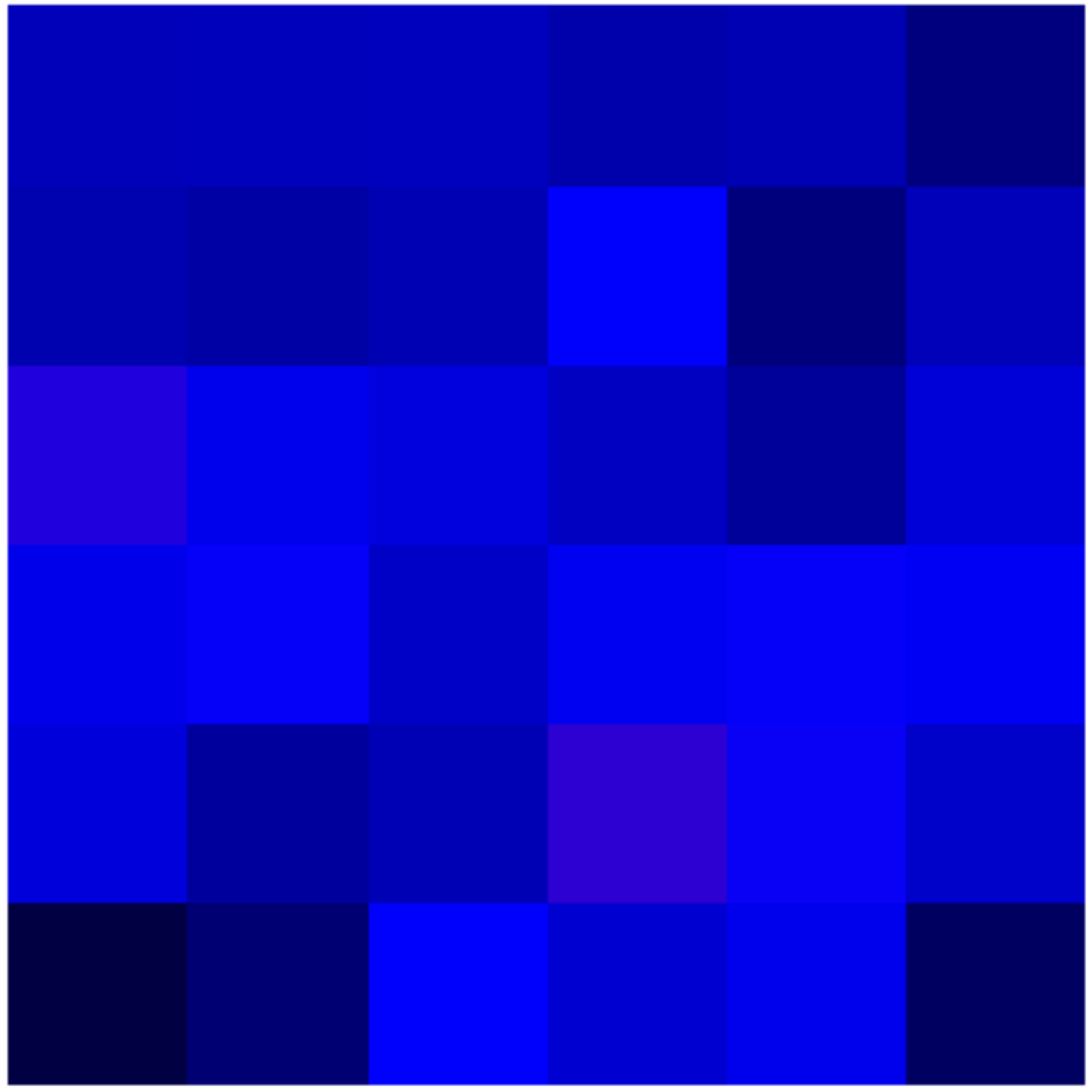}\\
\vspace*{0.3cm}
\hspace*{-1.3cm}
\includegraphics[width =40mm, bb=0 0 525 85]{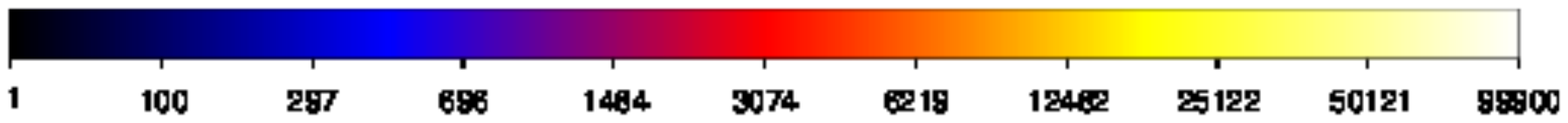}
\hspace{4.00cm}
\includegraphics[width =40mm, bb=0 0 525 85]{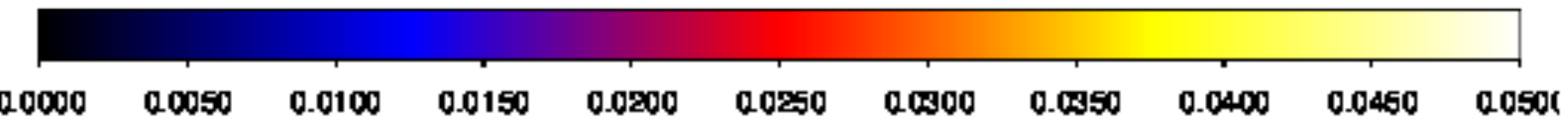}\\
-- Raytracing outputs -- \hspace{4cm} -- Ground measurements --
 \end{center}
\caption{Event distribution of the point-like source at 8\farcm6 off-axis positions. The region IDs for the
images correspond to those defined in figure\,\ref{fig:fov_pos}. }
\label{fig:hayashi03}
\end{figure}

\begin{ack}
The authors appreciate all people who contributed to the ASTRO-H mission, which made this work
possible. The authors are grateful to all the full-time
engineers and part-time workers in the GSFC/NASA laboratory
for support in  mass production of the Soft X-ray
Telescope reflectors. 

The authors thank the support from the JSPS Core-to-Core Program. The authors acknowledge all the JAXA members who have contributed to the ASTRO-H mission. All U.S. members gratefully acknowledge support 
through the NASA Science Mission Directorate.

R.I., M.T. and Y.M. acknowledge Support
from the Grants-in-Aid for Scientific Research (numbers
25870744, 25105516, 23540280, JP15H03642, JP16K05309) by the Ministry of Education,
Culture, Sports, Science and Technology, Japan., and the
Grant-in-Aid for Scientific Research on Innovative Areas Nuclear matter in neutron stars
investigated by experiments and astronomical observations.
TS is supported by the Grant-in-Aid for Japan Society for the Promotion of Science Fellows　(Grant Number 16J03448). 

\end{ack}

\bibliographystyle{pasj} 
\bibliography{sxtsbib} 

\end{document}